\documentclass[12pt]{article}
\usepackage[utf8]{inputenc}
\usepackage{amsmath,amsfonts,amssymb}
\usepackage{psfrag}
\usepackage{enumerate}
\usepackage{mathrsfs}
\usepackage{graphicx}
\usepackage{wrapfig}
\usepackage{xcolor}
\usepackage{caption}
\usepackage{amsthm}
\usepackage{subfig}
\usepackage{xcolor}
\usepackage{esint}

\usepackage{comment}

\usepackage{jheppub}
\allowdisplaybreaks
 \newcommand{\be}{\begin{equation}}
\newcommand{\beq}{\begin{equation}}
 \newcommand{\ee}{\end{equation}}
 \newcommand{\bea}{\begin{align}}
 
 \newcommand{\eea}{\end{align}}

 \def\pg#1{\textcolor{blue}{#1}}
\newcommand{\HL}[1]{{\textcolor{magenta}{#1}}}

\def\nn{\nonumber}

\global\long\def\mA{\mathcal{A}}%

\global\long\def\mC{\mathcal{C}}%
 
\global\long\def\mD{\mathcal{D}}%

\global\long\def\mF{\mathcal{F}}%
 
\global\long\def\mG{\mathcal{G}}%

\global\long\def\mI{\mathcal{I}}%
 
\global\long\def\mJ{\mathcal{J}}%
 
\global\long\def\mK{\mathcal{K}}%
 
\global\long\def\mL{\mathcal{L}}%
 
\global\long\def\mM{\mathcal{M}}%

\global\long\def\mO{\mathcal{O}}%
 
\global\long\def\mP{\mathcal{P}}%

\global\long\def\mT{\mathcal{T}}%

\global\long\def\e{\epsilon}%
 
\global\long\def\ra{\rightarrow}%

\global\long\def\avg#1{\left\langle #1\right\rangle }%

\global\long\def\f#1#2{\frac{#1}{#2}}%
 
\global\long\def\del{\partial}%
 
\global\long\def\t{\theta}%
 
\global\long\def\a{\alpha}%
 
\global\long\def\b{\beta}%
 
\global\long\def\g{\gamma}%
 
\global\long\def\G{\Gamma}%
 
\global\long\def\s{\sigma}%
 
\global\long\def\r{\rho}%
 
\global\long\def\d{\delta}%
 
\global\long\def\Tr{\text{Tr}}%

\global\long\def\ket#1{\left|#1\right\rangle }%
 
\global\long\def\N{\mathbb{N}}%
 \global\long\def\bT{\mathbb{T}}%
\global\long\def\Z{\mathbb{Z}}%

\global\long\def\C{\mathbb{C}}%
 
\global\long\def\p{\varphi}%

\global\long\def\T{\text{T}}%
 
\global\long\def\w{\omega}%
 
\global\long\def\D{\Delta}%

\global\long\def\S{\Sigma}%

\global\long\def\l{\ell}%

\global\long\def\sD{\mathscr{D}}%

\global\long\def\app{\approx}%
\global\long\def\sgn{\text{sgn}}%

\def\b{{\beta}}
\newcommand\vev[1]{{\ensuremath{\left\langle{#1}\right\rangle}}}
\newcommand{\bega}{\begin{gather}}
\newcommand{\eega}{\end{gather}}

\newcommand\al{{\alpha}}
\newcommand\ep{\epsilon}

\newcommand\lam{\lambda}

\newcommand\om{\omega}

\newcommand\vp{\varphi}

\newcommand{\ben}{\begin{enumerate}}
\newcommand{\een}{\end{enumerate}}
\newcommand{\bca}{\begin{cases}}
\newcommand{\eca}{\end{cases}}

\def\rt{{\mathfrak t}}

\newcommand\half{{\ensuremath{\frac{1}{2}}}}
\newcommand\ov{\over}
\newcommand\ha{{\half}}
\def\ri{\right}

\preprint{MIT-CTP/5511}

\title{An effective field theory for non-maximal quantum chaos}

\author{Ping Gao and Hong Liu}%

\affiliation{Center for Theoretical Physics,\\ Massachusetts Institute of Technology,
Cambridge, MA 02139, USA}

\emailAdd{pgao@mit.edu}
\emailAdd{hong\_liu@mit.edu}

\abstract{In non-maximally quantum chaotic systems, the exponential behavior of out-of-time-ordered correlators (OTOCs) results from summing over exchanges of an infinite tower of higher ``spin'' operators. We construct an effective field theory (EFT) to capture these exchanges in $(0+1)$ dimensions. The EFT generalizes the one for maximally chaotic systems, and reduces to it in the limit of maximal chaos. The theory predicts the general structure of OTOCs both at leading order in the $1/N$ expansion ($N$ is the number of degrees of freedom), and after resuming over an infinite number of higher order $1/N$ corrections. These general results agree with those previously explicitly obtained in specific models. We also show that the general structure of the EFT can be extracted from the large $q$ SYK model.
  
}

\begin{document}

\maketitle

\section{Introduction}



Information injected into a small subsystem of a quantum many-body system eventually spreads under time evolution across the entire system. Such scrambling of quantum information can be described in terms of growth of operators 
under Heisenberg evolution. More explicitly, consider a quantum mechanical system with $N$ degrees of freedom
and few-body interactions among them. The growth of operators can be 
probed by the so-called out-of-time-ordered-correlators (OTOC)~\cite{Shenker:2013pqa,Shenker:2013yza,Roberts:2014isa,Shenker:2014cwa,kitaevtalk15,Roberts:2014ifa}
\bega\label{otcs}
F (t) =\vev{W(t) V(0) W(t) V(0)}_\beta = \vev{\Psi_2 (t)|\Psi_1 (t)}  ,\\
\ket{\Psi_1 (t)} \equiv W(t) V(0) \ket{\Psi_\beta}, \qquad
 \ket{\Psi_2 (t)} \equiv V(0) W(t)\ket{\Psi_\beta} \ .
 \label{fev}
\end{gather}
Here  $V$ and $W$ are generic few-body operators which we will take to be Hermitian, and $\vev{\cdots}_\b$ denotes the thermal average at an inverse temperature $\beta$. 
In~\eqref{fev}, $\ket{\Psi_\beta}$ denotes the thermal field double state the expectation values with respect to which give the thermal averages. 

In the large $N$ limit, the degrees of freedom involved in generic few-body operators $V (0)$ and $W(0)$ do not overlap with each other. For small $t$,  $V(0)$ and $W(t)$ almost commute, and $\ket{\Psi_{1,2}}$ are almost identical, which means that $F (t)$ should be $O(1)$. As time increases, $W(t)$ grows, and $\Psi_{1,2}$ become more and more different, which decreases $F (t)$. It is expected  for chaotic systems~\cite{kitaevtalk15}
  \begin{equation} \label{ejn}
F (t) \sim c_1- {c_2 \ov N} e^{\lam t}  + \cdots 
\end{equation}
where $c_{1,2}$ are some constants and $\lam$ is the quantum Lyapunov exponent. In contrast, 
\bega \label{ejn1} 
H_1 (t) = \vev{\Psi_1 (t)|\Psi_1 (t)}  = \vev{V(0) W(t) W(t) V(0)}_\b, \\
 H_2 (t) = \vev{\Psi_2 (t)|\Psi_2 (t)}= 
\vev{W(t) V(0) V(0) W(t)}_\b , 
 \label{ejn2} 
\end{gather}
the so-called time-ordered correlators (TOCs), always remain $O(1)$. The exponential behavior in~\eqref{ejn} says that 
in a chaotic system, the slight difference in the initial preparation of $\ket{\Psi_{1,2}}$ will be quickly magnified during time evolution, which is the essence of the butterfly effect. This quantum butterfly effect was first \cite{larkin1969quasiclassical,Gu:2017ohj} characterized by the exponential growth of $\avg{[W(t),V(0)]}_\b$, which is motivated from the exponential growth of $\{x(t),p(0)\}_{\text{P.B.}}$ in the classical butterfly effect.\footnote{See \cite{Kolganov:2022mpe} for a model comparing the quantum and classical Lyapunov exponents. }

\begin{figure}
\begin{centering}
\subfloat[\label{fig:maxchaos}]{\begin{centering}
\includegraphics[height=2cm]{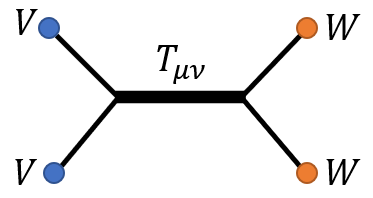}
\par\end{centering}

}\subfloat[\label{fig:reg}]{\begin{centering}
\includegraphics[height=2cm]{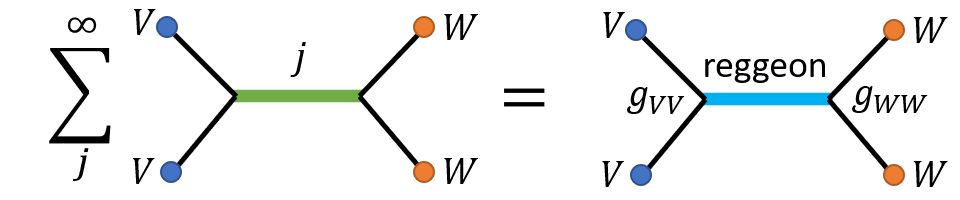}
\par\end{centering}
}
\par\end{centering}
\caption{(a) The maximal chaos can be described by exchange of the stress tensor between $W$ and $V$. (b) Sum over infinitely many spin $j$ particles exchange between $W$ and $V$ can be viewed as the exchange of a Reggeon.}

\end{figure}

Quantum Lyapunov exponent $\lam$ is state-dependent, describing operator growths 
``moderated'' by the state under consideration. It has an upper bound~\cite{Maldacena:2015waa}
\be
 \lam \leq {2 \pi \ov \b} \ .
 \ee
The bound is saturated by various systems, including holographic systems in the classical gravity limit, and SYK-type systems in the low temperature limit. 
These ``maximally'' chaotic systems are special: the exponential time dependence in~\eqref{ejn}  can be attributed to the exchange of the stress tensor  between $W$ and $V$ (see Fig.~\ref{fig:maxchaos}), and can be described by a hydrodynamic effective theory with a single effective field $\vp$ that plays the dual role of ensuring energy conservation and characterizing operator growth~\cite{Blake:2017ris,Blake:2021wqj}.

For non-maximal chaotic systems with $\lam < {2 \pi \ov \b}$, the origin of~\eqref{ejn} is more intricate, arising from exchanging an infinite number of operators. For example, in the SYK system, exchange of an operator characterized by some quantum number $j$ 
leads to an exponential decrease in $F (t)$ proportional to $e^{{2 \pi \ov \b} (j-1) t}$, which violates the bound for any $j> 2$.
Summing over exchanges of an infinite tower of such operators with increasingly larger values of $j$ leads to an effective $\lam$ satisfying the bound.  We will loosely refer to $j$ as ``spin'' in analogue with higher dimensional systems even though there is no spin for SYK.  Another example  is four-point correlation functions of a large $N$ CFT in the vacuum state in the so-called conformal Regge regime~\cite{Costa:2012cb,Cornalba:2007fs,Kravchuk:2018htv}, which can be interpreted as a thermal OTOC in terms of Rindler time (with $\b =2 \pi$). 
Here contribution from a spin-$j$ operator in the OPE of $WW$ (and $VV$) gives a contribution proportional to $e^{(j-1) t}$ ($t$ is now the Rindler time) and  summing over an infinite number of higher spin operator exchanges gives an $\lam < 1$.

It is natural to ask whether there exists an effective description with a small number of degrees of freedom that can capture
the sum over exchanges of the infinite tower of operators, with $j_{\rm eff} = 1 + \lam {\b \ov 2 \pi}$ interpreted as the effective spin of the  effective fields.  

We should stress that such an effective description is conceptually and philosophically different from that is usually used in effective field theory (EFT). Usually 
an EFT is used to describe the dynamics of a small number of ``low energy''  (or ``slow'') degrees of freedom whose contributions dominate over others in the regime of interests. Their effective actions can be formally defined from path integrals by integrating out 
other ``high energy'' (or ``fast'') modes. There is, however, no such decoupling of ``high energy'' (or ``fast'') degrees of freedom here. 
Spin $j$ (with $j >2$) exchanges  give important contributions to $F (t)$;  they cannot be integrated out in the usual sense. 
We are merely asking whether there is a way to capture the effect of the infinite sum. The effective fields here may not correspond to genuine physical collective degrees of freedom. The philosophy is also very different from the EFT for maximal chaos described in~\cite{Blake:2017ris}; there the stress tensor exchange dominates and the EFT is used to capture the most essential part of the stress tensor exchange.


%

The question of an effective description for non-maximal chaotic system is closely related a well-known problem in QCD, the formulation of  EFTs for Reggeons (see e.g.~\cite{forshaw_ross_1997} for a review). Consider a scattering process in some quantum system (say in QCD or string theory) 
\be \label{sejn}
V + W \to V + W
\ee
where $V, W$ denote different particles. We denote the scattering amplitude by $\mA (s, \rt)$, with $s$ the standard variable characterizing the total center of mass energy, and $\rt$ characterizing the momentum exchange between $V$ and $W$ particles. 
In the regime $s \to \infty$ with $\rt$ finite, each spin $j$ particle exchange between $V$ and $W$ 
gives a contribution to $\mA (s,\rt)$ proportional to $s^{j-1}$. When the Sommerfeld-Watson transform  is used to summing over all the higher spin exchanges,  the scattering amplitude can be written in a form 
\be
\mA (s,\rt) \propto g_{WW} (\rt) g_{VV} (\rt) s^{\al (\rt)-1} 
\ee
which can be interpreted as the exchange of a single effective particle, called reggeon,  with an effective spin $\al (\rt)$. 
 $g_{WW}, g_{VV}$ can be interpreted as couplings the Reggeon to $W$ and $V$. See Fig.~\ref{fig:reg} for an illustration. 
  

For systems with a gravity dual, the OTOC~\eqref{otcs} maps to the gravity side a scattering process precisely of the form~\eqref{sejn} in a black hole geometry~\cite{Shenker:2013pqa,Shenker:2013yza,Roberts:2014isa,Shenker:2014cwa}, with $V, W$ the corresponding bulk particles dual the boundary operators. The center of mass energy square $s$ for the scattering process is related to time separation $t$ in~\eqref{otcs} by $s \propto e^{{2 \pi \ov \b} t}$.
In the bulk language,  $\lam$ arises from summing over  exchanges of an infinite number stringy modes with increasingly higher spins. 

In the $\al' \to 0$ limit, the contributions of higher spin stringy modes decouple, with only graviton exchange remaining, and the system becomes maximally chaotic. In this limit, the maximal value $\lam_{\rm max} = {2 \pi \ov \b}$ 
is universal for all holographic systems, independent of the details of the
black hole geometries~\cite{Maldacena:2015waa}, and can be argued as a direct consequence of existence of a sharp horizon. 

Having an effective  description away from the $\al' \to 0$ limit that can capture an infinite number of stringy modes exchanges is clearly valuable. Such an effective description can also potentially give insights into what becomes of the event horizon in the stringy regime.

In this paper we make a proposal to formulate an EFT for a non-maximal chaotic system.  For simplicity, we will restrict to a quantum mechanical system with no spatial dependence. Generalization to having spatial dependence should be straightforward, and will be left elsewhere. Lacking at the moment a first-principle understanding  of the nature of the effective chaos field(s) or their effective action, our approach is 
phenomenological. We try to identify a minimal set of fields and a minimal set of conditions on their action, such that the following criteria are met:

\ben 

\item With $\lam$ being an input parameter, the EFT gives rise to exponential behavior~\eqref{ejn} for OTOCs, but no exponential for TOCs~\eqref{ejn1}--\eqref{ejn2}. 

\item It captures all the KMS properties of thermal 4-point functions.

\een
We will see that the above conditions are rather constraining, and the resulting EFT can be used to make a general prediction on the structure of OTOCs, which is consistent with that previously postulated in~\cite{Kitaev:2017awl,Gu:2018jsv}, and agree with the explicit expressions in large $q$ SYK model, holographic systems (obtained from stringy scattering), and the conformal Regge theory. 
 Furthermore, we show that in this framework it is possible to sum higher order terms in equation~\eqref{ejn} 
 of the form ${e^{k \lam t} \ov N^k}$ (with $k$ an integer)\footnote{Such terms are of the same order and dominate in the regime $N \to \infty$ and $t \sim {1 \ov \lam} \log N$.} in an exponential, which can again be viewed as a general prediction, and agrees with those previously obtained in specific systems~\cite{Shenker:2014cwa,Stanford:2021bhl, Gu:2021xaj}.
 
We also show that the structure proposed for the non-maximal chaos EFT can be in fact extracted in the large-$q$ SYK model, 
where it is possible to identify explicitly the chaos effective fields, and make much finer comparison between the EFT and the microscopic theory than the structure of TOCs and OTOCs.

\begin{figure}
\begin{centering}

\includegraphics[height=1.6cm]{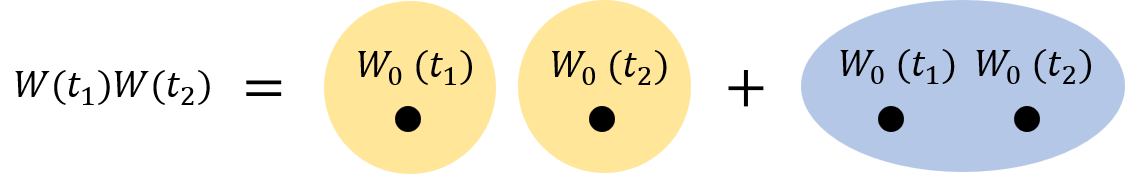}
\par\end{centering}

\caption{Effective description of $W(t_1)W(t_2)$ in non-maximal chaotic systems.  The black dots are bare operators $W_0$.  
There are now two types of dressing: one type dresses each local operator separately (yellow ``clouds''), and the other type dresses both operators together (blue ``clouds''). Maximal chaos case contains only the first type of dressing.
 }
\label{fig:scrambling}
\end{figure}

It is worth mentioning here a key difference between the non-maximal EFT to be discussed in this paper and that for maximal chaotic systems of~\cite{Blake:2017ris,Blake:2021wqj}. For maximal chaotic systems, $W(t)$ can be viewed as $W (t) = W [W_0 (t), \vp (t)]$ where ``bare'' operator $W_0 (t)$ describes $W(t)$ in the large $N$ limit (with no overlap with $V_0$). $\vp (t)$ is an effective field ``attached'' to $W_0$, i.e. $W$ is obtained by dressing $W_0$ with $\vp$. 
$\vp$ captures effectively the overlap between $W(t)$ and $V(0)$ due to scrambling, and its dynamics leads to $1/N$ corrections indicated in~\eqref{ejn}. For non-maximal chaotic systems, the chaos fields involve multiple components: (i) one component dresses each bare local operator as in the maximal chaotic case (and indeed it reduces to $\vp$ in the maximal chaos limit). This component carries only one time argument. (ii) There are other components which dress both $W$'s in~\eqref{otcs}, i.e. it has two time arguments (see Fig. \ref{fig:scrambling} for an illustration). 
Existence of such components leads to many new elements which are not present in the maximal case.

The paper is organized as follows. In Section \ref{sec:The-structure-of}, we construct the effective field theory of non-maximal chaos with two effective fields $\phi_{1,2}$ and show that the TOC does not have exponential growth and OTOC has exponential growth. 
In Section~\ref{sec:motoc}  we compare the general structure of OTOCs obtained in Sec.~\ref{sec:The-structure-of} with various known examples.  In Section \ref{sec:4}, we show that the two effective fields $\phi_{1,2}$ reduce to a single field in the maximal chaos limit and the EFT becomes the same as the EFT constructed for maximal chaos. In Section \ref{sec:5}, 
 we show how the general structure of the EFT discussed in Sec.~\ref{sec:The-structure-of} arises in the large $q$ SYK model, and 
 obtain the explicit form of the EFT action.
In Section \ref{sec:3}, we include higher order coupling to effective mode $\phi_{1,2}$ and show that 
certain higher-order terms of 
the four-point function can be resummed and exponentiated. We conclude in Section~\ref{sec:Discussion-and-conclusion} with a summary and a discussion of future directions.

\section{The structure of EFT\label{sec:The-structure-of}}


In this section we discuss the general formulation of an EFT for non-maximally chaotic systems. 
For simplicity we will consider quantum mechanical systems with no spatial dependence. Generalization to systems with spatial dependence can be readily made, although technically more intricate. We will also use the unit such that  $\b=2\pi$ . 

\subsection{General setup} \label{sec:2.0}

Consider a generic four-point Wightman function in thermal state
\be 
\mF_{abcd}(t_1,t_2,t_3,t_4)=\Tr \left(e^{-2\pi H}\mO_a(t_1)\mO_b(t_2)\mO_c(t_3)\mO_d(t_4) \ri) 
\equiv \vev{\mO_a(t_1)\mO_b(t_2)\mO_c(t_3)\mO_d(t_4) }
\ee
where the subscript refers to the ordering of operators and the time argument should be understood as corresponding to each subscript in the same order. We will treat time variables as complex, and $\mF_{abcd}(t_1,t_2,t_3,t_4)$ is analytic in the domain 
\be\label{timd}
\mD: \quad \Im t_4 -2\pi<\Im t_{1}<\Im t_{2}<\Im t_{3}< \Im t_4 \ .
\ee 
$\mF_{abcd}$ obeys the KMS condition 
\be 
\mF_{abcd}(t_1,t_2,t_3,t_4)=\mF_{bcda}(t_2,t_3,t_4,t_1 + 2\pi i) \label{kmsF}
\ee
which can be iterated cyclically to shift other time variables. 
It is convenient to introduce a time-ordered function 
\be\label{yhef}
\hat \mF_{abcd} (t_1, t_2, t_3, t_4) \equiv \vev{\mT \mO_a(t_1)\mO_b(t_2)\mO_c(t_3)\mO_d(t_4) }
\ee
where  $\mT$ denotes operators should be ordered from left to right according to the ascending order of their corresponding $\Im t_i$. 
Moreover, it should always be understood that 
\be\label{uhe}
(\Im t_i)_{\rm min} - (\Im t_i)_{\rm max}  \in (-2 \pi, 0), i=1,2,3,4  \ . 
\ee
The KMS condition~\eqref{kmsF} can then be written as 
\be \label{kmsa}
\hat \mF_{abcd} (t_1, t_2, t_3, t_4) = \hat \mF_{abcd}  (t_1', t_2', t_3', t_4'), \quad  t_i' = t_i + 2 \pi i m_i
\ee
where $m_i$ are integers, and should be such that $t_i'$'s obey~\eqref{uhe}.


Now consider 
\be \label{henw}
\hat \mF_{WWVV}(t_1,t_2;t_3,t_4)=\vev{\mT W(t_1)W(t_2)V(t_3)V(t_4)}
\ee
which by definition is symmetric under swapping $t_1\leftrightarrow t_2$ and $t_3\leftrightarrow t_4$
\be\label{swap}
\hat \mF_{WWVV}(t_1,t_2;t_3,t_4) = \hat \mF_{WWVV}(t_2,t_1;t_3,t_4)
= \hat \mF_{WWVV}(t_1,t_2;t_4,t_3) \ .
\ee
Depending on $\Im t_i$, $\hat \mF_{WWVV}(t_1,t_2;t_3,t_4)$ can correspond to TOC or OTOC. For example,  
\bega
\hat \mF_{WWVV}(t_1,t_2;t_3,t_4) = \vev{V (t_3) W(t_1)W(t_2) V(t_4)} , \quad \Im t_3 < \Im t_1< \Im t_2 < \Im t_4 \\
\hat \mF_{WWVV}(t_1,t_2;t_3,t_4) = \vev{ W(t_1) V (t_3) W(t_2) V(t_4)} , \quad   \Im t_1 < \Im t_3 < \Im t_2 < \Im t_4 
\end{gather} 
A TOC and OTOC cannot change into each other under cyclic permutations, so KMS conditions~\eqref{kmsF} relate functions within each type.

Our goal is to develop an effective description for obtaining $\hat \mF_{WWVV}$ for large $N$. To motivate the structure of our proposed EFT for non-maximally chaotic systems, it is useful to recall some key elements of that for maximally chaotic systems introduced in~\cite{Blake:2017ris}. 
One imagines  the scrambling of  $W(t)$  allows a coarse-grained description in terms of building up an ``effective cloud,'' i.e. 
\be \label{ngco}
W(t) = W [W_0 (t), \p (t)]  \ .
\ee  
Here $W_0 (t)$ is a ``bare'' operator involving the original degrees of freedom of $W$, and $\p (t)$ is an effective chaos mode that
describes macroscopically the growth of the operator in the space of degrees of freedom. $W(t)$ in~\eqref{ngco} is taken to be linear in $W_0$ but can in principle have any dependence on the effective field $\p (t)$. 
The dynamics of $\vp$ is governed by a chaos effective theory, with two-point function of $\vp$ scaling with $N$ as $1/N$. 
Thus $W_0$ can also be viewed as giving the leading part of $W (t)$ in a $1/N$ expansion. 

When $\p (t)$ is small, it can be expanded to linear order as 
\be 
W(t)=W_0(t)+L_t[W_0(t)]\p(t)+O(\p^2) \label{expans}
\ee
where $L_t [W_0]$ is a $W_0$-dependent differential operator acting on $\p$. More explicitly,
\be 
L_t[W_0(t)]\p(t)=\sum_{m,n=0}^\infty c_{mn} \partial_t^m W_0(t)  \partial^n_t \p(t) \label{n-uen}
\ee
Below it should always be understood that $L_t$ acts on the corresponding $\p (t)$ even when they are not written adjacently. 
It then follows that  
\be \label{nyhn1} 
W(t)W(t')=W_{0}(t)W_{0}(t')
+ L_{t} [W_0 (t)]  \p (t) W_0 (t') +W_0 (t) L_{t'} [W_0 (t')]  \p (t') + O(\p^2) , 
\ee
which for $\Im(t-t')\in[-2\pi,0]$ gives
\bega 
\avg{\mT W(t)W(t')} = g_W (t-t') + O(1/N), \quad g_W (t-t') \equiv \vev{\mT W_0 (t) W_0 (t')} , 
\end{gather} 
where we have assumed that one-point function of $\p$ is zero. The 
 $O(1/N)$ piece in the above equation comes from $O(\p^2)$ term in~\eqref{nyhn1}, and is proportional to two-point function of $\phi$.
 The KMS condition for $g_W$ is
 \be 
 g_W(t)=g_W(-t-2\pi i),\quad \Im t\in[-2\pi,0]
 \ee
 Note that $\vev{W(t) V(0)} \sim O(1/N)$ with $\vev{W_0 V_0} =0$, as for generic few-body operators $V, W$, their two-point function should vanish at the leading order in $1/N$ expansion. 

Plugging~\eqref{nyhn1} and the corresponding expression for $V$ into~\eqref{henw}, $\hat \mF_{WWVV}$
reduces to the two-point function of effective mode $\p(t)$ at leading order
\be 
\hat \mF_{WWVV} (t_1,t_2;t_3,t_4)=g_W g_V  +
 \sum_{i=1,2,j=3,4} L_{t_i}\tilde L_{t_j} [g_W g_V \avg{\mT \p(t_i)\p(t_j)}_{\rm EFT}] \label{ntimeO}
\ee
where $\tilde L_t$ is the differential operator from a similar expansion of $V$ with $c_{mn}\ra \tilde c_{mn}$, and 
\be \label{defg}
g_W \equiv g_W(t_{12})=\avg{\mT W_0(t_1)W_0 (t_2)},\quad g_V \equiv g_V(t_{34})=\avg{\mT V_0(t_3)V_0 (t_4)} \ .
\ee
Here the $\avg{\cdot}_{\rm EFT}$ means expectation value evaluated in the effective field theory of $\p$; $\mT$ in $\avg{\mT \p(t_i)\p(t_j)}_{\rm EFT}$  follows from the relative magnitude of $\Im t_i$ and $\Im t_j$. 
Equation~\eqref{ntimeO} has a very restrictive structure:  the two-point function of $\p$ in each term only depends 
on the locations of two operators. For example, the $\p$ correlation function $\avg{\mT \p(t_1)\p(t_3)}_{\rm EFT}$ has no knowledge of $t_2, t_4$ at all.  In other words, the four-point function essentially reduces to pairwise two-point functions of $\p$. 
This structure leads to various features of $\hat \mF$ that are consistent with a maximally chaotic system~\cite{Blake:2021wqj}, including the Lyapunov exponent $\lambda =1$ (after imposing a shift symmetry in the EFT of $\p$), but are not present in a non-maximally chaotic system. Since~\eqref{ntimeO} is a direct consequence of~\eqref{expans}, for non-maximally chaotic systems, we must generalize~\eqref{expans}.

\subsection{Two-component effective mode and constraints from KMS symmetries} \label{sec:2.1}

We will now propose a formulation for non-maximally chaotic systems which may be considered a minimal generalization of the EFT in \cite{Blake:2017ris} for maximal chaos. 
The formulation is partially motivated from features of the large $q$ SYK theory, and as we will show in Sec.~\ref{sec:5}, fully captures  
the physics of that theory. The general structure of OTOCs resulting from it is also compatible with the conclusions of~\cite{Shenker:2014cwa,Costa:2012cb,Mezei:2019dfv,Kravchuk:2018htv,Cornalba:2007fs}, as we will describe later.

In this formulation $\hat \mF$ still reduces to two-point functions of some effective fields, but the main new ingredient we would like to incorporate is that now  two-point functions of effective field(s) have knowledge of the locations of all four operators, not just two of them.   
For this purpose we consider the following generalization of~\eqref{expans}\footnote{We should emphasize that~\eqref{eq:n13} and \eqref{nyhn1} should not be viewed as OPEs. If there are $V$ insertion(s) between $W$'s we cannot do OPE, while these equations are supposed to be valid for any configurations of orderings.}
\be
W(t)W(t')=W_{0}(t)W_{0}(t')+\sum_{i=1}^2 \mD^{(i)}_{W}(t,t')\phi_i (t, t')+O(\phi{}^{2}) , 
\label{eq:n13}
\ee
where there are two fields $\phi_{1,2}$ who depend on both $t, t'$, and $\mD^{(1,2)}_{W}(t,t')$ are some $W_0$-dependent differential operators to be specified more explicitly below.  Now $\phi_{1,2}$ depend on both $t$ and $t'$ of 
$W(t)$ and $W (t')$, which means that we cannot view $\phi_{1,2}$ as the ``dressing'' of each individual operator, as in the case of~\eqref{expans}. 
Rather they should be interpreted as an effective description of  the sum over
an infinite number of higher spin operator exchanges that are known to 
contribute to $\hat \mF$ at the leading order in non-maximal systems~\cite{Costa:2012cb,Kravchuk:2018htv,Cornalba:2007fs,Shenker:2014cwa}. 
There is a parallel equation with $W$ replaced by $V$'s.


The effective theory of $\phi_i$ should satisfy the following criteria:

\ben 

\item Exponential growth in OTOCs with an arbitrary Lyapunov exponent $\lam$.

\item No such exponential growth in TOCs.

\item All the KMS conditions and analytic properties of $\hat \mF_{WWVV}$ are satisfied. 

\een
We will show that the above goals can be achieved with a minimal generalization of~\eqref{nyhn1}. 
In this subsection we first present the prescription for~\eqref{eq:n13}, and work out the constraints on two-point functions of $\phi_{1,2}$ 
from the KMS conditions of $\hat \mF_{WWVV}$, which provide the basic inputs for formulating the theory of $\phi_{1,2}$. 

We will take $\phi_{1,2}$  to  ``mainly'' couple to one of the $W$'s. A definition which respects the swap symmetry~\eqref{swap} of $\hat \mF$ is that $\phi_1$ ($\phi_2$) 
couples mainly to the $W$ with the smaller~(larger) $\Im t$. Denoting $t_S$ ($t_L$) with the smaller (larger) value of $\Im t, \Im t'$, 
by ``mainly'' we mean: 
\ben 

\item  $\phi_1 (t, t') = \phi_1 (\bar t; t_{S})$ depends weakly on $\bar t =  {t + t' \over 2}$ such that it can be expanded in terms of $\bar t$-derivatives. The dependence on $\bar t$ encodes the nonlocal information of the theory.
 Similarly, $\phi_2 (t, t') = \phi_2 (\bar t; t_{L})$.  

\item The action of $\mD^{(i)}_W (t, t')$ on $\phi_i$ can be expanded similarly as in~\eqref{n-uen}
\bega \label{new-uen1}
\mD^{(1)}_W (t, t') \phi_1 (t, t')  = W_0 (t_L) L_{t_S} [W_0] \phi_1 (\bar t; t_S) 
\\
\label{new-uen2}
\mD^{(2)}_W (t, t') \phi_2 (t, t')  = W_0 (t_S) L_{t_L} [W_0] \phi_2 (\bar t; t_L)
\\
L_t [W_0] \equiv \sum_{m,n=0}^\infty c_{mn} \partial_{t}^m W_0 (t) \partial^n_{t}  \label{x2.22}
 \ .
\end{gather} 
In other words, $\phi_1$ couples directly only to $W (t_S$), but does feel the presence of $W (t_L)$ through weak dependence on $\bar t$. Note that equation~\eqref{new-uen1} should be understood to be valid within time-ordered correlation functions, thus there is no need to worry about orderings between $W_0 (t_L)$ and $W_0 (t_S)$. Equation~\eqref{new-uen1} contains no derivative with respect to $\bar t$; it can be viewed as the leading term in a derivative expansion of $\bar t$. 

\een
For notational simplicity we take the coefficients $c_{nm}$ in~\eqref{new-uen1} to be the same for $\mD^{(2)}_W (t, t')$, but our discussion can be straightforwardly generalized to the cases that they are not the same. The vertex for $V$ will be denoted as $\tilde L_t$ with $c_{mn}\ra\tilde c_{mn}$.  The above prescription is a minimal nontrivial generalization of~\eqref{expans} that satisfies the aforestated criteria. In Appendix~\ref{app:a} we show that a few other simpler prescriptions cannot work. 

\begin{figure}
\begin{centering}
 
\includegraphics[height=7cm]{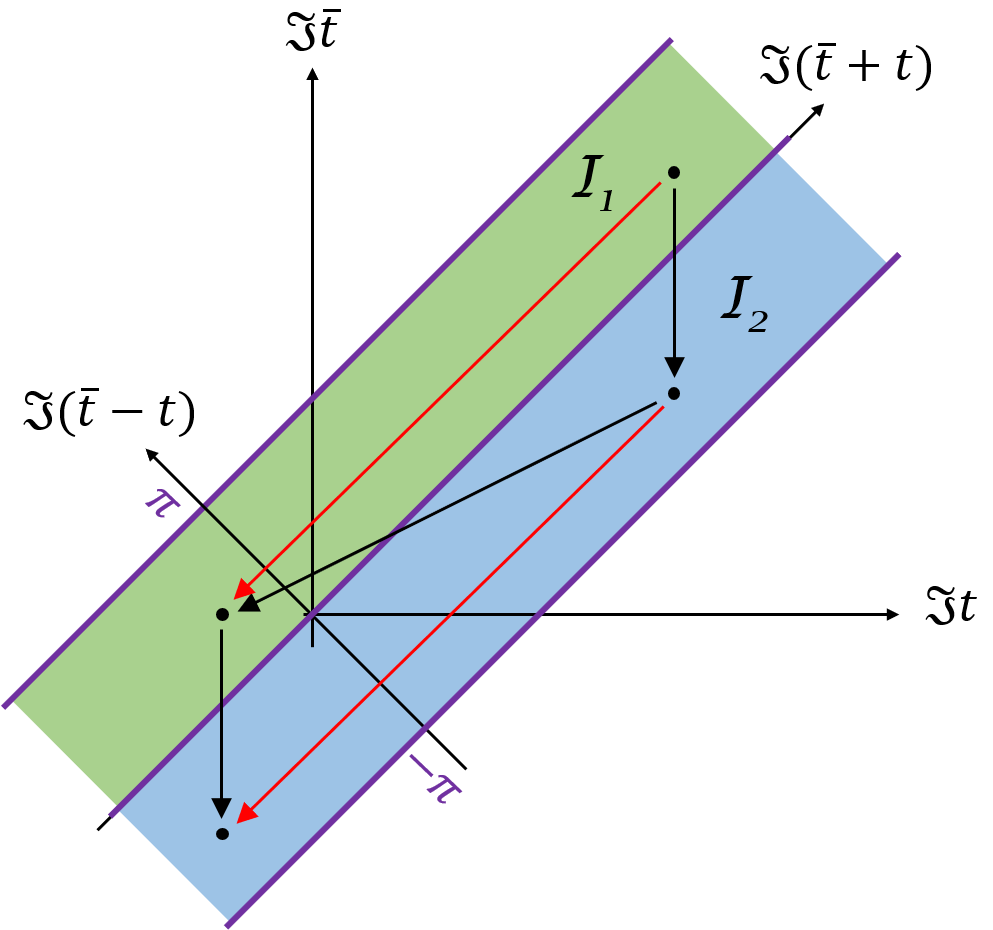} 

\par\end{centering}
\caption{The green region is the domain $\mI_1$ for $\phi_1(\bar t;t)$ and the blue region is the domain $\mI_2$ for $\phi_2(\bar t;t)$. The KMS conditions \eqref{pkms1}--\eqref{pkms2} relate $\phi_1$ with $\phi_2$ through identifying points between $\mI_1$ and $\mI_2$ as indicated by the black arrows. This generates the periodicity \eqref{pperid} on $\mI_{1,2}$ (red arrows). }
 \label{fig:nphifund}
\end{figure}

Since in~\eqref{eq:n13} $|\Im t-\Im t'|<2\pi$, by definition, $\phi_1(\bar t;t)$ is defined {for $\Im \bar t-\Im t\in(0,\pi)$}, while 
$\phi_2(\bar t;t)$ is defined for  $\Im \bar t-\Im t\in (-\pi,0)$. We refer to their domains as $\mI_1$ and $\mI_2$ respectively, 
see Fig. \ref{fig:nphifund}. 
Due to symmetries in exchanging $t_1$ and $t_2$, we will take $\Im t_1<\Im t_2$, and similarly take $\Im t_3<\Im t_4$. Therefore, $\phi_{1}$ always mainly couples to $W(t_1)$ and $V(t_3)$, $\phi_{2}$ always mainly couples to $W(t_2)$ and $V(t_4)$. 
Substituting~\eqref{eq:n13}--\eqref{new-uen2} into $\hat \mF_{WWVV}$ we find 
\be 
\hat \mF_{WWVV} (t_1,t_2;t_3,t_4)=g_W g_V+
 \sum_{i,j=1,2} L_{t_i}\tilde L_{t_{j+2}} \left[g_W g_V \avg{\hat \mT \phi_i(\bar t_W;t_i)\phi_j(\bar t_V; t_{j+2})} \ri] \label{n2.16}
\ee
where $\bar t_W=(t_1+t_2)/2$, $\bar t_V=(t_3+t_4)/2$, and the expectation value of $\phi_{1,2}$ should be understood as being evaluated in an effective theory. 
Unlike in~\eqref{ntimeO}, where the time-ordering $\mT$ follows from that of $\hat \mF$, here in~\eqref{n2.16}
the effective fields $\phi_{1,2} (\bar t;t)$ have  two time variables, and time-ordering in $\hat \mF$ no longer leads to a unique choice of orderings of $\phi_i$ and $\phi_j$. {We will specify the precise meaning of $\avg{\hat \mT \phi_i(\bar t_W;t_i)\phi_j(\bar t_V; t_{j+2})}$ 
in Section \ref{sec:diag}}. Here we will just list some properties they should satisfy:

\ben

\item  Since in time ordered correlation function $\hat \mF$ we can exchange $V$ and $W$ arbitrarily, the ordering of $\phi_i, \phi_j$ in the correlation function should  not matter, i.e.
\be
 \avg{\hat \mT \phi_i(\bar t_W;t_i)\phi_j(\bar t_V; t_{j+2})} = \avg{\hat \mT \phi_j(\bar t_V; t_{j+2}) \phi_i(\bar t_W;t_i)} \ .
\ee


\item From time translation invariance of the system,  $\hat \mF$ is invariant under shifts of all $t_i$ by the same constant, which implies 
the following translation invariance of two-point functions of $\phi_{1,2}$, 
\be\label{tras}
\avg{\hat \mT \phi_i (\bar t; t)\phi_j (\bar t';t')} = \avg{\hat \mT \phi_i(\bar t +c; t +c)\phi_j (\bar t' +c;t'+c)} \ .
\ee

\item 
The KMS conditions satisfied by $\hat \mF$ 
imply that these two-point functions of $\phi_{1,2}$ should satisfy the following constraints 
\begin{align}
\avg{\hat \mT \phi_1(\bar t; t)\phi_i (\bar t';t')}&\simeq\avg{\hat \mT \phi_2 (\bar t + \pi i ;t + 2 \pi i)  
\phi_i(\bar t';t')}, \quad \Im t < \Im t' 
  \label{pkms1}\\
\avg{\hat \mT \phi_2(\bar t;t)\phi_i(\bar t';t')} &\simeq \avg{\hat \mT \phi_1(\bar t +\pi i;t)\phi_i(\bar t';t')}  , 
\label{pkms2}
\end{align}
where $\simeq$ means equal up to zero modes, defined as functions $n_{ij}(t_1,t_2;t_3,t_4)$ satisfying 
\be
\sum_{i,j=1,2} L_{t_i}\tilde L_{t_j} [g_W g_V n_{ij}(t_1,t_2;t_3,t_4)] =0 \ . \label{eq:n2.26L}
\ee
 The zero modes can be viewed as field redefinition freedom of effective fields  that does not cause any difference in the original four-point function $\hat \mF$. From now on, we will set $n_{ij}=0$.
Combining~\eqref{pkms1}--\eqref{pkms2}, we also get the following periodicity
\be 
\avg{\hat \mT \phi_i (\bar t;t) \phi_j (\bar t';t')} \simeq \avg{\hat \mT \phi_i(\bar t+2\pi i;t+2\pi i)\phi_j(\bar t';t')} ,
\quad \Im t < \Im t' \ . 
\label{pperid}
\ee
{See Fig~\ref{fig:nphifund} for a diagrammatical depiction of~\eqref{pkms1}--\eqref{pkms2}.}

\item Four-point function $\hat \mF_{WWVV} (t_1,t_2;t_3,t_4)$ can have potential non-smoothness 
when the imaginary parts of two or more time arguments coincide, as these are the locations where ordering of operators change. 
There are two cases: 
\ben 

\item $\Im t_1=\Im t_2$,  which corresponds to order changes of $W$'s within themselves. 
In terms of $\phi_{1,2}(\bar t;t)$, this corresponds to $\Im \bar t- \Im t = 0$, where  the couplings of $W$'s to $\phi_{1,2}$ are switched.\footnote{For example, as we cross from $\Im t_1- \Im t_2 < 0$ to $\Im t_1- \Im t_2 > 0$, $W(t_1)$ switches from mainly coupled to $\phi_1 (\bar t_W; t_1)$ to mainly coupled to $\phi_2 (\bar t_W; t_1)$.} Similar statements apply to $t_3, t_4$. 
As stated earlier, we will restrict to $\Im t_1<\Im t_2$ and $\Im t_3<\Im t_4$ throughout, {so such potential non-smoothness will not be relevant for our discussion of $\avg{\hat\mT \phi_i(\bar t;t)\phi_j(\bar t';t')}$.}

\item  One of $\Im t_1, \Im t_2$ coinciding with one of $\Im t_3, \Im t_4$, which is a boundary between the domains of $t_i$ corresponding to TOCs and OTOCs; crossing such a boundary a pair of $W$ and $V$ will exchange order. 
In terms of two-point function $\avg{\hat\mT \phi_i(\bar t;t)\phi_j(\bar t';t')}$, this corresponds to potential non-smoothness at $\Im t= \Im t' $. In the domain $\mD$, we should not have any other singularities.  \label{item:4b}

\een

\een

\subsection{Diagonalize the KMS conditions} \label{sec:diag}


We will now proceed to formulate an effective field theory (EFT) that can be used to obtain correlation functions of $\phi_i$ in~\eqref{n2.16}.

There is an immediate  difficulty in directly formulating an EFT for $\phi_{1,2}$, due to that they are defined in different 
domains~(recall Fig.~\ref{fig:nphifund}). So they cannot appear in the same Lagrangian, but they transform to each other under the constraints~\eqref{pkms1}--\eqref{pkms2}
from the KMS conditions. To address this difficulty, we introduce two new fields,
\be 
\eta_\pm(\bar t;t)=\f {1}{ \sqrt{2}} (\phi_1(\bar t;t-i\pi)\pm \phi_2(\bar t;t)) \label{n2.24}
\ee
which are both defined in the domain $\mI_2: \Im (\bar t- t) \in (-\pi,0)$.  Conversely, we have 
\be \label{petr}
\phi_1(\bar t;t) = \f {1}{ \sqrt{2}}  (\eta_+ (\bar t;t+i\pi) +\eta_- (\bar t;t+i\pi)), \quad
\phi_2(\bar t;t) = \f {1}{ \sqrt{2}}  (\eta_+ (\bar t;t) - \eta_- (\bar t;t)) \ .
\ee
We will define two-point functions of $\phi_{1,2}$ in~\eqref{n2.16} in terms of those of $\eta_{\pm}$ using~\eqref{petr}. For example, 
\be
\avg{\hat \mT \phi_1(\bar t;t)\phi_2(\bar t';t')} \equiv \f 1 2 \sum_{s, s'=\pm} s' \avg{\hat \mT \eta_s(\bar t;t+i\pi)\eta_{s'} (\bar t';t')},
\label{nphi12}
\ee
where on the right hand side the time ordering $\hat \mT$ is defined in terms of that of $\Im t$, i.e. 
\bega
\avg{\hat \mT \eta_s(\bar t;t)\eta_{s'} (\bar t';t')} \equiv \begin{cases}
\avg{\eta_s(\bar t;t)\eta_{s'} (\bar t';t')},\quad \Im t < \Im t' \\
\avg{\eta_{s'}(\bar t';t')\eta_s(\bar t;t)},\quad \Im t > \Im t' 
\end{cases} , \quad s, s'= \pm \label{n2.28}  \ .
\end{gather} 
Now $\vev{\cdots}$ is understood as defined in the EFT of $\eta_\pm$, and the right hand side of~\eqref{n2.28} should be understood as Wightman functions in the EFT.  The motivations for choosing $\hat \mT$ ordering in terms of $\Im t$ are as follows. Firstly, 
as discussed in item~\ref{item:4b} at the end of last subsection, correlation functions of $\phi_{1,2}$ have potential non-smoothness at $\Im t = \Im t'$.  Ordering in $\Im t$ in $\eta$-correlators provides a simple way to realize that. Secondly, we assumed that the dependence of $\phi_{1,2}$ on $\bar t$ is weak, so should be $\eta_\pm$. Making the ordering independent of $\bar t$ is natural. 


Now consider the constraints~\eqref{pkms1}--\eqref{pkms2}. 
It can be checked that they are satisfied provided that 
\be\label{etakms}
\avg{\hat \mT \eta_s(\bar t;t)\eta_{s'}(\bar t';t')} =  s\avg{\hat \mT \eta_s(\bar t+ i \pi ;t + i \pi) \eta_{s'}(\bar t';t')},
\ee
which is diagonal in $\eta_\pm$. We see that introducing $\eta_\pm$ not only resolves the domain issue, but also diagonalize the constraints from KMS conditions. 
Equation~\eqref{etakms} implies 
\begin{align}. 
\avg{\hat \mT \eta_s(\bar t;t)\eta_{s'}(\bar t';t')} 
&=  s' s\avg{\hat \mT \eta_s(\bar t+ i \pi ;t + i \pi) \eta_{s'}(\bar t'+i \pi;t'+i\pi)} \ .
\label{eta2.26}
\end{align}
From~\eqref{tras}, we have 
\be  
\avg{\hat \mT \eta_s(\bar t;t) \eta_{s'}(\bar t';t')}=\avg{\hat \mT \eta_s(\bar t+c;t+c) \eta_{s'}(\bar t'+c ;t'+c)},\quad \forall c\in\C \label{n2.27} \ .
\ee
It then follows from \eqref{eta2.26} that 
\be \label{etakms1} 
\avg{\hat \mT \eta_+(\bar t;t)\eta_-(\bar t';t')}= -\avg{\hat \mT \eta_+(\bar t;t)\eta_-(\bar t';t')} = 0 ,
\ee
i.e. time-ordered functions of $\eta_\pm$ are also diagonal. 

\begin{figure}
\begin{centering}
\includegraphics[height=6cm]{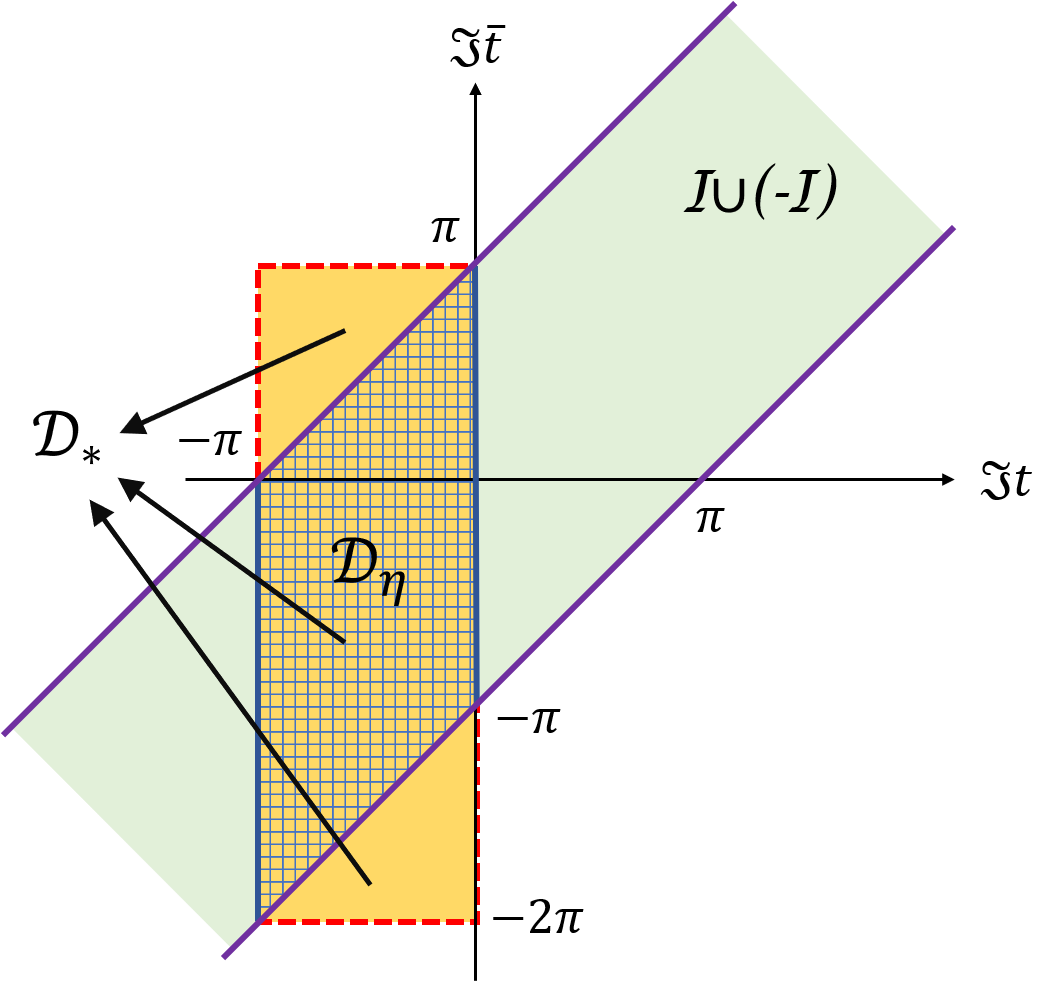} 
\par\end{centering}


\caption{The correlation functions of $\eta_s$ are defined on the green strip $\mI\cup(-\mI)$, on which a fundamental domain is the shaded parallelogram $\mD_\eta$. We analytically continue $\mD_\eta$ to the yellow rectangular domain $\mD_*$, which is bounded by the red dashed lines.
}
\label{pic:Dz}
\end{figure}

We now proceed to formulate an effective theory of $\eta_\pm$ with the following considerations in mind: 

\ben 

\item With the assumption of weak dependence on $\bar t$,  we assume that the effective action can be expanded in derivatives of $\bar t$. This leads to an immediate simplification: with only derivative dependence on $\bar t$, the EFT becomes translationally invariant in $\bar t$. Now given~\eqref{n2.27}, we also have translation invariance in $t$, i.e. 
\be  
\avg{\hat \mT \eta_s(\bar t;t) \eta_{s'}(\bar t';t')}=\avg{\hat \mT \eta_s(\bar t- \bar t'; t-t') \eta_{s'}(0;0)} \ .
\ee
The domain for function 
\be \label{defgf}
G_F (\bar t; t) \equiv \avg{\hat \mT \eta_s(\bar t; t) \eta_{s'}(0;0)}
\ee
is then given by the shaded stripe indicated in Fig.~\ref{pic:Dz}. 

At quadratic order in $\eta_\pm$, the effective action should be translationally invariant in both $\bar t$ and $t$. 

\item We would like to interpret~\eqref{etakms} as the KMS conditions for the $\eta_\pm$-system at a finite temperature. 
Given the definition of $\hat \mT$ in terms of $t$, it is natural to interpret the temperature as being associated with $t$. 
However, the condition \eqref{etakms} shifts both $\bar t$ and $t$ simultaneously, which is not of the conventional form. 
In next subsection, we will discuss how to convert it into the standard form.

\item So far the time variables $\bar t, t$ are complex. To write down an effective action we need to choose a real section in the  complex $\bar t, t$ planes. From Fig.~\ref{pic:Dz} it is convenient to choose the section to be that of 
imaginary $\bar t$ and real $t$, i.e. we will let $\bar t = - i \bar \tau$ and write down an action for $\eta_\pm (\bar \tau; t)$. 
It can be viewed as a two-dimensional field theory with $\bar \tau$ being a ``spatial'' coordinate and (real) $t$ being time. 
 Behavior of correlation functions for $\eta_\pm$ elsewhere are obtained by analytic continuations.

\een

\subsection{Reformulating the KMS conditions} \label{sec:refkms}

In this subsection we reformulate~\eqref{etakms} as the KMS conditions for 
$\eta_\pm$ at a finite temperature (associated with $t$) with $\bar \tau = - \Im \bar t$ as a spatial direction. 

Let us first recall the standard story. 
For a quantum field $\chi$  in a two-dimensional spacetime $(t, \bar \tau)$ at a nonzero inverse temperature $\b$, the KMS condition for Wightman functions are $\vev{\chi (\bar \tau_1, t_1) \chi ( \bar \tau_2, t_2)}
= \vev{ \chi ( \bar \tau_2, t_2) \chi (\bar \tau_1, t_1 + i \b)}$, and the Feyman functions 
$G_F (\bar \tau, t) \equiv \vev{\hat \mT \chi (\bar \tau,t) \chi (0,0)} = G_F (\bar \tau, t + i \b)$, with its fundamental domain  
being $\Im t \in (-\b,0)$. $G_F (\bar \tau, t)$ may have non-analytic behavior such as branch cuts at $\Im t =0$ and $\Im t = -\b$.  

Now consider $G_F (\bar t; t)$ defined in~\eqref{defgf}. From~\eqref{etakms}, the fundamental domain of $G_F (\bar t; t)$ can be chosen to be the region  $\mD_\eta$ in Fig. \ref{pic:Dz}. Equation~\eqref{etakms} is not quite the KMS condition with $\b = \pi$ (note that this is $\ha$ of the temperature we started with) due to the shift in $\bar \tau$. We can resolve this issue by extending the region $\mD_\eta$ to 
the larger region $\mD_*$ in Fig. \ref{pic:Dz}. Region $\mD_\eta$ is bounded above and below by lines $\bar \tau - \Im t = \pm \pi$, which are part of the boundary of the analytic domain of the original four-point function $\hat \mF$. The behavior of $G_F$ at these boundaries are system-dependent and depend on UV physics.  In other words, in principle  for different systems different boundary conditions should be imposed there. In the spirit of effective field theories we expect that the general structure of the effective action should not depend on the specific UV physics, although the coefficients in the effective action will. Since we are only interested in the general structure of the effective action, we can choose a most convenient boundary condition: we extend the domain to 
$\mD_*$,  and identify the values of $G_F (\bar t; t)$ at $\bar \tau = - \pi$ and $\bar \tau = 2 \pi$. 
In other words, we have periodic boundary conditions in $\bar \tau$ direction. Note that later we will only need to use the behavior of $
G_F (\bar \tau; t)$ in region $\mD_\eta$. 

Denote the conjugate momentum for $ \bar \tau$  as $P$, then $P = {2 \ov 3} m$ with $m$ an integer. 
We can decompose $\eta_s$ into three part 
\be
\eta_s(\bar \tau ;t)=\eta_{s,0}(\bar \tau;t)+\eta_{s,+}(\bar \tau;t)+\eta_{s,-}(\bar \tau;t) \label{x2.41}
\ee
where $\eta_{s, p}$ contains only $\bar \tau$-momenta $P = {2 \ov 3} (3n + p)$ with $n$ an integer and $p =-1,0,1$. 
$\eta_{s, p}$ has the behavior 
\be
\eta_{s,p}(\bar \tau + \pi;t)=e^{2\pi i p/3}\eta_{s,p}(\bar \tau;t),\quad p=0,\pm \ .
 \label{etasp}
\ee
Because of the additional phase $e^{2\pi i p/3}$, we should regard $\eta_{s,\pm}$ as complex scalar fields. Since the original $\eta_s$ is a real scalar, we need to identify them as hermitian conjugate to each other, i.e.
\begin{equation}
\eta_{s,p}^{\dagger}(\bar{\tau};t)=\eta_{s,-p}(\bar{\tau};t)
\end{equation}

Given the translation symmetry,  the following correlation functions vanish 
\be 
\avg{\hat \mT \eta_{s,0}(\bar \tau;t)\eta_{s,\pm}(0;0)}=\avg{\hat \mT \eta_{s,+}(\bar \tau;t)\eta_{s,+}(0;0)}=\avg{\hat \mT \eta_{s,-}(\bar \tau;t)\eta_{s,-}(0;0)}=0
\ee
and only  $\avg{\hat \mT \eta_{s,0}(\bar \tau;t)\eta_{s,0}(0;0)}$ and $\avg{\hat \mT \eta_{s,\mp}(\bar \tau;t)\eta_{s,\pm}(0;0)}$ could survive.
In terms of these three modes, the KMS conditions~\eqref{etakms} become
\begin{align} 
\avg{\hat \mT\eta_{s,p}(\bar  \tau;t-i\pi)\eta_{s,-p}(0;0)} & = s e^{-2\pi ip/3}\avg{\hat \mT\eta_{s,p}(\bar  \tau;t)\eta_{s,-p}(0;0)},   \label{kmseta}
\end{align}
Up to a phase these conditions are exactly the ordinary KMS conditions for inverse temperature $\b = \pi$. 
Equation~\eqref{kmseta} can also be interpreted as that $\eta_{s,p}$ have the following periodic conditions in the imaginary $t$ direction 
\be \label{yhi}
\eta_{s,p} (\bar  \tau;t- i\pi) = s e^{-2\pi ip/3} \eta_{s,p}(\bar  \tau;t) \ .
\ee 

Below we will also use Wightman functions 
\begin{align}
G_{s,p}^{>}(\bar{ \tau};t)=\avg{\eta_{s,p}(\bar{ \tau};t)\eta_{s,-p}(0;0)},\quad \Im t \in (-\pi, 0) \\
 G_{s,p}^{<}(\bar{ \tau};t)=\avg{\eta_{s,-p}(0;0)\eta_{s,p}(\bar{ \tau};t)} , \quad \Im t \in (0, \pi) \ .  \label{n2.36}
\end{align}
By translation symmetry, we have the relation $G_{s,p}^{<}(\bar{ \tau};t)=G_{s,-p}^{>}(-\bar{ \tau};-t)$, and the KMS condition \eqref{kmseta} can be written in terms of Wightman functions as
\begin{align}
G_{s,p}^{>}(\bar{ \tau};t-i\pi) & =s e^{-2\pi ip/3} G_{s,p}^{<}(\bar{ \tau};t),\quad \Im t\in(0,\pi) \ . \label{eq:x17}
\end{align}

We can now express $\avg{\hat\mT\phi_i(\bar t;t)\phi_j(0;0)}$ in terms of thermal correlation functions of $\eta_{s,p}$. 
From~\eqref{n2.16}, the relevant range for  $t$  
is $\Im t \in (-2 \pi, 0)$. 
As mentioned earlier, thermal correlation functions $\avg{\hat \mT \eta_{s,p}(\bar \tau;t)\eta_{s, p'}(0;0)}$ can have 
 discontinuities at $\Im t= 0, \pm \pi, \cdots$, which can potentially lead to discontinuity in $\avg{\hat\mT\phi_i(\bar t;t)\phi_j(0;0)}$
 at $\Im t=-\pi$, which would be unphysical.\footnote{As mentioned in item \ref{item:4b}, the only physical singularity for $\phi_i$ correlation function is at $\Im t=0$.} To make clear the potential discontinuity, it is convenient to 
write the two-point function of $\phi_i$ using Wightman functions of $\eta_{s, p}$, 
\begin{align}
\avg{\hat\mT\phi_{1}(\bar{ t};t)\phi_{1}(0;0)} &=\avg{\hat\mT\phi_{2}(\bar{ t};t)\phi_{2}(0;0)}\nn\\
&= \begin{cases}
\f 12\sum_{s,p}G_{s,p}^{>}(i\bar{ t};t), &\Im t\in[-\pi,0]\\
\f 12\sum_{s,p}se^{-2\pi i p/3}G_{s,p}^{>}(i\bar{ t};t+i\pi ), &\Im t\in[-2\pi,-\pi]
\end{cases}\label{eq:x29-2}  \\
\avg{\hat\mT\phi_{1}(\bar{ t};t)  \phi_{2}(0;0)}&= \begin{cases}
\f 12\sum_{s,p}e^{2\pi ip/3}G_{s,p}^{>}(i\bar{ t};t), &\Im t\in[-\pi,0]\\
\f 12\sum_{s,p}s G_{s,p}^{>}(i\bar{ t};t+i\pi), &\Im t\in[-2\pi,-\pi]
\end{cases} \label{eq:phi12}\\
\avg{\hat\mT\phi_{2}(\bar{ t};t)  \phi_{1}(0;0)} &= \begin{cases}
\f 12\sum_{s,p}e^{-2\pi ip/3}G_{s,p}^{>}(i\bar{ t};t), &\Im t\in[-\pi,0]\\
\f 12\sum_{s,p}se^{2\pi i p/3}G_{s,p}^{>}(i\bar{ t};t+i\pi), &\Im t\in[-2\pi,-\pi]
\end{cases} \label{eq:phi21}
\end{align}
where we have used~\eqref{kmseta} to shift the $t$ argument of $G^>_{s,p}$ such that it lies in the analytic domain of $G^>_{s,p}$. 

In~\eqref{eq:x29-2}, in order to avoid potential discontinuity at $\Im t=-\pi$ we need 
\begin{equation}
\sum_{s,p}G_{s,p}^{>}(\bar \tau, t)=\sum_{s,p}se^{-2\pi i p/3}G_{s,p}^{>}(\bar{ \tau};t + i\pi),\quad \Im t = - \pi, \quad \bar \tau\in[0,2\pi]  \ . \label{eq:x26-1}
\end{equation}
It is important to stress that with $\Im t = -\pi$, we have $\bar \tau \in [0,2 \pi]$ for two-point functions of $\phi_1$ or $\phi_2$ as indicated in the above equation. Now considering~\eqref{eq:phi12} and keeping in mind that for $\Im t = -\pi$, we have $\bar \tau \in [-\pi , \pi]$. In order to compare with~\eqref{eq:x26-1} we can shift $\bar \tau$ of~\eqref{eq:phi12} by $\pi$ using periodicity~\eqref{etasp}, after which we again find equation~\eqref{eq:x26-1}. Similarly in~\eqref{eq:phi21}, we have $\bar \tau \in [\pi, 3 \pi]$, and after shifting $\bar \tau$ by $-\pi$ we obtain the same equation. Equation~\eqref{eq:x26-1} should be understood as two equations, one for $\tau \in [0, \pi]$, and 
the other for $\tau \in [\pi, 2 \pi]$ which can in turn be shifted to the range $\tau \in [0, \pi]$ using periodicity~\eqref{etasp}. 
Applying~\eqref{eq:x17} to the left hand side of~\eqref{eq:x26-1} we find 
\begin{align}   
\sum_{p} e^{-2\pi i p/3}\mG_{+,p}(\bar\tau,t)&=\sum_{p} e^{-2\pi i p/3}\mG_{-,p}(\bar\tau,t) , \quad \Im t =0, \; \bar \tau \in [0, \pi]
\label{sm1}\\
\sum_{p} \mG_{+,p}(\bar\tau,t)&=\sum_{p} \mG_{-,p}(\bar\tau,t) , \quad \Im t =0, \; \bar \tau \in [0, \pi]
\label{sm2}
\end{align}
where we have defined for $\Im t =0$ 
\begin{equation}
\mG_{s,p}(\bar{ \tau};t)\equiv\avg{[\eta_{s,p}(\bar{ \tau};t),\eta_{s,-p}(0;0)]}=G_{s,p}^{>}(\bar{ \tau};t)-G_{s,p}^{<}(\bar{ \tau};t)=\mG_{s,-p}(-\bar{ \tau};-t) \label{n2.43}
\end{equation}
Equations~\eqref{sm1} and \eqref{sm2} can also be written in a more compact form
\be  
\mG_{+,\pm}(\bar\tau;t)-\mG_{-,\pm}(\bar\tau;t)=-e^{\pm \pi i/3}(\mG_{+,0}(\bar\tau;t)-\mG_{-,0}(\bar\tau;t))
\quad \Im t =0, \; \bar \tau \in [0, \pi] \ .
 \label{sm}
\ee


\subsection{The quadratic effective action} \label{sec:2.5action}

In this section, we will construct an effective action for $\eta_{s,p} (\bar \tau; t)$ defined in last subsection. We treat Euclidean time $\bar \tau$ as spatial coordinate in the range $\bar \tau \in [0, \pi]$ and $t$ as real time. 
$\eta_{s,p}(\bar\tau;t)$ satisfy the boundary conditions  \eqref{etasp} in $\bar \tau$ direction.
Real-time action for excitations in a thermal state can be formulated using the Schwinger-Keldysh formalism. We will follow the non-equilibrium EFT approach developed in~\cite{Crossley:2015evo, Glorioso:2017fpd, Liu:2018kfw, Blake:2017ris}.  

\begin{figure}
\begin{centering}

\includegraphics[height=3cm]{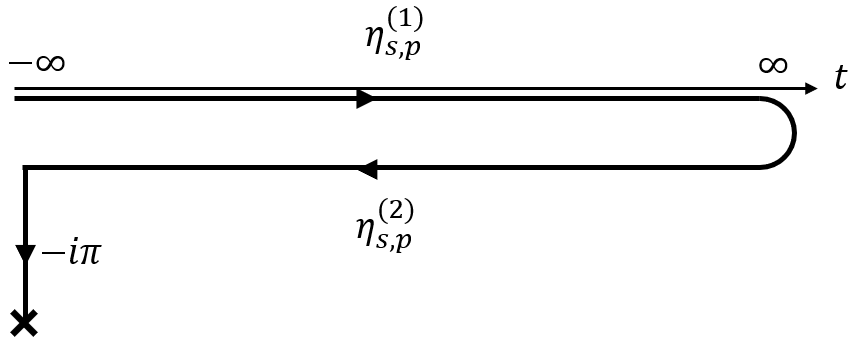} 

\par\end{centering}
\caption{The Keldysh contour of $t$ for $\eta_{s,p}(\bar \tau,t)$. The cross means multiplied with $se^{-2\pi ip/3}$ to respect KMS condition \eqref{kmseta}. \label{fig:keldysh}}

\end{figure}

To write down a real-time action we need to double the degrees of freedom on a two-way Keldysh contour for $t$, where the fields $\eta_{s,p}^{(1)}$
and $\eta_{s,p}^{(2)}$ are on the first and second contour respectively (see Fig. \ref{fig:keldysh}). For each $\eta_{s,p}$ we also have the so-called $r$-$a$ variables $\eta_{s,p}^r, \eta_{s,p}^a$ defined as
\begin{equation}
\eta_{s,p}^{a}(\bar{ \tau};t)=\eta_{s,p}^{(1)}(\bar{ \tau};t)-\eta_{s,p}^{(2)}(\bar{ t};t),\quad\eta_{s,p}^{r}(\bar{ \tau};t)=(\eta_{s,p}^{(1)}(\bar{ \tau};t)+\eta_{s,p}^{(2)}(\bar{ \tau};t))/2
\end{equation}
The effective action should satisfy various unitary constraints and the dynamical KMS condition (to ensure local thermal equilibrium). 
We derive these conditions in detail in Appendix \ref{app:dyna}, and just briefly present them here. 
\begin{enumerate}
    \item The action $S[\eta^r_{s,p},\eta^a_{s,p}]$ should contain terms in the form of
\be  
\int K^{\a_1\cdots \a_k}_{s_1,p_1,\cdots,s_k,p_k}(\del_{\bar\tau},\del_t)\eta^{\a_1}_{s_1,p_1}(\bar\tau,t)\cdots \eta^{\a_k}_{s_k,p_k}(\bar\tau,t), \quad (\a_1,\cdots\a_k\in\{a,r\})
\ee  
with $\prod_{i=1}^k s_i=1$ and $\sum_{i=1}^k p_i=3\Z$. 
\item Each term in above form must contain at least one $\eta^a_{s,p}$.
\item The imaginary part of effective is nonnegative $\Im S[\eta_{s,p}^{r},\eta_{s,p}^{a}] \geq 0$. \item For the terms with odd numbers of $\eta^a_{s,p}$, the action needs to be real, which means
\be  
K^{\a_1\cdots \a_k}_{s_1,p_1,\cdots,s_k,p_k}(\del_{\bar\tau},\del_t)=\left(K^{\a_1\cdots \a_k}_{s_1,-p_1,\cdots,s_k,-p_k}(\del_{\bar\tau},\del_t)\right)^* 
\ee 
for $\a_1,\cdots,\a_k$ contain odd numbers of $a$.
\item The action needs to obey dynamical KMS condition $S[\eta_{s,p}^{r},\eta_{s,p}^{a}]=S[\tilde{\eta}_{s,p}^{r},\tilde{\eta}_{s,p}^{a}]$, where $\tilde\eta_{s,p}^{r,a}$ are defined by \eqref{eq:d17} and \eqref{eq:d18-1}.
\end{enumerate}

At quadratic order, the effective action $S_{\rm EFT}$ can then be written as 
\be
S_{\rm EFT} = \sum_{s,p}\int_{0}^{\pi} d\bar\tau \int_{-\infty}^\infty dt \left[\eta_{s,-p}^{a}K_{s,p}^{ar}(\del_{\bar{\tau}},\del_t)\eta_{s,p}^{r}+\f 1 2\eta_{s,-p}^{a}K_{s,p}^{aa}(\del_{\bar{\tau}},\del_t)\eta_{s,p}^{a}\right]  \label{eq:2.53}
\ee
where from the above conditions we have
\bega  
K^{ar}_{s,p}(\del_{\bar\tau},\del_t)=\left(K^{ar}_{s,-p}(\del_{\bar\tau},\del_t) \ri)^*,\\
 \Im \sum_{s,p}\eta_{s,-p}^{a}K_{s,p}^{aa}(\del_{\bar{\tau}},\del_{t})\eta_{s,p}^{a}\geq 0 \ .
\end{gather} 
In Appendix \ref{app:dyna}, we derive the following dynamical KMS condition for the quadratic action \eqref{eq:2.53}
\begin{equation}
K_{s,p}^{ar}(\del_{\bar{\tau}},\del_{t})-K_{s,-p}^{ar}(-\del_{\bar{\tau}},-\del_{t})=-2is\left(\tan\pi(p/3-\del_{t}/2)\right)^{s}K_{s,p}^{aa}(\del_{\bar{\tau}},\del_{t}) \label{dkms}
\end{equation}
As shown in \cite{Glorioso:2016gsa}, setting $K^{aa}_{s,p}=0$ means the local entropy current is conserved and the system is non-dissipative. In this case, \eqref{dkms} reduces to
\be
K_{s,p}^{ar}(\del_{\bar{\tau}},\del_{t})=K_{s,-p}^{ar}(-\del_{\bar{\tau}},-\del_{t})=\left(K^{ar}_{s,p}(-\del_{\bar\tau},-\del_t) \ri)^* ,\label{nondis}
\ee
and the resulting action can be factorized~\cite{Blake:2021wqj,Glorioso:2016gsa,Liu:2018kfw}, i.e. $S_{\rm EFT}=S_f[\eta^{(1)}_{s,p}]-S_f[\eta^{(2)}_{s,p}]$ with 
\be  
S_f[\eta_{s,p}]=\f 1 2 \sum_{s,p}\int_{0}^{\pi} d\bar\tau \int_{-\infty}^\infty dt \, \eta_{s,-p}K_{s,p}^{ar}(\del_{\bar{\tau}},\del_t)\eta_{s,p}  \ .\label{faction}
\ee  
Taking $t \ra -i\tau$ in the above action we obtain a Euclidean action defined for both Euclidean times $\bar \tau, \tau$. 
We stress that the factorization and thus the Euclidean action are not possible when dissipations are included.
For simplicity, in this paper we will only consider the non-dissipative case with constraint \eqref{nondis} though the generalization to dissipative case should be straightforward.

As discussed earlier, we assume that the action can be expanded in derivatives of $\bar \tau$. 
As in~\cite{Blake:2017ris,Blake:2021wqj}, we cannot, however, expand the action in derivatives in $t$, since we are interested in time scales of order $1/\lam$ so as to be able to probe the exponential growth $e^{\lam t}$.  The Lyapunov exponent $\lam$ could be comparable to the inverse temperature $\b$, and thus there is no scale separation in $t$. 

Since $\eta_{s,p}(\bar\tau;t)$ have different boundary conditions~\eqref{etasp}, they allow 
different lowest order of $\del_{\bar\tau}$ in the action. For $\eta_{s,0}$, which is periodic in $\bar\tau$, the lowest order of $\del_{\bar\tau}$ in $K_{s,0}^{ar}$ is just constant, namely
\bega  \label{k0}
K_{s,0}^{ar}(\del_{\bar\tau},\del_t)=K_{s,0}(i\del_t)+O(\del_{\bar\tau}) \ .
\end{gather}
For $\eta_{s,\pm}$, which gains a nontrivial phase after shift $\bar\tau\ra\bar\tau+\pi$, the lowest order of $\del_{\bar\tau}$ in  $K_{s,\pm}^{ar}$ must be nontrivial, and we will keep to the linear order 
\be 
K_{s,\pm}^{ar}(\del_{\bar\tau},\del_t)=\del_{\bar\tau}K_{s,\pm}(i\del_t)+O(\del_{\bar\tau}^2)  \ .
\label{kpm}
\ee
It follows from \eqref{nondis} that for all $p$, 
\be 
K_{s,p}(i \del_t)=(-)^pK_{s,-p}(-i \del_t)  = (-)^p \left(K_{s,p}(-i \del_t)\ri)^* \ .
\label{Kp}
\ee
Thus $K_{s,0}(x)$ is an even function of $x$ with real coefficients (when expanded in power series), while $K_{s,\pm}(x)$ are functions of pure imaginary coefficients.

Keeping only leading orders, we can reduce $K^{ar}_{s,0}$ piece to one dimension of $t$ and write the leading order quadratic effective action as
\be 
S_{\rm EFT}=\sum_{s=\pm}\left[\int_{-\infty}^{\infty}dt \eta^a_{s,0}(t)K_{s,0}(i\del_t)\eta^r_{s,0}(t)+\sum_{p=\pm}\int_0 ^\pi d\bar\tau \int_{-\infty}^{\infty}dt \eta^a_{s,-p}(\bar\tau;t)\del_{\bar\tau}K_{s,p}(i\del_t)\eta^r_{s,p}(\bar\tau;t)\right]  \ .
\label{leadS}
\ee
With the leading order effective action \eqref{leadS}, we have
\begin{align} 
K_{s,0}(i\del_t)G_{s,0}^{ra}(t)&=-\d(t) \label{greeneq0}\\ \del_{\bar{\tau}}K_{s,\pm}(i\del_t)G_{s,\pm}^{ra}(\bar{\tau};t)&=- \d(\bar{\tau})\d(t) \label{greeneqp}
\end{align}
where $G^{ra}_{s,p}$ are retarded functions of $\eta_{s,p}$, i.e. 
\be
G_{s,p}^{ra}(\bar{ \tau};t)  = i \t(t)\mG_{s,p}(\bar{ \tau};t)  \ . 
\ee
Give the periodic boundary condition \eqref{etasp}, we can write 
\be  
\mG_{s,0}(\bar\tau;t)=\D_{s,0}(t),\quad \mG_{s,\pm}(\bar\tau;t)=\D_{s,\pm}(t)\left(e^{\mp 2\pi i /3}+\t(\bar\tau) (1-e^{\mp 2\pi i/3})\right) \label{2.73x}
\ee   
where $\D_{s,p}(t)$ can be written in Fourier space as
\begin{align}  
\t(t)\D_{s,0}(t)&=i\int_\mC d\w \f{e^{-i\w t}}{2\pi K_{s,0}(\w)}\label{2.74x-1}\\
\t(t)\D_{s,\pm}(t)&=i\int_\mC d\w \f{e^{-i\w t}}{2\pi K_{s,\pm}(\w)(1-e^{\mp 2\pi i/3})} \label{2.74x}
\end{align}
which holds for $t>0$. Here the integral contour $\mC$ must be above all poles of integrand on the complex $\w$ plane because $G^{ra}_{s,p}(\bar\tau;t)$ is proportional to $\theta(t)$. Note that $\theta (\bar \tau)$ in~\eqref{2.73x} comes from $\del_{\bar \tau}$ in \eqref{greeneqp}.

Equations~\eqref{2.73x} imply that except for certain jumps at $\bar\tau=0$, correlation functions have no dependence on $\bar \tau$. 
From item \ref{item:4b}, however,  such branch cut 
should not be present in the four-point function $\hat \mF$, and thus should be cancelled in ~\eqref{n2.16}, i.e. 
\be  
 \sum_{i,j=1,2} L_{t_i}\tilde L_{t_{j+2}} \left[g_W g_V \avg{\hat \mT \big(\phi_i(i \e;t_i)\phi_j(0; t_{j+2})-\phi_i(-i \e;t_i)\phi_j(0; t_{j+2})\big)} \ri]=0 \label{mcond}
\ee 
for infinitesimal positive $\e$.  Also note that the above condition is relevant only  for TOC of types $\avg{WVVW}$ and $\avg{VWWV}$ for which $\Im \bar t_W-\bar t_V$ could have either sign without changing the order of four fields. All other four-point functions have definite sign for $\Im \bar t_W-\bar t_V$.





Now recall the smoothness conditions \eqref{sm}, which upon using~\eqref{2.73x}-\eqref{2.74x} leads to
\be   
\f 1 {K_{+,\pm}(\w)}-\f 1 {K_{-,\pm}(\w)}=\mp\sqrt{3}i \left(\f 1 {K_{+,0}(\w)}-\f 1 {K_{-,0}(\w)}\right) \label{eq:2.76xx}
\ee 
which shows that the terms in \eqref{leadS} are not independent. It can also be checked that \eqref{eq:2.76xx} is consistent with constraints \eqref{Kp}.

\subsection{Shift symmetry and exponential growth in correlation functions} \label{sec:corr}

Similar to the maximal chaos case~\cite{Blake:2017ris}, we will postulate that the action and the vertices~\eqref{new-uen1}--\eqref{new-uen2} possess a shift symmetry.
The choice of the shift symmetry is motivated from that OTOCs should have exponential growth, but not TOCs.

It turns out the requirement can be achieved by the following two conditions 
\ben

\item The action and the vertices~\eqref{new-uen1}--\eqref{new-uen2}  are invariant under 
\begin{equation}
\eta_{-}^{r}\ra\eta_{-}^{r}+\a_+ e^{\lambda t }+ \a_- e^{-\lambda t }
\label{eq:x33}
\end{equation}
where $\a_\pm$ are constants.

\item There is no exponential growth in the symmetric correlation functions of $\eta_+$, i.e. 
\be 
G^{rr}_+(\bar\tau;t)= \sum_p G^{rr}_{+,p} (\bar\tau;t) 
= \f 12 \sum_p \left(G_{+,p}^{>}(\bar{ \tau};t)+G_{+,p}^{<}(\bar{ \tau};t)\right) 
=0 e^{\lam t} + \cdots  \ .
\label{Grr=0}
\ee
Note that the KMS condition \eqref{eq:x17} leads to the fluctuation-dissipation relation
\begin{equation}
G_{s,p}^{rr}(\bar{\tau};t)=\f 1 2 \f {1+s e^{2\pi ip/3}e^{-i\pi \del_t}}{1-s e^{2\pi ip/3}e^{-i\pi \del_t}}\mG_{s,p}(\bar{\tau};t),\quad \Im t\in(0,\pi) 
\ .
\label{eq:x18}
\end{equation}

\een

In terms of $\phi_{1,2}$, the shift symmetry~\eqref{eq:x33} can be written as
\begin{equation}
(\phi_{1},\phi_{2})\ra(\phi_{1},\phi_{2})+(e^{\pm\lambda(t+i\pi)},-e^{\pm\lambda t}) \ .
\label{eq:x44-1}
\end{equation}
We also require that the vertices in~\eqref{new-uen1}--\eqref{new-uen2}  be compatible with the shift symmetry~\eqref{eq:x33},
i.e. $L_t$ satisfies 
\begin{equation}
L_{t_{1}}[g_{W}(t_{12})e^{\pm\lambda(t_{1}+i\pi)}]=L_{t_{2}}[g_{W}(t_{12})e^{\pm\lambda t_{2}}]\label{eq:x44}
\end{equation}

Since $\eta_-$ is a sum of $\eta_{-,0}, \eta_{-,\pm}$, the invariance under \eqref{eq:x33} means that at least one of $K_{-,p}$ has a factor of $\del_t^2-\lam^2$. For convenience, we will write
\begin{equation}
K_{-,p}(i\del_{t})=(\del_{t}^{2}-\lambda^{2})k_{-,p} (i\del_{t}) \ .
\end{equation}
The constraint~\eqref{eq:2.76xx} implies that  $K_{+,p}$ may also contain a factor $\del_t^2-\lam^2$, and we can similarly write\footnote{Note that having a factor $(\del_{t}^{2}-\lambda^{2})$ in $K_{+,p}$ does not imply there is a shift symmetry in $\eta_+$. An important part of the shift symmetry is that vertices should also be invariant. 
}
\be
K_{s,p}(i\del_{t})=(\del_{t}^{2}-\lambda^{2})k_{s,p} (i\del_{t}) \ .
 \label{Ksp256}
\ee
From~\eqref{Kp} we have $k_{s,p}(x)=(-)^p k_{s,-p}(-x)$.\footnote{Note that it is enough to impose the invariance under a shift proportional to $e^{\lam t}$ in non-dissipative case. The parity property~\eqref{Kp} will then lead to invariance under a shift proportional to $e^{-\lam t}$.\label{fn6}} The general structure of our discussion will not depend on the specific forms of $k_{+, p}$ and $k_{-,p}$.

\begin{figure}
\begin{centering}
\includegraphics[height=4cm]{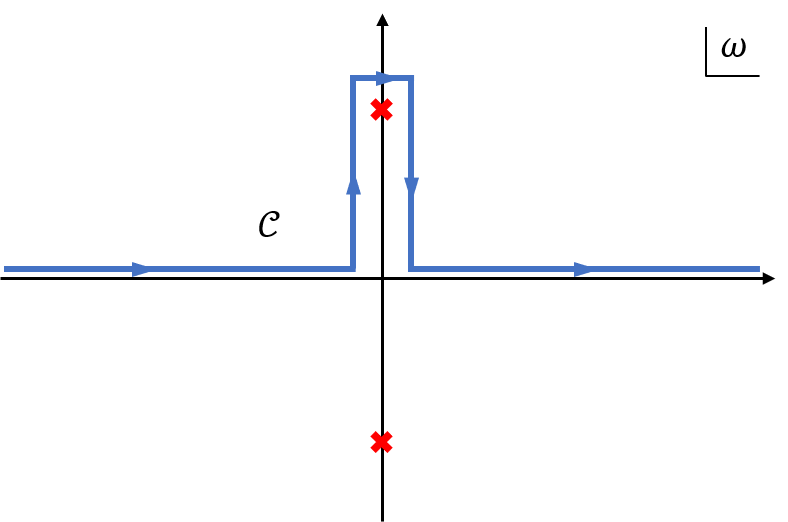}
\par\end{centering}
\caption{The contour $\mC$ on the $\protect\w$ plane for retarded propagator
$G^{ra}_{s,p}(\bar\tau;t)$. The two red crosses are poles at $\protect\w=\pm i \lambda$
due to shift symmetry. \label{fig:The-contour-choice}}
\end{figure}

Following from \eqref{2.74x-1}, we have
\be 
\t(t)\D_{s,0}(t)=\int_{\mC}\f{d\w}{2\pi i}\f{e^{-i\w t}}{(\w^{2}+\lambda^{2})k_{s,0}(\w)} \label{x2.103}
\ee
where the contour $\mC$ is chosen to be above all poles (see Fig.
\ref{fig:The-contour-choice}) of the integrand because the LHS is proportional to $\t(t)$. 
We then find\footnote{Note that when $k_{s,0} (x)$ has no zero at $x = i \lam$, the coefficient of the exponential pieces below vanishes.} 
\begin{align}
\D_{s,0}(t)&=\f{i}{2\lam k_{s,0}(i\lam)}(e^{\lam t}-e^{-\lam t})+\cdots \label{Ds0}
\end{align}
where we used the parity of $k_{s,0}$, and $\cdots$ denote possible contributions from other singularities. 
From now on we will suppress $\cdots$ and only write exponential terms. 
Using~\eqref{eq:x18} and 
\be
G_{s,p}^{>}(\bar{ \tau};t)  =G_{s,p}^{rr}(\bar{ \tau};t)+\f 12\mG_{s,p}(\bar{ \tau};t) \label{eq:x15}
\ee
we find from~\eqref{Ds0} 
\be 
G^{>}_{s,0}(\bar\tau;t)=\f{i}{2\lam k_{s,0}(i\lam)(1-se^{-i\lam \pi})}(e^{\lam t}+se^{-i\lam \pi}e^{-\lam t}) \equiv 
h_{s,0}(t)  \ .
\label{2.108}
\ee

Similarly, from \eqref{2.74x} we have
\begin{equation}
\t(t)\D_{s,\pm}(t)=\int_{\mC}\f{d\w}{2\pi i}\f{e^{-i\w t}}{(\w^{2}+\lambda^{2})k_{s,\pm}(\w)(1-e^{\mp 2\pi i/3})} \label{eq:x2.75}
\end{equation}
which leads to
\be 
\D_{s,\pm}(t)=\pm \f {e^{\pm i \pi/3}}{2\sqrt{3}\lambda}\left(\f{e^{\lambda t}}{k_{s,\pm}(i\lambda)}+\f{e^{-\lambda t}}{k_{s,\mp}(i\lambda)}\right) \label{Dspm}
\ee
and 
\begin{equation}
G_{s,\pm}^{>}(\bar{\tau};t)=\begin{cases}
h_{s,\pm}(t) & \bar{\tau}\in[0,\pi]\\
e^{\mp 2\pi i/3}h_{s,\pm}(t) & \bar{\tau}\in[-\pi,0]
\end{cases}\label{eq:x26}
\end{equation}
where $h_{s,\pm}$ are given by
\begin{align}
h_{s,\pm}(t)&=\pm \f {e^{\pm i \pi/3}}{2\sqrt{3}\lambda}\left(\f{(k_{s,\pm} (i\lambda))^{-1}e^{\lambda t}}{1-se^{-i\pi \lam}e^{\pm2\pi i /3}}+\f{(k_{s,\mp} (i\lambda))^{-1} e^{-\lambda t}}{1-se^{i\pi \lam}e^{\pm2\pi i /3}}\right) \ .
 \label{2.109}
\end{align}

The constraint \eqref{Grr=0} implies that, up to non-exponential pieces,
\be
G_{+}^{rr}(\bar \tau;t)=\sum_p \left( h_{+,p}(t)+e^{2\pi i p /3}h_{+,-p}(-t) \ri)=0 ,
\label{G+rr}
\ee
which upon using \eqref{2.108} and \eqref{2.109}, further implies 
\begin{equation}
k_{+,\pm}(i\lam)=\mp \f{ik_{+,0}(i\lam)}{\sqrt{3}}\tan\left(\f{\pi}{2}(\lam\pm 1/3)\right)\tan \f {\pi \lam}{2}    \ .
 \label{eq:x49}
\end{equation}
Note that~\eqref{eq:x49} is consistent with \eqref{Kp}. Notice that the factor 
$\tan\left(\f{\pi}{2}(\lam - 1/3)\right)$ on the right hand side of~\eqref{eq:x49} for $k_{+,-}(i\lam)$ becomes zero 
 for $\lam=1/3$, which cannot happen as a zero for $k_{+,-}(i\lam)$ would lead to divergences in~\eqref{2.109}.  
 This means that  $k_{+,0}$ must have a pole at $\lam = {1 \ov 3}$, i.e. $k_{+,0}(i\lam)\sim (\lam-1/3)^{-1}$, which in turn means that 
the prefactor in~\eqref{Ds0} vanishes and that the factor $\del_t^2-\lam^2$ in $K_{+,0}$ is in fact not there (it cancels with a factor hidden in $k_{+,0}$). For $\lam=2/3$, the right hand side of~\eqref{eq:x49} is divergent for $k_{+,+}(i\lam)$, which means that the factor $\del_t^2-\lam^2$ should also cancel for  $K_{+,+}(i\del_t)$ at $\lam=2/3$. The divergence of the factor $\tan \f {\pi \lam}{2}$ for $\lam =1$ will be commented on later in Sec.~\ref{sec:4}.  


\subsection{Summary of the effective field theory} 

We have now discussed all elements of the EFT formulation, which we summarize here in one place: 
\begin{enumerate}
\item The product $W(t_1)W(t_2)$ is written in terms of an expansion in terms of two effective fields $\phi_1(\bar t_W;t_S)$ and $\phi_2(\bar t_W;t_L)$ through a vertex. Similar expansion applies to $V(t_3)V(t_4)$. At leading nontrivial order in the $1/N$ expansion, 
we have 
\be \label{re1}
\hat \mF_{WWVV} (t_1,t_2;t_3,t_4)=g_W g_V+
 \sum_{i,j=1,2} L_{t_i}\tilde L_{t_{j+2}} \left[g_W g_V \avg{\hat \mT \phi_i(\bar t_W;t_i)\phi_j(\bar t_V; t_{j+2})} \ri] \ .
\ee
The domain of $\phi_i$ is given by $\mI_i$ in Fig. \ref{fig:nphifund}. 


\item The KMS conditions of $\hat \mF$ impose constraints on correlation functions of $\phi_i$, which can in turn 
be obtained in terms of those a new pair of fields 
\be\label{re11}
\eta_\pm(\bar t;t) =\f {1}{ \sqrt{2}} (\phi_1(\bar t;t-i\pi)\pm \phi_2(\bar t;t))
\ee
defined in the domain $\mI_2$. $\hat \mT$ in~\eqref{re1} is defined in terms of ordering of $\Im t$ for $\eta_\pm$.

\item  The effective action of $\eta_\pm$ is written for pure imaginary $\bar t=-i\bar \tau$ and real $t$. Correlation functions of $\eta_\pm$ for general complex $\bar t$ and $t$ are obtained from analytic continuation. 
We also assume that the effective action can be expanded in terms of derivatives of $\bar \tau$, which in turn implies that the action is translation invariant for both $\bar \tau$ and $t$. Two-point functions of $\eta_\pm$ are then defined in the domain $\mD_\eta$
of Fig.~\ref{pic:Dz}. 

The domain $\mD_\eta$ is irregular and inconvenient to work with. It is extended to $\mD_*$ of Fig.~\ref{pic:Dz}.  
$\eta_\pm$ is then further decomposed into  
\be \label{re2}
 \eta_s(\bar \tau ;t)=\eta_{s,0}(\bar \tau;t)+\eta_{s,+}(\bar \tau;t)+\eta_{s,-}(\bar \tau;t),
 \ee
 in terms of periodicity conditions in $\bar \tau$-direction 
\be\label{re3}
 \eta_{s,p}(\bar \tau + \pi;t)=e^{2\pi i p/3}\eta_{s,p}(\bar \tau;t), \qquad 
 s = \pm , \; p =0, \pm \ .
\ee

\item With the decomposition~\eqref{re2}--\eqref{re3}, the KMS conditions of the original four-point function $\hat \mF$ can be formulated in terms of KMS conditions for $\eta_{s,p}$ at the inverse temperature $\pi$ (half of the original inverse temperature), and
can be written as periodic conditions in the imaginary $t$ direction 
\be \label{re4}
\eta_{s,p} (\bar  \tau;t- i\pi) = s e^{-2\pi ip/3} \eta_{s,p}(\bar  \tau;t) \ .
\ee 
The leading actions in the $\bar \tau$-derivative expansion for $\eta_{s,0}$ contain no $\bar \tau$ derivative and thus 
$\eta_{s,0}$ can be thought as $\bar \tau$-independent, while the leading actions for $\eta_{s, \pm}$ contains one $\bar \tau$-derivative. 

\item From~\eqref{re11}, ~\eqref{re2} and~\eqref{re4}, we can write $\phi_{1,2}$ as
\bega \label{petr11}
\phi_1(\bar t;t) = \vp (t) +  e^{2\pi i \ov 3} \vp_+ (\bar t; t) +   e^{-{2\pi i \ov 3}} \vp_- (\bar t; t), \\
\label{petr12}
\phi_2(\bar t;t) = \vp (t) + \vp_+ (\bar t; t) +  \vp_- (\bar t; t) , \\
 \vp (t) \equiv {1 \ov \sqrt{2}}(\eta_{+,0} ( t) - \eta_{-,0} (t)), \quad
\vp_\pm (\bar t; t) \equiv  {1 \ov \sqrt{2}} (\eta_{+,\pm} (i \bar t;t) - \eta_{-,\pm} (i \bar t;t)),
 \label{petr13}
\end{gather} 
where we have used that $\eta_{\pm,0}$ can be viewed as being independent of $\bar \tau$. 
Note from~\eqref{re4} 
\be \label{rev5}
\vp (t - i \pi) = \tilde \vp (t) = {1 \ov \sqrt{2}}(\eta_{+,0} ( t) + \eta_{-,0} (t)), \quad \vp (t - 2 \pi i) = \vp (t)  \ .
\ee 

\item For OTOCs to have exponential dependence on $t$,  we impose the shift symmetry \begin{equation}
\eta_{-}^{r}\ra\eta_{-}^{r}+\a_+ e^{\lambda t }+ \a_- e^{-\lambda t }
\end{equation}
on both the action and the vertex. We also require no-exponential growth in $G_+^{rr}(\bar\tau;t)$, which is needed such that TOCs do not have exponential $t$-dependence. This condition requires that various terms in the action should obey~\eqref{eq:x49}. 

\item The effective action is further constrained by~\eqref{Kp}, and two continuity conditions~\eqref{mcond}--\eqref{eq:2.76xx}.

\end{enumerate}

\subsection{TOC and OTOC \label{sec:2.4}}

\begin{figure}
\begin{centering}
\subfloat[\label{pic:f4}]{\begin{centering}
\includegraphics[height=2cm]{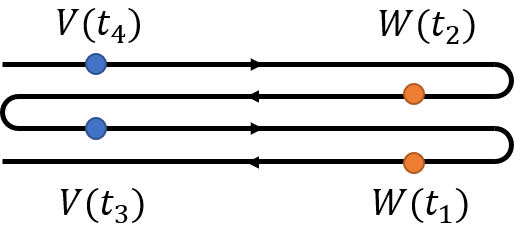} 
\par\end{centering}

}
\subfloat[\label{pic:g4}]{\begin{centering}
\includegraphics[height=2cm]{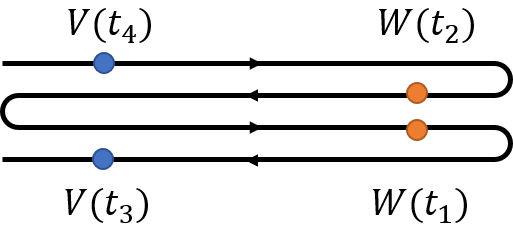} 
\par\end{centering}
}
\subfloat[\label{pic:h4}]{\begin{centering}
\includegraphics[height=2cm]{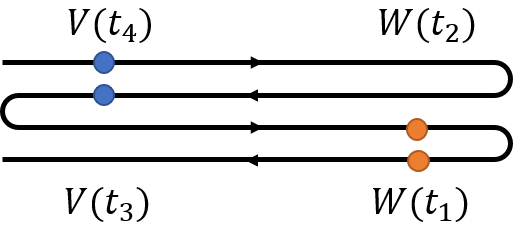} 
\par\end{centering}
}

\par\end{centering}
\caption{(a) The 4-way contour for $F_4$ (b) The 4-way contour for $G_4$ (c) The 4-way contour for $H_4$. \label{fig:F4G4}}

\end{figure}

Now consider the following two four-point functions
\begin{equation}
F_{4}=\avg{W(t_{1})V(t_{3})W(t_{2})V(t_{4})},\quad G_{4}=\avg{V(t_{3})W(t_{1})W(t_{2})V(t_{4})} \label{eq:FG2.63}
\end{equation}
where $\Re  t_{1},  \Re t_{2}\gg \Re t_{3},\Re t_{4}$ or $\Re t_{1},\Re t_{2}\ll \Re t_{3},\Re t_{4}$, i.e. $F_{4}$ is OTOC and $G_{4}$ is TOC. 
We suppose  each $t_i$ has a small imaginary part  such that the orderings in~\eqref{eq:FG2.63} follow that defined in~\eqref{yhef} (see Fig. \ref{fig:F4G4} as an illustration). For $F_4$, the small imaginary part for each $t_i$ leads to $\Im \bar t_W<\Im \bar t_V$, but for $G_4$, depending on the relative value of the imaginary part of each $t_i$, we may have either $\Im \bar t_W< \Im \bar t_V$ or $\Im \bar t_W>\Im \bar t_V$. For definiteness, we consider the former case $\Im \bar t_W< \Im \bar t_V$.
 From~\eqref{n2.16} and \eqref{eq:x29-2}-\eqref{eq:phi21}, we find that
\begin{align}
F_{4}-G_{4} & =\f 1 2 \sum_{s,p} L_{t_{1}}\tilde{L}_{t_{3}}\left[g_{W}g_{V}\left(G^>_{s,p}(i(\bar t_W-\bar t_V),t_{13})-G^<_{s,p}(i(\bar t_W-\bar t_V),t_{13})\right)\right]\nonumber \\
 & =\f 1 2 \sum_{s,p}L_{t_{1}}\tilde{L}_{t_{3}}\left[g_{W}g_{V} \avg{[\eta_{s,p}(i\bar t_W;t_1),\eta_{s,-p}(i\bar t_V;t_{3})]}\right] \nn\\
 & =L_{t_{1}}\tilde{L}_{t_{3}}\left[g_{W}g_{V}\D(t_{13})\right]
  \label{eq:F-G}
\end{align}
where we have used \eqref{sm2}, and 
\begin{align} 
\D(t)&\equiv\sum_p \D_{+,p}(t)=\sum_p \D_{-,p}(t)\nn\\
&=\f{3}{4\lam (1/2-\cos\pi\lam)\sin \f{\pi\lam}{2}k_{+,0}(i\lam)}(e^{i\pi\lambda/2}e^{\lambda t}+e^{-i\pi\lam/2}e^{-\lambda t}) \ .
\label{delta2.91}
\end{align}
In the second line of the above equation we have used \eqref{eq:x49}. We thus find that the difference between OTOC and TOC has exponential growth. Note that there is no divergence in~\eqref{delta2.91} at $\lam = {1 \ov 3}$; as mentioned earlier below~\eqref{eq:x49}, $k_{+,0}(i\lam)$ has a pole at $\lam = {1 \ov 3}$, which is canceled by $1/2-\cos \pi \lam$.

We  will now show that TOC $G_{4}$ does not have exponential
growth. In terms of Wightman functions \eqref{eq:x29-2} to \eqref{eq:phi21}, $G_4$ can be
written as
\begin{align}
G_{4}-g_{W}g_{V}=&\f 1 2  \sum_{s,p}L_{t_{1}}\tilde{L}_{t_{3}}\left[g_{W}g_{V}G_{s,p}^>(+;t_{31})\right]+e^{2\pi ip/3}L_{t_{2}}\tilde{L}_{t_{3}}\left[g_{W}g_{V}G_{s,p}^>(+;t_{32})\right]\nonumber \\
 & +e^{2\pi ip/3}L_{t_{1}}\tilde{L}_{t_{4}}\left[g_{W}g_{V}G_{s,p}^>(-;t_{14})\right]+L_{t_{2}}\tilde{L}_{t_{4}}\left[g_{W}g_{V}G_{s,p}^>(-;t_{24})\right] \label{eq:g4}
\end{align}
where we assume $\Im \bar{t}_W<\Im \bar{t}_V$ and $\pm$
sign in the first time argument means $\Im \bar{t}>0$ or $\Im \bar{t}<0$.
Using \eqref{2.108}, (\ref{eq:x26}), and (\ref{eq:x44}) (and the
counterpart for $\tilde{L}_{t}$), we can simplify \eqref{eq:g4} as 
\begin{equation}
G_{4}-g_{W}g_{V}=L_{t_{2}}\tilde{L}_{t_{4}}\left[g_{W}g_{V}\left(C_1e^{\lambda t_{24}}+C_2e^{-\lambda t_{24}}\right)\right] \label{eq:g4-2.87}
\end{equation}
where
\begin{align}
C_1=\f 1 2 \sum_{s,p}(1+e^{2\pi i p/3}e^{-i\pi \lam})A_{s,p}+(e^{-2\pi i p/3}+e^{i\pi \lam})B_{s,p}\label{C1} \\
C_2=\f 1 2 \sum_{s,p}(e^{-2\pi i p/3}+e^{-i\pi \lam})A_{s,p}+(1+e^{2\pi i p/3}e^{i\pi \lam})B_{s,p}\label{C2}
\end{align} 
and $A_{s,p}, B_{s,p}$ are defined from~\eqref{2.108} and~\eqref{2.109} as
\be 
h_{s,p} \equiv A_{s,p}e^{\lam t}+B_{s,p}e^{-\lam t} \ . \label{exphsp}
\ee
It can be checked using~\eqref{eq:x49} that, $C_1=C_2=0$. 

We can similarly examine $G_{4}$ for $\Im \bar{t}_W>\Im\bar{t}_V$
and another type of TOC
\begin{equation}
H_{4}=\avg{W(t_{1})W(t_{2})V(t_{3})V(t_{4})}
\end{equation}
with $\Im \bar{t}_W<\Im\bar{t}_V$. We again find their exponential growth pieces vanish due to (\ref{eq:x49}). In particular, the condition \eqref{mcond} is automatically satisfied up to non-exponential pieces.

Given that TOCs do not have exponential terms, equation~\eqref{delta2.91} implies that the exponential terms in 
OTOCs  depend only on $k_{+,0}(i\lam)$. Note that $k_{+,\pm}(i\lam)$ are determined from $k_{+,0}(i\lam)$ by \eqref{eq:x49}, and $k_{-,p}$ are also constrained from $k_{+,p}$ from~\eqref{eq:2.76xx}.


In \eqref{x2.103} and \eqref{eq:x2.75}, we assumed for simplicity that the integrand only has simple poles at $\pm i \lambda$.
This assumption can be relaxed to have higher order poles. In fact, it can be shown that at most double poles are allowed due to the condition \eqref{mcond}. These double poles lead to linear-exponential terms $te^{\pm\lam t}$ in the correlation functions of $\eta_\pm$. Interestingly, the contributions from the double poles to any four-point function cancel out. 
So what we discussed in fact gives the most general form for four-point functions. See Appendix~\ref{app:b} for details. 

\subsection{General structure of OTOCs for non-maximal chaos}

We have seen that the shift symmetry \eqref{eq:x44-1} and requirement~\eqref{Grr=0} guarantee exponential growth of OTOC and the absence of exponential growth of TOC. We will now examine the general structure of OTOCs as predicted by the theory. 

Using \eqref{n-uen}, we can expand \eqref{eq:x44} explicitly as 
\begin{equation}
\sum_{mn}c_{mn}\left(-e^{\pm \lambda(t+i\pi)}+(-1)^{m}\right)\del_{t}^{m}g_{W}(t)(\pm \lambda)^{n}=0\label{eq:92}
\end{equation}
Similar to \cite{Blake:2021wqj}, we define
\begin{align}
G^W_{even}(\pm\lambda,t)=\sum_{m\text{ even}}c_{mn}\del_{t}^{m}g_{W}(t)\sum_{n}(\pm \lambda)^{n}\label{eq:91-0}\\
G^W_{odd}(\pm\lambda,t)=\sum_{m\text{ odd}}c_{mn}\del_{t}^{m}g_{W}(t)\sum_{n}(\pm \lambda)^{n}\label{eq:91}
\end{align}
and (\ref{eq:92}) becomes
\begin{equation}
\f{G^W_{even}(\pm\lambda, t)}{G^W_{odd}(\pm\lambda, t)}=\mp\coth\f{\lambda( t+i\pi)}2\label{eq:94}
\end{equation}
KMS transformation of $g_{W}( t)$ is $ t\ra- t-2\pi i$ that
leads to
\begin{align}
G^W_{even}(\mp\lambda, t) & \ra G^W_{even}(\mp\lambda,- t- 2\pi i)=G_{even}(\mp\lambda, t)\\
G^W_{odd}(\mp\lambda, t) & \ra G^W_{odd}(\mp\lambda,- t-2\pi i)=-G_{odd}(\mp\lambda, t)
\end{align}
where we used invariance of $g_{W}$ under KMS transformation. This
is compatible with (\ref{eq:94}) without any restriction on $\lambda$,
unlike the EFT of \cite{Blake:2017ris}, where the KMS condition of $g_W$ restricts
$\lambda=\lambda_{\max}=1$~\cite{Blake:2021wqj}.

Using the definition in \eqref{eq:91-0} and \eqref{eq:91},
we can write the OTOC $F_4$ in a more symmetric way. Since TOC $G_4$ does not have exponential piece, OTOC $F_4$ has the same exponential piece as \eqref{eq:F-G}. Using the shift symmetry of vertex \eqref{eq:x44}, we can write each exponential in a symmmetric way
\begin{align}
L_{t_1}[g_W e^{\pm\lam t_1}]&=\f 1 2\left(L_{t_1}[g_W e^{\pm\lam t_1}]+L_{t_2}[g_W e^{\pm\lam (t_2-i\pi)}]\right) \\
\tilde L_{t_3}[g_V e^{\pm\lam t_3}]&=\f 1 2\left(\tilde L_{t_3}[g_W e^{\pm\lam t_3}]+\tilde L_{t_4}[g_W e^{\pm\lam (t_4-i\pi)}]\right)
\end{align}
Using \eqref{eq:91-0} and \eqref{eq:91}, we can write the connected piece
\begin{align}
F_4=& \a e^{\lambda(t_{1}+t_{2}-t_{3}-t_{4}+i\pi)/2} \left(G^W_{even}(\lambda,t_{12})\cosh\f{\lambda(t_{12}+i\pi)}2+G^W_{odd}(\lambda,t_{12})\sinh\f{\lambda(t_{12}+i\pi)}2\right)\nonumber \\
 & \times\left(G^V_{even}(-\lambda,t_{34})\cosh\f{\lambda( t_{34}+i\pi)}2-G^V_{odd}(-\lambda,t_{34})\sinh\f{\lambda(t_{34}+i\pi)}2\right)+(\lambda\leftrightarrow-\lambda)\nonumber \\
= & \a e^{\lambda(t_{1}+t_{2}-t_{3}-t_{4}+i\pi)/2} \f{G^W_{even}(\lambda,t_{12})G^V_{even}(-\lambda,t_{34})}{\cosh\f{\lambda(t_{12}+i\pi)}2\cosh\f{\lambda(t_{34}+i\pi)}2}+(\lambda\leftrightarrow-\lambda)\label{eq:151}
\end{align}
where we have used $\al$ to denote the prefactor of~\eqref{delta2.91}, i.e.~\eqref{delta2.91} becomes 
\be 
\D(t) \equiv \a(e^{i\pi\lambda/2}e^{\lambda t}+e^{-i\pi\lam/2}e^{-\lambda t}), \label{deltanew}
\ee
and used (\ref{eq:94}) in the last line. Equation~\eqref{eq:151} has the same structure as that assumed in \cite{Kitaev:2017awl,Gu:2018jsv}, including the phase $e^{i\pi\lam/2}$. In~\cite{Kitaev:2017awl,Gu:2018jsv}, in the place of $G^W_{even}(\lam,t)$ and $G^V_{even}(-\lam,t)$ are certain advanced and retarded vertices, which are invariant under KMS transformation $t\ra-t-2\pi i$. For a specific microscopic system, these two vertices may obey certain differential equations  \cite{Kitaev:2017awl,Gu:2018jsv}, which in our language translate into conditions on the effective vertices.  In \eqref{eq:151}, there is a second term, obtained from $\lam\ra -\lam$, which exponentially decays for $t_1,t_2\gg t_3,t_4$, and was not present in~\cite{Kitaev:2017awl,Gu:2018jsv}. Here it is a consequence of shift symmetry for both signs in \eqref{eq:x33}, which are needed in order for TOCs to not have exponential growth.  In the non-dissipative case, this term should also exist even if we only assume shift symmetry for just plus sign in \eqref{eq:x33} as explained in footnote \ref{fn6}.


Now consider the double commutator defined as 
\be 
C^\t_4(t_1,t_2;t_3,t_4)=\avg{[W(t_1-i\t),V(t_3-i\t)][W(t_2),V(t_4)]},\quad \t\in[0,2\pi]
\ee
which can be rewritten  in terms of four-point functions leads to
\be
C_4^\t=F_4^\t-G_4^\t+\tilde F_4^\t-\tilde G_4^\t
\ee
where $F_4^\t$ and $G_4^\t$ are the four-point functions in \eqref{eq:FG2.63} with $t_1\ra t_1-i\t$ and $t_3\ra t_3-i\t$, and $\tilde F_4^\t$ and $\tilde G_4^\t$ are respectively $F_4^\t$  and $G_4^\t$ with exchange $W\leftrightarrow V$, $t_{1}\leftrightarrow t_{3}$ and $t_{3}\leftrightarrow t_{4}$. 
It follows from~\eqref{eq:151} that
\be 
C_4^\t=2\a \cos\f{\lam \pi}{2} e^{\lambda(t_{1}+t_{2}-t_{3}-t_{4})/2} \f{G^W_{even}(\lambda,t_{12}-i\t)G^V_{even}(-\lambda,t_{34}-i\t)}{\cosh\f{\lambda(t_{12}+i(\pi-\t))}2\cosh\f{\lambda(t_{34}+i(\pi-\t))}2}+(\lambda\leftrightarrow-\lambda) \ .
\ee
The factor $\cos\f{\lam \pi}{2} \to 0$ as $\lam \to 1$, consistent with the result of~\cite{Blake:2021wqj}
and those of SYK and holographic systems in the maximal chaos limit~\cite{Shenker:2014cwa,Kitaev:2017awl,Gu:2018jsv,kitaevtalk}.

\section{Comparisons with OTOCs in various theories}  \label{sec:motoc}

In this section we compare the general structure of OTOCs obtained in last section with various known examples. 

\subsection{The large $q$ SYK model}
We first look at the large $q$ SYK model \cite{sachdev1993gapless,kitaev2015simple,Maldacena:2016hyu,Cotler:2016fpe}. OTOC $F_4$ of fundamental fermions was obtained in \cite{Choi:2019bmd} and has the form (after analytic continuation to Lorentzian signature)
\be  
F^{\rm SYK}_{4}=-\f 2 N \f {g_\psi(t_{12})g_\psi(t_{34})\cosh\f{\lam(t_1+t_2-t_3-t_4+i\pi)} {2} }{\cos \f {\pi \lam} 2 \cosh \f {\lam(t_{12}+i\pi)}2 \cosh \f {\lam(t_{34}+i\pi)}2} \label{sykotoc1}
\ee  
where $G_\psi(t)$ is the two-point function fundamental Majorana fermions in the large $q$ SYK model (see more details in Section \ref{sec:syk4.1}) given by
\be\label{pio-1}
g_\psi(t)\equiv\f 1 2 \left(\f {\cos \f {\lam \pi}{2}}{\cosh\f{\lambda(t+i\pi)}{2}}\right)^{2\D}
\ee
where $\D=1/q$ is the conformal weight of the fundamental fermion $\psi$. The 
OTOC~\eqref{sykotoc1} has exactly the form of~\eqref{eq:151}, with
the $\cosh\lam(t_1+t_2-t_3-t_4+i\pi)/2$ term containing exponentially growing and decaying terms that are both present in \eqref{eq:151}, including the phases. We can further identify 
\be  
 G^W_{even} (\lam, t)=G^V_{even} (\lam, t) = C_0 (\lam)  g_\psi (t),
\quad \a=-\f{1}{C_0 (\lam) C_0 (-\lam) N\cos(\pi\lam/2)} \label{alpha4.18}
\ee
where $C_0 (\lam)$ is a constant. From~\eqref{eq:91-0} we find that 
$C_0 (\lam) = \sum_n c_{0,n} \lam^n$. From~\eqref{eq:94} we find 
\be
G^\psi_{odd}(\pm\lam,t) = \del_t g_\psi (t) C_1 (\lam), \quad C_1 (\lam) = \sum_n c_{1,n} \lam^n, \quad C_0 (\lam) = \lam \Delta C_1 (\lam) \ .
\ee
As the simplest possibility we take $c_{1,n}=\d_{n,0}$ and $c_{0,n}=\D\d_{n,1}$ which gives
\be  \label{pio-2}
L_t[W]\phi = \del_t W \phi+\D W \del_t \phi,\quad \a=\f{1}{\lam ^2 N \D^2\cos(\pi \lam /2)}
\ee

\subsection{Stringy scattering in a AdS black hole}
The next example is the scattering in the AdS black hole background with stringy correction \cite{Shenker:2014cwa}. Assume the bulk spacetime dimension is $d+1$ and the $d$-dimensional boundary coordinate is $x^\mu=(t,\vec x)$. When $t_1,t_2\gg t_3,t_4$ or $t_1,t_2 \ll t_3,t_4$, the boundary out-of-time-ordered four-point function $F_4$ is dominated by the $W+V\ra W+V$ scattering near the horizon of the AdS black hole. At $G_N\sim 1/N$ order, the OTOC $F_4$ is given by\footnote{The notation here differs from \cite{Shenker:2014cwa} by switching $2\leftrightarrow 3$.} 
\be  
F_4^{\rm string}= \f{i a_0^4}{(2\pi)^4}\int \d(s,|\vec z-\vec z'|)\left[p^u \psi_1^*(p^u,\vec z;x^\mu_1)\psi_2(p^u,\vec z;x^\mu_2)\right]\left[p^v \psi_3^*(p^v,\vec z';x^\mu_3)\psi_4(p^v,\vec z';x^\mu_4)\right] \label{stringF4}
\ee
where $a_0$ is a number depending on the background, $s=a_0 p^u p^v $ is the total energy of the scattering in the unboosted frame, $\psi_{1,2}(p^u,\vec z; x^\mu_{1,2})$ are wavefunctions of $W(x^\mu_{1,2})$ expanded in the null momentum basis $p^u$, $\psi_{3,4}(p^v,\vec z;x^\mu_{3,4})$ are wavefunctions of $V(x^\mu_{3,4})$ expanded in the orthogonal null momentum basis $p^v$, $\vec z$ and $\vec z'$ are $d-1$ dimensional bulk transverse coordinates, and the integral runs over $p^u,p^v,\vec z,\vec z'$. With stringy correction, the scattering amplitude $\d(s,|\vec z|)$ for $t_1+t_2\ll t_3+t_4$ is given by
\be  
\d(s,|\vec z|)\sim G_N s\int \f {d^{d-1}k}{(2\pi)^{d-1}}\f {e^{i\vec k\cdot \vec z}}{\vec k ^2+\mu^2}(e^{-i\pi/2}\a' s/4)^{-\a'(\vec k ^2+\mu^2)/2r_0^2} \label{stringd}
\ee 
where $r_0$ is the horizon radius, $\a'=\l_s^2$ (with $\l_s$ the string length) and $\mu^2=\f{d(d-1)r_0^2}{2\l_{AdS}^2}$. 

By translation symmetry of the AdS-Schwarzschild black hole in transverse directions, $\psi_i(p,\vec z;x^\mu)$ are functions of $\vec z-\vec x$. By time translation symmetry, one can show that the wave function $\psi_i(p,\vec z;x^\mu)$ in the unboosted frame has the following property
\begin{align}  
\psi_{1,2}(p^u,\vec z;x^\mu)&=e^{-2\pi a/\b}\psi_{1,2}(p^ue^{-2\pi a/\b},\vec z;(t+a,\vec x)) \\
\psi_{3,4}(p^v,\vec z;x^\mu)&=e^{2\pi a/\b}\psi_{3,4}(p^ve^{2\pi a/\b},\vec z;(t+a,\vec x))
\end{align} 
which implies that $\psi_{1,2}=e^{2\pi t/\b}f_{1,2}(p^u e^{2\pi t/\b})$ and $\psi_{3,4}=e^{-2\pi t/\b}f_{3,4}(p^v e^{-2\pi t/\b})$ for some functions $f_i$. Here $\b$ is the inverse temperature of the black hole. Switching to the boosted frame by redefining $p^u\ra p^ue^{-\pi(t_1+t_2)/\b}$ leads to
\be  
dp^u p^u \psi_{1}^*(p^u,\vec z;x_1^\mu)\psi_{2}(p^u,\vec z;x_2^\mu)\ra dp^u p^uf_{1}^*(p^u e^{\pi (t_1-t_2)/\b})f_{2}(p^u e^{\pi (t_2-t_1)/\b})
\ee 
which is a function of $t_{12}$. Similarly we can redefine $p^v\ra p^v e^{\pi(t_3+t_4)/\b}$ and find the wavefunctions for $V$ become a function of $t_{34}$. Taking this into \eqref{stringF4}, we can derive
\be  
F_4^{\rm string}= i\int \d(s,|\vec z-\vec z'|)f_W(p^u,\vec z-\vec x_1,\vec z-\vec x_2;t_{12})f_V(p^v,\vec z'-\vec x_3,\vec z'-\vec x_4;t_{34})\label{stringF4-1}
\ee 
with $s=a_0 p^u p^v e^{-\pi (t_1+t_2-t_3-t_4)/\b} $ and two functions
\begin{align}  
f_W&=p^uf_{1}^*(p^u e^{\pi (t_1-t_2)/\b})f_{2}(p^u e^{\pi (t_2-t_1)/\b})\label{3.15fW}\\
f_V&=p^vf_{3}^*(p^v e^{\pi (t_4-t_3)/\b})f_{4}(p^v e^{\pi (t_3-t_4)/\b})
\end{align}

In the regime that the scattering amplitude $\d(s,|\vec z|)$ is of order one and slowly varies with respect to $p^u$ and $p^v$, we can assume the integral over $p^u$ and $p^v$ in \eqref{stringF4-1} can be approximated by their characteristic values $p_c^u$ and $p_c^v$, which only depend on the wave functions. Moreover, the spatial dependence of wave function $\psi_i$ should be peaked around $\vec z\sim \vec x$. To compare with our 0+1 dimensional EFT, we should integrate over all $\vec x$, which basically sets $\vec x\sim\vec z$. Since wave functions are translational invariant along tranverse directions, we can integrate over $\vec z$ directly in \eqref{stringd}, which fixes $\vec k=0$ and leads to
\be  
\d(s,0)\sim -iG_N c_d e^{-\lam(t_1+t_2-t_3-t_4+i\b/2)/2}\quad (t_3+t_4\gg t_1 +t_2)
\ee 
where $c_d$ is a real constant and the Lyapunov exponent $\lam$ is
\be  
\lam=\f{2\pi}\b \left(1-\f{d(d-1)\a'}{\l_{AdS}^2}\right)
\ee 
It follows that we can write \eqref{stringF4} as 
\be  
F^{\rm string}_4\sim  G_N c_d e^{-\lam(t_1+t_2-t_3-t_4+i\b/2)/2} f^c_W(t_{12}) f^c_V(t_{34}) \label{f4string}
\ee 
where $f^c_{W,V}$ means $f_{W,V}$ taking value at $p^{u,v}=p^{u,v}_c$ and we have suppressed all transverse coordinates. Comparing \eqref{f4string} with \eqref{eq:151}, we see that they both have non-maximal exponential growth 
 and the phase $-i\lam\b/4$ in \eqref{f4string} exactly matches with $-i\lam\pi/2$ (of the $-\lam$ term) in \eqref{eq:151} by $\b=2\pi$. In the case of $t_1+t_2\gg t_3+t_4$, we need to flip the phase $e^{-i\pi/2}$ to $e^{i\pi/2}$ in \eqref{stringd} and will find consistency with \eqref{eq:151} as well.
 
 Matching \eqref{f4string} with \eqref{eq:151} also leads to 
 \be  
 f^c_W(t)=\f{G^{W}_{even}(\lam,t)}{\cosh\f{\lam(t+i\pi)}{2}},\quad  f^c_V(t)=\f{G^{V}_{even}(-\lam,t)}{\cosh\f{\lam(t+i\pi)}{2}} \ .
  \label{matchf}
 \ee  
For example, in AdS$_3$ \cite{Shenker:2014cwa}, we have ($\b=2\pi$)
 \be  
  f_W(p^u,0,0;t_{12})=\f{4^{\D_W}\pi^2 c_W^2}{\G(\D_W)^2}(p^u)^{2\D_W-1}e^{-4ip^u\sinh(t_{12}/2)} \label{2.144wf}
 \ee  
 where we have set all transverse coordinates as zero due to the integral over these directions. For large $\D_W$, the characteristic $p^u$ is at $p^u_c=\f{2\D_W-1}{4i \sinh(t_{12}/2)}$, which leads to
 \be 
f^c_W(t)\sim \f {g_W(t_{12})} {\cosh\f{t+i\pi}{2}}, \quad g_W(t)\equiv \f 1 {(\cosh\f{t+i\pi}{2})^{2\D_W}} \label{3.15}
 \ee  
where $g_W(t)$ is the boundary two-point function in a thermal state. 
Note that the wave function \eqref{2.144wf} is computed in the pure gravity background and does not include any stringy corrections that should introduce non-maximal $\lam$ to the wave function. Thus~\eqref{3.15} should be compared with the right hand side of~\eqref{matchf} at leading order in $\al'$ expansion, i.e. we can identify $G^W_{even}(1,t)=g_W(t)$. 
It is clearly of interests to understand $\al'$ corrections of~\eqref{2.144wf}. 
In general dimension, the explicit forms of wave functions are not known, but~\eqref{matchf} gives a constraint on their general structure. 



\subsection{Conformal Regge theory}

\begin{figure}
\begin{centering}
\includegraphics[height=6cm]{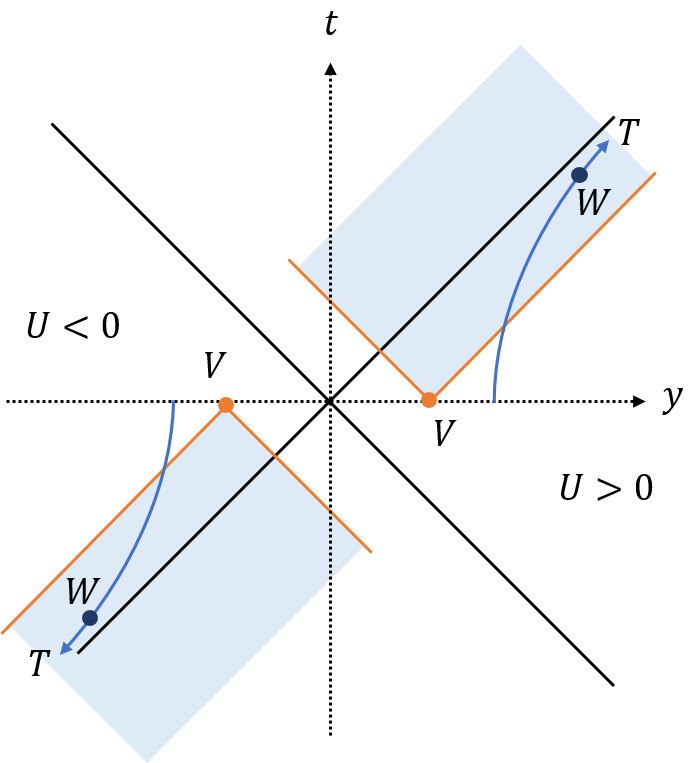}
\par\end{centering}
\caption{The locations of four operators in Regge limit. \label{fig:regge}}
\end{figure}

The conformal Regge theory was developed in \cite{Cornalba:2007fs,Costa:2012cb} and analyzed for Lyapunov exponent and butterfly effect in \cite{Mezei:2019dfv}. Consider a four-point function $\avg{W(x_1)V(x_3)V(x_4)W(x_2)}$ in the CFT in $d$-dimensional Minkowski spacetime. Assume their locations are assigned in the $(t,y)$ plane with
\be  
x_i=(t_i,y_i,0,\cdots,0)
\ee 
On this plane, we can define Rindler coordinate \cite{Mezei:2019dfv}
\be  
t=U \sinh T,\quad y=U\cosh T
\ee 
where $U>0$ is for right Rindler wedge and $U<0$ is for left Rindler wedge (see Fig. \ref{fig:regge}). These two wedges are related by analytic continuation $T\ra T+i\pi$ and the vacuum in Minkowski spacetime is equivalent to the thermal state in one of the Rindler wedge with inverse temperature $2\pi$. We can consider each pair of $W$ and $V$ located in two different wedges, by which we can construct an OTOC in one Rindler wedge with inverse temperature $2\pi$. For example, we can take $T_1=T_2=T$, $T_3=T_4=0$, $U_1=-U_2=U>0$ and $U_3=-U_4=1$ (see Fig. \ref{fig:regge}), which leads to
\be  
\avg{W(T,U)V(0,1)V(0,-1)W(T,-U)}_{\rm M}=\avg{W_R(T,U)V_R(0,1)W_R(T+i\pi,U)V_R(i\pi,1)}_{\rm R}
\ee  
where the LHS is the correlation function in Minkowski vacuum and the RHS is a thermal OTOC in right Rindler wedge. The conformal Regge theory studies this four-point function under Regge limit with $T\ra \infty$ but fixed $U$. It has been shown \cite{Mezei:2019dfv} that it has a non-maximal exponential growth $e^{\lam T}$ with $\lam < 1$. This non-maximal Lyapunov exponent comes from summing over infinitely many higher spin channels in the four-point function.



To compare with our 0+1 dimensional result, we should perform dimensional reduction by restricting to zero momentum mode along all spatial directions, which is a bit intricate. Here we will simply  take $U_i=U$ for all $i$ and compare the $T$-dependence with~\eqref{eq:151}.
 
 By conformal symmetry, the four-point function $\avg{W(x_1)V(x_3)V(x_4)W(x_2)}$ can be written as 
\be  
\avg{W(x_1)V(x_3)V(x_4)W(x_2)}=\f{1}{(x^2_{12})^{\D_W} (x^2_{34})^{\D_V}}\mA(u,v) \label{3.19}
\ee 
where the conformal invariant cross ratios are 
\be  
u=\f{x_{12}^2x_{34}^2}{x_{13}^2x_{24}^2}=\f{\sinh^2 \f {T_{12}}{2}\sinh^2 \f {T_{34}}{2}}{\sinh^2 \f {T_{13}}{2}\sinh^2 \f {T_{24}}{2}},\quad v=\f{x_{14}^2x_{23}^2}{x_{13}^2x_{24}^2}=\f{\sinh^2 \f {T_{14}}{2}\sinh^2 \f {T_{23}}{2}}{\sinh^2 \f {T_{13}}{2}\sinh^2 \f {T_{24}}{2}}
\ee 
The Regge limit is for $T_1,T_2\gg T_3,T_4$ limit, and we have 
\be  
u\ra 16e^{-T}\sinh^2 \f {T_{12}}{2}\sinh^2 \f {T_{34}}{2},\quad v\ra 1-8 e^{-T/2}\sinh\f {T_{12}}{2}\sinh \f {T_{34}}{2} \label{3.21}
\ee 
where we define $T=T_1+T_2-T_3-T_4\gg 0$. From \cite{Cornalba:2007fs,Costa:2012cb}, $\mA(u,v)$ under this limit is given by\footnote{To get the correct phase $e^{-i\pi/2}$, one needs to be careful about how $u$ and $v$ circle around origin after we set infinitesimal imaginary part $T_i\ra T_i+i\e_i$ with $\e_1<\e_3<\e_2<\e_4$ for OTOC. Unlike \cite{Costa:2012cb}, both $u$ and $v$ circle around origin clockwise for $2\pi$ when $T\gg 0$ in our case.}
\be  
\mA \app 2\pi \int d\nu \tilde \a(\nu)  \left(\f{e^{-i\pi/2}}{2}\log v\right)^{1-j(\nu)}\Omega_{i\nu}(0) \label{mA}
\ee 
where $\Omega_{iv}(\rho)$ is a harmonic function on $(d-1)$-dimensional hyperbolic space (here by assuming $U_i=U$ we have $\rho=0$), $j(\nu)$ is the leading Regge trajectory, and $\tilde \a(\nu)$ is a slowly varying function of $\nu$. In large $T$ limit, 
the $\nu$-integral can be evaluated using saddle point approximation with the saddle point given by $\nu=0$~\cite{Mezei:2019dfv},
which gives
\be  
\avg{W(x_1)V(x_3)V(x_4)W(x_2)}\app C_R\f {g_W(T_{12})g_V(T_{34})}{\left(\cosh \f {T_{12}+i\pi}{2}\right)^\lam\left(\cosh \f {T_{34}+i\pi}{2}\right)^\lam}e^{\lam(T+i\pi)/2}  \ .
\label{regotoc}
\ee 
Here $C_R$ is a constant, $\lam=j(0)-1$, and $g_{W, V}$ is the conformal correlator in Rindler wedge with no spatial separation
\be  
g_{W,V}(T)=\f{1}{\left(\cosh\f{T+i\pi}{2}\right)^{2\D_{W,V}}}
\ee 


Now comparing~\eqref{regotoc} with~\eqref{eq:151}, we see the downstairs of~\eqref{regotoc} is proportional to 
$(\cosh \f {T_{12}+i\pi}{2})^\lam$ while in~\eqref{eq:151} we have $\cosh \f {\lam (T_{12}+i\pi)}{2}$. It is not clear whether this difference is due to we are comparing a $d$-dimensional theory with a $(0+1)$-dimensional system. 
Assuming not, we can match~\eqref{regotoc} with~\eqref{eq:151} with the identification 
\begin{align}  
G^{W,V}_{even}(\pm\lam,T)&= \f{K_{W,V}(\pm\lam){\cosh\f{\lam(T+i\pi)}{2}}}{\left(\cosh \f {T+i\pi}{2}\right)^{2\D_{W,V}+\lam}}\\
\a&=\f{C_R}{K_W(\lam) K_V(-\lam)}=\f{C_R}{K_W(-\lam) K_V(\lam)}
\end{align} 
up to arbitrary $K_{W,V}(\lam)$.
 Since this equation is essentially the same for $W$ and $V$ except conformal dimensions, we will suppress $W,V$ labels in the following. Using \eqref{eq:94}, we have 
\be  
G_{odd}(\pm\lam,T)=\mp\f{K(\pm\lam){\sinh\f{\lam(T+i\pi)}{2}}}{\left(\cosh \f {T+i\pi}{2}\right)^{2\D+\lam}}
\ee 
which leads to
\begin{align}  
&G(\pm\lam,T)\equiv \sum_{mn}c_{mn}\del^m_T G(T) (\pm\lam)^n \label{dfGT}\\
=&G_{even}(\pm\lam,T)+G_{odd}(\pm\lam,T)=\f{K(\pm\lam) e^{\mp\f{\lam(T+i\pi)}{2}}}{\left(\cosh \f {T+i\pi}{2}\right)^{2\D+\lam}}\label{3.34x}
\end{align} 
Let us define the Fourier transformation of $G(T)$ as 
\be 
I(\D;\w)=\int_{-\infty}^\infty dT e^{i\w T} G(T-i\e)
\ee 
where $\e$ is an infinitesimal positive number. It has been shown in \cite{Papadodimas:2012aq} that
\be  
I(\D;\w)=2^{2\D-1}\pi^2 e^{-i\pi\D+\pi\w}\G(\D+i\w)\G(\D-i\w)/\G(2\D)
\ee 
From \eqref{3.34x}, it is clear that the Fourier transformation of $G(\pm\lam,T)$ is $K(\pm\lam)  e^{\mp i\pi\lam /2}I(\D+\lam/2;\w\pm i\lam/2)$, which leads to
\begin{align}  
G(\pm\lam,T)&=\f{K(\pm\lam)  e^{\mp i\pi\lam /2}}{2\pi}\int d\w e^{-i\w T} \f{I(\D+\lam/2;\w\pm i\lam/2)}{I(\D;\w)} I(\D;\w) \nn\\
&= \f{2^\lam K(\pm\lam)  e^{-i\pi\lam /2}\G(2\D)}{2\pi\G(2\D+\lam)}\int d\w e^{-i\w T} \f{\G(\D+\lam\mp i \w)}{\G(\D\mp i\w)} I(\D;\w) \nn\\
&=\f{2^\lam K(\pm\lam)   e^{-i\pi\lam /2}\G(2\D)}{\G(2\D+\lam)}\f{\G(\D+\lam\pm \del_T)}{\G(\D\pm \del_T)}G(T)
\end{align}
From this equation and the definition \eqref{dfGT}, there is no unique solution for the coefficient $c_{mn}$ of the differential operator $L_T$. A convenient choice is 
\begin{align}
 K(\pm\lam)&=\pm\f{i\G(2\D+\lam)}{2^\lam \G(2\D)}\\
L_T= \sum_{mn}c_{mn}\del^m_T (\del_T^\phi)^n&=ie^{-i\pi\lam/2}\f{\G \left(\lam+\lam^{-1}\del_T^\phi(\D\lam^{-1}\del_T^\phi+\del_T)\right)}{\lam^{-1}\del_T^\phi\G\left(\lam^{-1}\del_T^\phi(\D\lam^{-1}\del_T^\phi+\del_T)\right)}
\end{align} 
where $\del_T$ acts on bare operator and $\del_T^\phi$ acts on the effective mode. Note that this $L_T$ can be expanded in power series in both $\del_T$ and $\del_T^\phi$. Moreover, we can expand $L_T$ near maximal chaos limit $\lam= 1$ and find
\be  
L_T= (\del_T+\D \del_T^\phi)+(1-\lam)\left[(\gamma+ \f{i \pi}{2})(\del_T+\Delta  \del_T^\phi)+(\D-\frac{\pi^2}{6} \del_T^2)\del_T^\phi \right]+O((1-\lam)^2,(\del_T^\phi)^2)
\ee 
where $\g$ is the Euler's constant. The first term is the same as \eqref{pio-2} and the subleading terms can be regarded as perturbative corrections of higher spins to the vertex.


\section{Relation to the EFT of maximal chaos} \label{sec:4}

The effective field theory for maximal chaos was constructed in \cite{Blake:2017ris},
which contains just one effective mode $\p$ on a Keldysh contour
in the thermal state with inverse temperature $\b=2\pi$. This effective
mode $\p$ has correlation function with exponential growth that explains
the behavior of OTOC. In this section, we will show that our effective
field theory of non-maximal chaos at maximal chaos $\lambda=1$ can
be equivalently connected to the theory in \cite{Blake:2017ris}.
In particular, our two-component effective mode $\phi_{1,2}$ reduces to
one mode $\p$.

Taking the maximal chaos limit $\lam\ra 1$ in \eqref{eq:x49}, we see that both $k_{+,\pm}(i\lam)$ diverge. This implies that the two operators $K_{+,\pm}(i\del_t)$ do not contain the factor $\del_t^2-\lam^2$ at maximal chaos. Physically, the two component fields $\eta_{+,\pm}(\bar \tau;t)$ decouple from the dynamics of quantum chaos. Now let us examine the behavior of $k_{-, p} (i \lam)$ in this limit. Consider $\w=i\lam$ in~\eqref{eq:2.76xx}; there are two equations but three parameters $k_{-,\pm}(i\lam)$ and $k_{-,0}(i\lam)$, whose general solution is 
\be  
\f 1 {k_{-,\pm}(i\lam)}=\pm \sqrt{3} i \left[ \left(1+\f 1 {\tan\left(\f{\pi}{2}(\lam\pm 1/3)\right)\tan \f {\pi \lam}{2}}   \right) \f 1 {k_{+,0}(i\lam)}-\f 1 {k_{-,0}(i\lam)} \right] \ .
\ee  
In the $\lam \to 1$ limit, with $k_{+,0} (i \lam)$ finite,\footnote{See Section \ref{sec:sykeffact} for a different case. }  we have 
\be  
\f 1 {k_{-,\pm}(i)}=\pm \sqrt{3} i \left[ \f 1 {k_{+,0}(i)}-\f 1 {k_{-,0}(i)} \right]
\ee 

In the EFT of maximal chaos \cite{Blake:2017ris}, the effective mode $\p$ is local and only depends on one time variable. This implies that our two effective modes $\phi_{1,2}$ at $\lam=1$ should not have any nontrivial dependence on $\bar t$. In other words, consistency requires that $\eta_{-,\pm}$ must also decouple at maximal chaos, which implies
\be 
k_{+,0}(i)=k_{-,0}(i) \label{4.3kk}
\ee 
From~\eqref{petr11}--\eqref{petr13} we then find that in the $\lam \to 1$ limit, $\vp_\pm$ decouple (i.e. they are not relevant for the exponential behavior), and 
\be 
\phi_1 (\bar t; t) = \phi_2 (\bar t; t) = \vp (t) \label{4.4id}
\ee
That is, in this limit, $\bar t$-dependences drop out and $\phi_{1,2}$ become the same field. Furthermore, from~\eqref{rev5}, $\vp$ has periodicity $2 \pi$ in imaginary $t$ direction, i.e. it satisfies the standard KMS condition with inverse temperature $2 \pi$. We have thus fully recovered the setup of the EFT for maximal chaos. 

The EFT action (i.e. the part relevant for exponential behavior) now becomes 
\begin{align}
S_{\rm EFT}&=\sum_{s=\pm}\int_{-\infty}^{\infty}dt \, \eta^a_{s,0}(t) K_{s,0}(i\del_t)\eta^r_{s,0}(t) \nn \\
 &=\f 1 2\sum_{s=\pm}\int_{-\infty}^{\infty}dt (\vp^a + s \tilde \vp^a)K_{s,0}(i\del_t) (\vp^r + s \tilde \vp^r)\nn\\
&= \int_{-\infty}^{\infty}dt\vp^a  K(i \del_t) \vp^r + \tilde \vp^a K(i \del_t)  \tilde \vp^r + \vp^a \tilde K(i \del_t)  \tilde \vp^r + \tilde \vp^a \tilde K(i \del_t)  \vp^r  \label{maxS4.5}
\end{align}
where $\tilde \vp$ was introduced in~\eqref{rev5} and 
\be  
K(i\del_t)  = \ha (K_{+,0}(i\del_t) + K_{-,0}(i\del_t)) , \quad \tilde K(i\del_t) = \ha (K_{+,0}(i\del_t) - K_{-,0}(i\del_t))  \ .
\ee 
Now recall from~\eqref{n2.16} that only $\vp$ is relevant for the four-point function (as $W$ and $V$ couple to $\phi_{1,2}$ which become $\vp$). Furthermore, from~\eqref{4.3kk}, $\vp$ and $\tilde \vp$ decouple at  $\om =i$. 
Thus for the purpose of understanding the exponential behavior of the four-point function, we can just keep the first term in the effective action~\eqref{maxS4.5}, reducing back to~\cite{Blake:2017ris}.

As discussed in Sec.~\ref{sec:2.0} and \ref{sec:2.1}, the differential operator $L_t$ in the vertex that couples the bare operators and effective fields has the same form in both  maximal and non-maximal cases. Moreover, in the $\lam \ra 1$ limit, the shift symmetry \eqref{eq:x44-1} of  $\phi_{1,2}$,  becomes 
\begin{equation}
(\phi_1,\phi_2)\ra(\phi_1,\phi_2)+(e^{\pm t},e^{\pm t}) \quad \to \quad \vp (t) \ra \vp (t) + e^{\pm t}  \ .
\label{eq:144}
\end{equation}
\eqref{eq:x44}~implies that the shift symmetry obeyed by vertex $L_t$ is 
\be  
L_{t_{1}}[g_{W}(t_{12})e^{\pm t_{1}}]=L_{t_{2}}[g_{W}(t_{12})e^{\pm t_{2}}]
\ee  
which also  matches with that in the EFT of maximal chaos~\cite{Blake:2017ris}.

To close this section, we note that the Wightman function $G^>_{-,0}(\bar\tau;t)$ given in~\eqref{2.108} diverges in the limit 
$\lam \to 1$. This divergence reflects that in the limit $G^>_{-,0}$ develops a $t e^{\lam t}$ term which is not present 
for $\lam < 1$. More explicitly, from~\eqref{Ds0} and~\eqref{eq:x18}, we find $G^{rr}_{-,0}(t)$ should satisfy 
\be  
(1+e^{-i\pi\del_t})G^{rr}_{-,0}(t)=\f 1 2 (1-e^{-i\pi\del_t})\left(\f i {2\lam \tilde k_0(i\lam)}(e^{t}-e^{-t})\right) \label{fdr3.5}
\ee  
whose general solution is
\be  
G^{rr}_{-,0}(t)=\f{t}{2\pi \tilde k_{0}(i)}(e^t-e^{-t})+c_0(e^t+ e^{-t})
\ee 
where $c_0$ is an arbitrary constant.  It then follows from \eqref{eq:x15} 
\be  
G^>_{-,0}(\bar\tau;t)=\f{t}{2\pi \tilde k_{0}(i)}(e^t-e^{-t})+\left(c_0+\f {i}{4\tilde k_{0}(i)}\right) e^t+\left(c_0-\f {i}{4\tilde k_{0}(i)}\right) e^{-t}  \ .\label{g-0max}
\ee 
The presence of linear-exponential term in Wightman function at maximal chaos was already observed in \cite{Blake:2017ris}. 


\section{Identifying the effective fields in the large $q$ SYK model} \label{sec:5}

In this section we examine the large $q$ SYK model in some detail.
In this model, four-point functions of fundamental fermions can be  computed analytically in the Euclidean signature. 
We show that in this theory it is possible to identify two Euclidean effective fields $\phi^E_{1,2}$ in terms of the microscopic description,
 which can be identified as 
$\phi_{1,2} (\bar t; t)$ of Sec.~\ref{sec:The-structure-of} evaluated in the Euclidean section with pure imaginary $\bar t$ and $t$.
It is possible to calculate Euclidean two-point functions of $\phi^E_{1,2}$ using the microscopic description. We show that 
the Lorentzian analytic continuation of these two-point functions can be fully captured by the EFT of Sec.~\ref{sec:The-structure-of}. This 
provides a stronger check on the EFT formulation than just matching the structure of OTOCs done in Sec.~\ref{sec:motoc}.


\subsection{OTOC of the large $q$ SYK model} \label{sec:syk4.1}

We start with a brief review of the essential aspects of the large $q$ SYK model. The SYK model \cite{sachdev1993gapless,kitaev2015simple} is a 0+1 dimensional quantum mechanical system which consists of $N$ Majorana fermions with an all-to-all and $q$-local Hamiltonian
\begin{align}
H &=i^{q/2}\sum_{1\leq j_{1}<\cdots<j_{q}\leq N}J_{j_{1}\cdots j_{q}}\psi^{j_{1}}\cdots\psi^{j_{q}}\nn\\
\overline{\left(J_{j_{1}\cdots j_{q}}\right)^{2}}&=\f{2^{q-1}\mJ^{2}(q-1)!}{qN^{q-1}}=\f{J^{2}(q-1)!}{N^{q-1}}
\end{align}
where $J_{j_{1}\cdots j_{q}}$ is a random coupling with Gaussian
distribution.  At an inverse temperature $\b$, it is possible to derive a bilocal
effective action in Euclidean signature \cite{Maldacena:2016hyu}
\begin{equation}
S_E=\f N2\log\det(\del_{\tau}-\S)+\f N2\iint_{0}^{\b}d\tau_{1}d\tau_{2}\left[\S(\tau_{1},\tau_{2})G(\tau_{1},\tau_{2})-\f{J^{2}}qG(\tau_{1},\tau_{2})^{q}\right]
\end{equation}
where $G(\tau_{1},\tau_{2})=\f 1N\sum_{j} \psi^{j}(\tau_{1})\psi^{j}(\tau_{2})$ is a bi-local field, 
 and $\S$ is an auxiliary bi-local field. In the double scaling limit with $N, q \ra\infty$ and
$N^{2}/q$ fixed, it is convenient to introduce $G(\tau_{1},\tau_{2})=\f 12\sgn(\tau_{1}-\tau_{2})e^{\s(\tau_{1},\tau_{2})/q}$
and $\S$ can be integrated out, resulting a Liouville action for $\s(\tau_{1},\tau_{2})$ \cite{Cotler:2016fpe}, 
\begin{equation}
S_{E}\app\f N{4q^{2}}\int d\tau_{1}d\tau_{2}\left[\f 14\del_{1}\s(\tau_{1},\tau_{2})\del_{2}\s(\tau_{1},\tau_{2})-\mJ^{2}e^{\s(\tau_{1},\tau_{2})}\right] \ . \label{eq:1}
\end{equation}
The large $q$ SYK model is obtained by further taking $N/q^{2}\ra\infty$, in which limit the Euclidean path integral over $\s$ is
dominated by saddle-point solution(s) to the equation of motion of~\eqref{eq:1}, i.e. by solutions to the Liouville equation
\begin{equation}
\del_{1}\del_{2}\s(\tau_{1},\tau_{2})=-2\mJ^{2}e^{\s(\tau_{1},\tau_{2})} \ .
\label{eq:2}
\end{equation}
By definition, $\s (\tau, \tau')$ has the following properties 
\begin{equation}
\s(\tau,\tau')=\s(\tau',\tau), \quad  \s(\tau,\tau)=0,
\label{eq:8}
\end{equation}
where the second equation can be interpreted as a UV boundary condition.
It is also periodic in $\b$ for both arguments 
\be
\quad\s(\tau,\tau')=\s(\tau'+\b,\tau) \ .
\label{eq:7-1}
\ee

The solution $\s_0$ to~\eqref{eq:2} with time translation symmetry can be interpreted as describing the equilibrium state, and can be written as
\begin{equation}
 \label{eq:03}
e^{\s_{0}(\tau_{12})}=\f{\cos^{2}\lam\b/4}{\cos^{2}\f \lam 2(|\tau_{12}|-\b/2)},\quad\tau_{12} = \tau_1 - \tau_2, 
\end{equation}
and $\lam$ is obtained from the solution to equation 
\be
 \lam=2\mJ\cos\lam\b/4 \ . 
 \label{eq:3}
\ee
The equilibrium two-point function is given by
\be \label{27}
G_0 (\tau) = \vev{G(\tau,0)}_\b = \f 1 2 e^{\s_0 (\tau)/q} , \quad \tau \in [0, \b] \ .
\ee
From now on, for notational simplicity we use the unit such that $\beta=2\pi$. 

Fluctuations around the thermal equilibrium can be described by  expanding $\s=\s_{0}+\e$ around $\s_{0}$.
To order $\e^{2}$, (\ref{eq:1})  becomes 
\begin{equation}
S_{E}\app S_{0}+\f N{8q^{2}}\int d\tau_{1}d\tau_{2}\left[\f 12\del_{1}\e(\tau_{1},\tau_{2})\del_{2}\e(\tau_{1},\tau_{2})-\mJ^{2}e^{\s_{0}(\tau_{12})}\e^2 (\tau_{1},\tau_{2})\right]\label{eq:4}
\end{equation}
where $S_0$ is a constant scaling as $N/q^2$. By definition we have $\avg{\e(\tau_{1},\tau_{2})}=0$ to this order, where 
$\vev{\cdots}$ denotes correlation functions of $\ep$ obtained from path integrals with Euclidean action~\eqref{eq:4}. 
Now consider the  Euclidean four-point function 
\begin{align}
\mF_\psi(\tau_{1},\tau_{2},\tau_{3},\tau_{4})&\equiv\avg{G(\tau_{1},\tau_{2})G(\tau_{3},\tau_{4})}\nn\\
&\app G_0(\tau_{12})G_0(\tau_{34})\left(1+\f 1{q^{2}}\avg{\e(\tau_{1},\tau_{2})\e(\tau_{3},\tau_{4})}\right) \label{n4.7}
\end{align}
where in the second line we have indicated that its leading connected part is given by two-point function of $\ep$. 

\begin{figure}
\begin{centering}

\includegraphics[height=5cm]{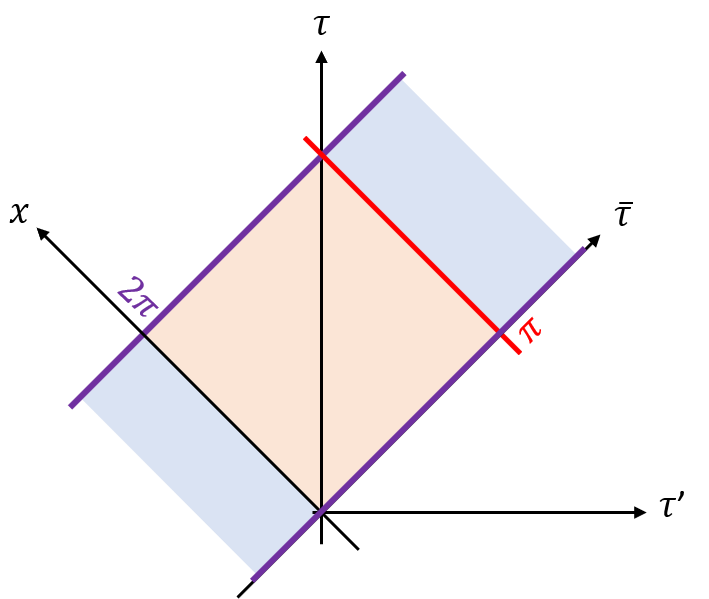} 

\par\end{centering}
\caption{The red region is the fundamental domain $\mD_\e$ for $\e(\bar\tau,x)$. \label{fig:fundamental}}

\end{figure}

It will be convenient to write~\eqref{eq:4} in a different form by introducing~\cite{Choi:2019bmd}
\begin{equation}
x=\tau-\tau',\quad\bar{\tau}=\f{\tau+\tau'}2
\end{equation}
and redefine the fields $\s(\bar\tau, x)=\s(\tau,\tau')$ and $\e(\bar\tau, x)=\e(\tau,\tau')$. From~ \eqref{eq:8}--\eqref{eq:7-1}, $\e(\bar\tau, x)$  satisfy the following conditions
\be 
\e(\bar\tau,x)=\e(\bar\tau,-x)=\e(\bar\tau+\pi,2\pi-x),\quad \e(\bar\tau,0)=\e(\bar\tau,2\pi)=0 \label{eq:4.11}
\ee
from which we can define the fundamental domain of $\e$ as $\mD_\e=\{(\bar\tau,x)|\bar\tau\in[0,\pi],x\in[0,2\pi]\}$ (see Fig. \ref{fig:fundamental}). On the fundamental domain $\mD_\e$, we can rewrite the Euclidean effective action (\ref{eq:4}) as
\be
S_{E}=\f N{8q^{2}}\int_0^\pi d\bar{\tau}\int_0^{2\pi} dx\left[\f 14(\del_{\bar{\tau}}\e)^{2}-(\del_{x}\e)^{2}-\f{\lam^{2}}{2\cos^{2}\f{\lam}2(x-\pi)}\e^{2}\right] \equiv \f N{8q^{2}} \int d \bar \tau dx \, \ep \mL \ep
\label{eq:10}
\ee
where the irrelevant constant $S_0$ is omitted, and $\mL$ is the differential operator corresponds to the quadratic action.

Two-point functions of $\ep$ can be obtained summing over the eigenfunctions of $\mL$~\cite{Choi:2019bmd} as follows. 
Eigenfunctions of $\mL$ can be labeled by two quantum numbers $n, m$, and separated into two groups 
\be 
\psi_{n,m}  (\tau, x) =
 e^{in\bar\tau}\psi_m(x), \quad (n, m) \in (\Z_\pm, \mM^\pm)
\ee
with 
\begin{equation}
\mL\psi_{n,m}(\bar\tau,x)\equiv\left(-\f 14\del_{\bar{\tau}}^{2}+\del_{x}^{2}-\f{\lambda^{2}}{2\cos^{2}\f{\lam(x-\pi)}{2}}\right)\psi_{n,m}(\bar\tau,x)=\f{n^{2}-m^{2}}4\psi_{n,m}(\bar\tau,x) \label{eigeneq}
\end{equation}
where $\Z_+$/$\Z_-$ denotes even/odd integers respectively and $\mM^\pm$ are two infinite discrete sets of real numbers with magnitude larger than 1. When $n$ takes values in $\Z_+$ ($\Z_-$) , $m$ takes value in $\mM_+$ ($\mM_-$).  
The two-point function of $\e$ can then be written as 
\begin{align} 
\avg{\e(\bar\tau_a,x_a)\e(\bar\tau_b,x_b)}&=\sum_{(n, m) \in (\Z_\pm, \mM^\pm)} \f {e^{in(\bar\tau_a-\bar\tau_b)}\psi_m(x_a)\psi_m(x_b)}{(n^2-m^2)/4} \nn\\
=-2\pi&\sum_{m\in\mM^{\pm}}\f{\psi_{m} (x_a)\psi_{m}(x_b)}{m\sin m\pi}\left[\cos m(\pi-|\bar{\tau}_a-\bar{\tau}_b|)\pm\cos m(\bar{\tau}_a-\bar{\tau}_b)\right] \label{4.14x}
\end{align} 
where we define the notation
\be 
\bar\tau_a=(\tau_1+\tau_2)/2,\quad  \bar\tau_b=(\tau_3+\tau_4)/2,\quad x_a=\tau_1-\tau_2,\quad x_b=\tau_3-\tau_4 \ .
\ee

The infinite sum of $m$ in \eqref{4.14x} can be evaluated in a closed form by Sommerfeld-Watson transformation~\cite{Choi:2019bmd}. For different ordering assignment for $\tau_k$, they lead to TOC or OTOC
 after analytic continuation $\tau_k\ra i t_k$. TOCs do not have exponential growth, while the OTOC with $2\pi >\tau_1>\tau_3>\tau_2>\tau_4\geq 0$ is given by
\be  
\mF_{\rm OTOC}=-\f 2 N \f {G_0(\tau_{12})G_0(\tau_{34})\cos\f{\lam(\tau_1+\tau_2-\tau_3-\tau_4-\pi)} {2} }{\cos \f {\pi \lam} 2 \cos \f {\lam(\pi-\tau_{12})}2 \cos \f {\lam(\pi-\tau_{34})}2}  \ . 
\label{sykotoc}
\ee  
After analytic continuation $\tau_k\ra i t_k$, it gives~\eqref{sykotoc1} mentioned earlier. 
Notice that, with $\tau_k\ra i t_k$, the cosine function in~\eqref{4.14x} leads to exponential growth $e^{\pm m (t_1+t_2-t_3-t_4)/2}$. 
Since every quantum number $m\in\mM^\pm$ has magnitude greater than 1, the exponential growth in each individual term in \eqref{4.14x} violates the chaos bound.  This infinite tower of quantum numbers $m$ is the analogue of the infinite tower of higher spins to be summed over in the higher dimensional Regge theory.


\subsection{Identifying the effective fields} \label{sec:eftsyk}

We now seek an alternative way to understand two-point function of $\ep$, without going through the infinite sums over $m, n$. We 
are interested only in the exponential part~(after analytic continuation) of the two-point function, and would like to 
identify a finite number of effective fields that can capture that.

For this purpose, consider general solutions to the saddle-point equation \eqref{eq:2}, which can be written in a form 
\begin{equation}
e^{\s (\tau_1, \tau_2)}=\f{f'(\tau_{1})g'(\tau_{2})}{\mJ^{2}(f(\tau_{1})-g(\tau_{2}))^{2}} \label{esigma}
\end{equation}
where $f, g$ are arbitrary functions. The above parameterization of $\s$ is not unique as the right hand side is invariant under an arbitrary $SL(2,\C)$ transformation
\be
f\ra \f{a+b f}{c+d f},\quad g\ra \f{a+b g}{c+d g},\quad bc-ad=1  \ . \label{sl2}
\ee
Now consider small perturbations $\ep_{\rm on-shell}$ around the equilibrium solution~\eqref{eq:03} in the space of solutions~\eqref{esigma}, which can be parameterized as\footnote{To see this, we can write the equilibrium solution (\ref{eq:03}) 
as $f_{0}(\tau)=-e^{i \lambda (\tau-\pi)},\, g_{0}(\tau)=e^{i \lambda \tau}$ and parameterize small perturbations around it as 
$f(\tau)=-e^{i\lambda (\tau-\pi+\chi_1(\tau))},\; g(\tau)=e^{i\lambda (\tau+\chi_2(\tau))}$. Equation~\eqref{esigma} results from expanding $\chi_{1,2}$ to  linear order. \label{ft11}
} 
\begin{equation}
\e_{\rm on-shell}(\tau_{1},\tau_{2})=\lambda(\chi_1(\tau_{1})-\chi_2(\tau_{2}))\tan\f{\lambda}2(\tau_{12}-\pi)+\chi_1'(\tau_{1})+\chi_2'(\tau_{2})\label{eq:43}
\end{equation}
where $\chi_{1,2}$ are two arbitrary (infinitesimal) functions, and parameterize the full set of solutions to equation of motion 
of the quadratic action~\eqref{eq:10}. 
The paramterization freedom~\eqref{sl2} translates to $\chi_{1,2}$ as $\ep_{\rm on-shell}$ being invariant under transformations 
\begin{equation}
(\d\chi_1 (\tau_1),\d\chi_2 (\tau_2))= \left(-e^{\pm i \lambda (\tau_1-\pi)}, e^{\pm i \lambda \tau_2}\ri),
 \label{eq:44}
\end{equation}
or a simultaneous shift by a constant. 

To obtain an effective description of the exponential behavior, 
we first rewrite~\eqref{eq:43} in a more convenient form 
\begin{align}  \label{uhw}
\e_{\rm on-shell} (\tau_{1},\tau_{2})&=\f 1{\D G_0(\tau_{12})} \sum_{i=1,2} L_{\tau_i}[G_0(\tau_{12})] \chi_i(\tau_i) \\
L_\tau[\mO(\tau)]\chi(\tau)&\equiv\del_{\tau}\mO(\tau)\chi(\tau)+\D \mO(\tau)\del_{\tau} \chi(\tau) \label{x4.47}
\end{align} 
where $G_0$ is the equilibrium two-point function of fundamental fermions given earlier in~\eqref{27},
and $\D =1/q$ is the conformal dimension of fundamental fermionic operator $\psi_i$. 
The differential operator $L_\tau$ can be interpreted as a vertex that couples $\ep_{\rm on-shell}$ to $\chi_i (\tau_i)$, and 
from~\eqref{eq:44}, it obeys the following symmetry
\be  
L_{\tau_1}[G_0(\tau_{12})e^{\pm i \lam (\tau_1-\pi)}]=L_{\tau_2}[G_0(\tau_{12})e^{\pm i \lam \tau_2}]  \ .
\label{Lshift}
\ee 

Motived from~\eqref{uhw}, we write $\e(\tau_1,\tau_{2})$ as
\be  
\e(\tau_1,\tau_{2}) =\f {1}{\D G_0(\tau_{12})}\left(L_{\tau_L}[G_0(\tau_{12})] \phi^E_1(\bar\tau;\tau_L)+L_{\tau_S}[G_0(\tau_{12}) ]\phi^E_2(\bar\tau;\tau_S)\right) , 
\label{epdec}
\ee  
where $\bar \tau = {\tau_1 + \tau_2 \ov 2}$. $\tau_L$ ($\tau_S$) is the larger (smaller) of $\tau_{1,2}$, whose usage ensures that~\eqref{epdec} respects the invariance of $\ep$ under switch of $\tau_{1,2}$. $\phi^E_{1,2}$ are dynamical counterparts of $\chi_{1,2}$, with $\chi_{1,2}$ parameterizing their classical solutions. 

Equation~\eqref{epdec} is exactly the Euclidean version of~\eqref{eq:n13}--\eqref{new-uen2} with the bi-fermionic field $\ep$ identified with the second term in~\eqref{eq:n13}, $L_\tau$ is Euclidean version of $L_t$, and $\phi^E_{1,2}$ are $\phi_{1,2}$ evaluated at Euclidean times. 
By construction, $\ep$ as defined in~\eqref{epdec} is invariant under transformations of $\phi_{1,2}^E$ of the form~\eqref{eq:44}, and thus whatever the action for $\phi_{1,2}^E$ is, it should be invariant under~\eqref{eq:44}, which is precisely the Euclidean version of the shift symmetry~\eqref{eq:x44-1}. 
And equation~\eqref{Lshift} can be identified with the Euclidean version of the shift symmetry~\eqref{eq:x44} satisfied by the effective vertex. 

Given that the action for $\phi_{1,2}^E$ must be invariant under the shift symmetry~\eqref{eq:44}, we can be certain that correlation functions of $\phi_{1,2}^E$ must contain exponential time-dependence. This establish $\phi_{1,2}$ as the effective fields which directly captures the exponential behavior of two-point function of $\ep$.\footnote{Note that two-point function of $\ep$ has exponential behavior  only when the ordering of time arguments corresponds to OTOC, while two-point functions of $\phi^E$ always have exponential behavior.}

\subsection{Two-point function of $\phi_i^E$} \label{sec:quantimode}

In principle we can try to find the (Euclidean) action of $\phi^E_{1,2}$ by plugging~\eqref{epdec} into~\eqref{eq:4} or~\eqref{eq:10}, and being careful about the Jacobian in changing variables from $\ep$ to $\phi^E_{1,2}$ in the path integral for $\ep$. 
It is, however, difficult to do in practice. In addition to having to understand the Jacobian, various complications discussed in Sec.~\ref{sec:2.1}--\ref{sec:diag}, including that $\phi_{1}^E$ and $\phi_{2}^E$ are defined on different domains, should also be faced here. 

Here we show that using~\eqref{eq:10} we can nevertheless find their Euclidean two-point functions, and confirm their 
exponential time-dependence. In next subsection, we show that these Euclidean two-point functions can be reproduced from the EFT formulation of Sec.~\ref{sec:The-structure-of} by a suitable choice of the action there. 

To compute two-point functions of $\phi^E_{1,2}$,  instead of considering~\eqref{epdec} as a change of variables in the path integral, we 
canonically quantize~\eqref{eq:10} by treating $\bar \tau$ as ``time'', and treat~\eqref{epdec} 
as an operator equation. Below we will use $\ep, \phi^E_{1,2}$ to denote the fields in the Euclidean path integral defined with~\eqref{eq:10}, and $\hat \ep, \hat \phi^E_{1,2}$ the corresponding operators in the canonical quantization. By definition, Euclidean correlation functions of $\phi_{1,2}^E$ are given by ``time-ordered'' correlation functions of $\hat \phi^E_{1,2}$, i.e. 
\be  
\avg{\phi^E_i(\bar\tau;\tau)\phi^E_j(\bar\tau';\tau')} = -i\avg{\bar\mT \hat \phi^E_i(\bar\tau;\tau)\hat\phi^E_j(\bar\tau';\tau')}_{\rm c.q.}
 \label{eq:4.41df0}
\ee 
where $\bar \mT$ denotes ordering in $\bar \tau$ and the subscript ``c.q.'' on the RHS is to distinguish the expectation value in Euclidean path integral on the LHS.  We outline the main steps here, leaving technical calculations to Appendix~\ref{app:d}:

\begin{enumerate}
    \item In canonical quantization of~\eqref{eq:10}, we can expand $\hat \ep$ in terms of a complete set of modes $\{g_m\}$ 
    as 
      \begin{equation}
\hat \e(\bar\tau,x)=\sum_m  
\left(g_{m}\hat a_{m}+g_{m}^{*}\hat a_{m}^{\dagger}\right) \label{ephat}
\end{equation}
where $g_m$ solve the equation of motion $\mL g_m(\bar\tau,x)=0$, and obey the conditions $g_m(\bar\tau,0)=g_m(\bar\tau,2\pi)=0$ (from \eqref{eq:4.11}). $m$ takes value in the same sets $\mM^\pm$ discussed below~\eqref{eigeneq}. 
$\{g_m\}$ are assumed to be properly normalized under the Klein-Gordon
inner product, 
\begin{equation}
(g_{1},g_{2})=-{i \ov 2} \int_0^{2\pi} dx \, (g_{1}\del_{\bar{\tau}}g_{2}^{*}-g_{2}^{*}\del_{\bar{\tau}}g_{1})  \label{eq:16}
\end{equation}
such that annihilation and creation
operators $\hat a_{m}$ and $\hat a_{m}^{\dagger}$ obey the standard commutation relation $[\hat a_{m},\hat a_{m'}^{\dagger}]=\d_{m,m'}$.  

Due to the nontrivial boundary condition~\eqref{eq:4.11} in $\bar \tau$ direction, the system is not in the vacuum state of $\hat a_m$, and it can be shown that $\hat a_m, \hat a_m^\dag$ have correlation functions\footnote{Two-point function of $\hat \ep$ then follows and can be checked to give the same answer as~\eqref{4.14x}.}
\be
\avg{\hat a_{m}^{s}\hat a_{m}^{s\dagger}}_{\rm c.q.} =\f{1}{1-se^{-im\pi}},\quad\avg{\hat a_{m}^{s\dagger}\hat a_{m}^{s}}_{\rm c.q.}=\f{se^{-im\pi}}{1-se^{-im\pi}} \ .
\label{eq:1172}
\ee
where $s=\pm$ labels two different sectors of $\hat a_m$ and $\hat a^\dag_m$ with $m\in\mM^\pm$ respectively.
    
\item Since the mode functions $g_m$ solve the equation of motion of $\ep$, from \eqref{uhw},
they can be written as 
\be  
g_m(\bar\tau,x)=\f 1{\D G_0(\tau_{12})} \sum_{i=1,2} L_{\tau_i}[G_0(\tau_{12})] \chi_{i,m}(\tau_i)  \label{chi4.32}
\ee 
where  $L_\tau$ is defined in \eqref{x4.47}, and $\{\chi_{i, m}\}$ is a complete set of basis functions for $\chi_{i}$ in~\eqref{uhw}.
Plugging~\eqref{chi4.32} into~\eqref{ephat}, then from~\eqref{epdec}, we can write $\hat \phi^E_{1,2}$ as 
\begin{align}
\hat \phi^E_i(\bar\tau;\tau_i) = \sum_m  
(\chi_{i,m}(\tau_i)\hat a_m+\chi_{i,m}^*(\tau_i)\hat a_m^\dag)  \ .
\label{4.34x}
\end{align}
Two-point functions of $\phi_i^E$ can then follow from~\eqref{eq:1172}. Notice from~\eqref{4.34x} that $\hat \phi^E_i(\bar\tau;\tau_i)$ does not have any $\bar \tau$ dependence. So the only $\bar \tau$-dependence in two-point functions of $\phi^E_{1,2}$ comes from $\theta (\bar \tau)$ on the right hand side of~\eqref{eq:4.41df0}. 
Such $\bar \tau$-dependence is precisely what we had in Sec.~\ref{sec:The-structure-of}, see e.g.~\eqref{2.73x}, except that there it came from our assumption of weak $\bar \tau$-dependence and derivative expansion in $\bar \tau$, but here for the large $q$ SYK model it is exact.

\end{enumerate}

From the calculation of Appendix \ref{app:d}, we have 
\begin{align}
\avg{\phi^E_i(\bar\tau;\tau)\phi^E_j(0;0)}&=\t(\bar\tau)M_{ij}(\tau)+\t(-\bar\tau)M_{ji}(-\tau)\label{phiEcor1}\\
M_{11}(\tau)&=M_{22}(\tau)=M_{12}(2\pi-\tau)=M_{21}(-\tau) \label{4.42chi1}\\
&=\f{8}{N\D^2}\times\begin{cases}
\hat m_{+}e^{i\lam \tau}+\hat m_1 \tau e^{i\lam \tau}+c.c., &\tau\in[0,2\pi] \\
\hat m_{-}e^{i\lam \tau}+\hat m_1 \tau e^{i\lam \tau}+c.c., &\tau\in[-\pi,0]
\end{cases} \label{4.43chi2}
\end{align}
where in the last equality we have kept only the exponential part, and $\hat m_{+}$, $\hat m_-$ and $\hat m_1$
are some constants given by
\begin{align}
\hat m_{+} & =i\f{2\pi\lambda-\sin2\pi\lambda+2(\pi\lambda+\sin\pi\lambda)(2\pi i\lambda+3-e^{-i\pi\lambda})}{32\lambda^{2}(\pi\lambda+\sin\pi\lambda)^{2}(1+e^{i\pi\lambda})},\\
\hat m_{-} & =\hat m_{+}+\f 1{4\lambda^{2}(1+e^{i\pi\lambda})},\quad \hat m_{1}=\f 1{8\lambda(\pi\lambda+\sin\pi\lambda)(1+e^{i\pi\lambda})}
\end{align}

\subsection{Effective action for large $q$ SYK} \label{sec:sykeffact}

We now work out the explicit form the EFT action which match with correlation functions~\eqref{4.42chi1}--\eqref{4.43chi2}. As our EFT is formulated in Lorentzian time, we need to first analytically continue the Euclidena correlation fuctions \eqref{phiEcor1}. The correct analytic continuation for large $q$ SYK model is $\phi^E_j(\bar\tau;\tau)\ra \phi_j(\bar t;t)=-i\phi^E_j(i\bar t;it)$ and $L_\tau\ra L_t$. Note that \eqref{4.43chi2} contains linear-exponential terms $t e^{\pm \lam t}$ after analytic continuation $\tau\ra it$.
From discussion of Appendix~\ref{app:b}, this means that we need to include quadratic order of  $\del_t^2-\lam^2$ in some $K_{s,p}(i\del_t)$, i.e. 
\be 
K_{+,p}(i\del_t)=(\del_t^2-\lam^2)k_{+,p}(i\del_t),\quad K_{-,p}(i\del_t)=(\del_t^2-\lam^2)^2 k_{-,p}(i\del_t) \label{sykact-1}
\ee
where $K_{-,p}(x)$ has a double zero at $\pm i\lam$. With appropriate choice of $k_{+,0}(i\lam)$, $k_{-,0}(i\lam)$ and $k'_{-,0}(i\lam)$ (the derivative to $k_{-,0}(x)$ at $i\lam$) that depend on the three parameters $\hat m_{+}$, $\hat m_-$ and $\hat m_1$, we find that this effective action completely reproduces the correlation functions~\eqref{phiEcor1}--\eqref{4.43chi2}. 

More explicitly, we find in Appendix \ref{app:d4} that 
\begin{align} 
k_{+,0}(i\lam)&=\f{3\lam N \D^2\cot \f {\pi \lam}{2}}{2(1-2\cos\pi \lam)} \label{k0sl}\\
k_{-,0}(i\lam)&=\frac{3 N \D^2 (\pi  \lambda +\sin \pi  \lambda )}{8\lambda (2  \cos \pi  \lambda -1)} \\
k'_{-,0}(i\lam)&=-\frac{3 i N \D^2\left(\left(2 \pi ^2 \lambda ^2-5\right) \sin \pi  \lambda +2 \sin 2 \pi  \lambda +8 \pi  \lambda  \cos \pi  \lambda -\pi  \lambda  \left(3 \pi  \lambda  \tan \frac{\pi  \lambda }{2}+7\right)\right)}{16 \lambda ^2 (1-2 \cos \pi  \lambda )^2} \label{km}
\end{align}
and other parameters $k_{s,\pm}(i\lam)$ and $k'_{-,\pm}(i\lam)$ 
are derived from the constraints \eqref{Grr=0}, \eqref{Kp} and two continuity conditions \eqref{mcond}-\eqref{eq:2.76xx} and explicitly given by \eqref{c22}-\eqref{c24}, which we copy as follows
\begin{align}  
k_{+,\pm}(i\lam)&=\mp \f{ik_{+,0}(i\lam)}{\sqrt{3}}\tan\left(\f{\pi}{2}(\lam\pm 1/3)\right)\tan \f {\pi \lam}{2} \label{kppm}\\
k_{-,\pm}(i\lam)&=\pm \f{i k_{-,0}(i\lam)}{\sqrt{3}} \\
k'_{-,\pm}(i\lam)&=\pm \f{\sqrt{3}}{3}\left[ \f{2\lam k_{-,0}(i\lam)^2}{k_{+,0}(i\lam)}\left(1+\f 1 {\tan\left(\f{\pi}{2}(\lam\pm 1/3)\right)\tan \f {\pi \lam}{2}}   \right)+i k'_{-,0}(i\lam) \right] \label{kmpmd}
\end{align}
 
There is a subtlety in this effective action when we take the maximal chaos limit $\lam \to 1$. Taking $\lam\ra 1$ in \eqref{k0sl}-\eqref{km}, we find that 
\be  
k_{+,0}(i\lam)\ra 0,\quad k_{-,0}(i\lam)\sim O(1),\quad  k_{-,0}'(i\lam)\ra \infty
\ee
Note that in \eqref{sykact-1} $K_{+,p}$ is linear in $\del_t^2-\lam^2$, which implies that correlation functions for $s=+$ in Section \ref{sec:corr} all hold. It follows that we will have singular $\D_{+,0}(t)$ by \eqref{Ds0}. This singularity implies that the maximal chaos limit in large $q$ SYK model should be taken with some care. 

Indeed, there is a scaling limit in SYK model when we take maximal chaos limit. The Lyapunov exponent $\lam$ is related to the inverse temperature $\b$ and coupling $\mJ$ by  \eqref{eq:3}. The maximal chaos limit corresponds to strong coupling limit $\mJ\ra \infty$ (or equivalently low temperature limit $\b\ra \infty$), under which we can solve \eqref{eq:3} perturbatively as ($\b=2\pi$)
\be  
\lam =1-\f1 {\pi \mJ}+O(1/\mJ^2)
\ee  
However, the low temperature limit should still be understood in the regime of validity of large $N$, i.e. $N\gg \mJ \gg 1$. Therefore, in maximal chaos limit, for \eqref{k0sl}-\eqref{km} we have
\be  
k_{+,0}(i\lam)\sim O(\mJ^{-1}),\quad k_{-,0}(i\lam)\sim O(1),\quad  k_{-,0}'(i\lam)\sim O(\mJ) \label{ksc1}
\ee 
which by \eqref{kppm}-\eqref{kmpmd} implies
\be  
k_{+,\pm}(i\lam)\sim O(1),\quad k_{-,\pm}(i\lam)\sim O(1),\quad k_{-,\pm}'(i\lam)\sim O(1) \label{ksc2}
\ee 
Given the action \eqref{sykact-1}, by \eqref{2.74x-1} and \eqref{2.74x} the exponential terms in retarded correlation functions scale as
\be  
\D_{+,p}(t)\sim \f{e^{\pm\lam t}}{k_{+,\pm p}(i\lam)},\quad \D_{-,p}(t)\sim \f{k'_{-,\pm p}(i\lam)e^{\pm \lam t}}{k_{-,\pm p}(i\lam)^2}+(c_0+c_1 t)\f{e^{\pm\lam t}}{k_{-,\pm p}(i\lam)} \label{5.50x}
\ee 
where $c_{1,2}$ are two $O(1)$ numbers. From \eqref{ksc1} and \eqref{ksc2} in maximal chaos, there is an enhancement of $\mJ$ to $\D_{s,0}(t)$ while $\D_{s,\pm}(t)$ are still $O(1)$. This means that $\eta_{s,\pm}$ decouple at maximal chaos at leading order of $\mJ$. Since $\bar\tau$ dependence only exists for $p=\pm$, this means that at leading order of $\mJ$ two effective modes $\phi_{1,2}$ reduce to a single field at maximal chaos following the same argument for \eqref{4.4id}. 

Note that the first term of $\D_{-,0}(t)$ in \eqref{5.50x} is $O(\mJ)$ but the second term is $O(1)$. Therefore, at leading order of $\mJ$, $\D_{-,0}(t)$ only has pure exponential terms.
Using the explicit expression for $\D_{s,p}(t)$ in \eqref{c25x}-\eqref{c28x}, one can show that
\be  
\D_{+,0}(t)=\D_{-,0}(t)=\f{2i\mJ}{N\D^2}(e^{t}-e^{-t}) +O(1/N)
\ee  
It follows that $K_{-,0}(i\del_t)$ in the effective action \eqref{sykact-1} can be reduced to be just linear in $\del_t^2-\lam^2$ and we can take ansatz
\be  
K^{\max}_{s,0}(i\del_t)=(\del_t^2-1)k^{\max}_{s,0}(i\del_t) \label{maxactsyk}
\ee 
where $k^{\max}_{s,0}(i\del_t)$ satisfies
\be  
k_{+,0}^{\max}(i)=k_{-,0}^{\max}(i)=\f{ N\D^2}{4\mJ} \label{kpm-i}
\ee
This is the same condition we impose in \eqref{4.3kk} for maximal chaos. Following the discussion in Section \ref{sec:4}, this EFT reduces back to case in \cite{Blake:2017ris}. Note that the overall scaling $N/\mJ$ in \eqref{kpm-i} reflects the fact that the effective action for SYK model in strong coupling/low temperature limit is proportional to $N/(\b \mJ)$, which is observed in the Schwarzian action \cite{Maldacena:2016hyu}.

\section{Higher order terms and exponentiation}\label{sec:3}

Higher order terms in equation~\eqref{ejn} are suppressed by higher powers of $1/N$. Here we show that a subset of terms 
of the form ${e^{k \lam t} \ov N^k}$ ($k$ an integer) can be rensummed and exponentiated. Such terms dominate in the 
regime $N \to \infty$ and $t \sim {1 \ov \lam} \log N$ such that ${e^{\lam t} \ov N}$ is finte. 
These contributions come from including higher powers of effective fields $\phi_i$ in the product~\eqref{eq:n13}, but in the effective action still keep only quadratic terms.\footnote{Including nonlinear terms in the EFT action leads to higher order terms of the form ${e^{k_1 \lam t} \ov N^{k_2}}$ with $k_2 > k_1$.} The full four-point functions of $V$ and $W$ 
then involve multi-point correlation functions of effective fields $\phi_i$, which factorize to products of two-point functions.
We show that the shift symmetry on the vertices that couple $W(t_1) W(t_2)$ to  higher powers of $\phi_i$ 
 implies that TOCs again do not have exponential growth, and the OTOC  (recall~\eqref{eq:FG2.63}) has the exponentiated form 
\be  \label{bsa}
F_4=\int_{0}^{\infty}d\tilde{y}\int_{0}^{\infty}dy \, e^{-\a y\tilde{y}e^{\lam(t_1+t_2-t_3-t_4+i\pi)/2}}h(t_{12},y)\tilde{h}(t_{34},\tilde{y})
\ee 
where various notations will be explained below. We now proceed to describe the derivation of~\eqref{bsa}.


\subsection{Towards a scattering formula\label{subsec:Towards-a-scattering}}


Without derivative on $\bar t$, all order generalization of the vertex can be written as
\begin{align} \label{eq:93}
\avg{W(t_{1})W(t_{2})}_0 &=  \sum_{\substack{i_s=1,2\\j,k_s,m\in\N\\s=1,\cdots,m}}\f {C_{j;k_{1}\cdots k_{m}}^{i_{1}\cdots i_{m}}}{m!}\del^{j}g_W(t_{12})\del^{k_{1}}_{t_{i_1}}\phi_{i_{1}}(\bar t_W;t_{i_1})\cdots\del^{k_{m}}_{t_{i_m}}\phi_{i_{m}}(\bar t_W;t_{i_m}) \\
& \equiv  \sum_{\{I_s\},m} \f 1 {m!} \mC_{I_1\cdots I_m} \phi^{I_1}\cdots\phi^{I_m}
\label{ecd}
\end{align}
where in the first line $\avg{\cdot}_0$ means taking the expectation value of bare operators, and we assume $\Im t_1 < \Im t_2$ so that 
the arguments for $\phi_i$ is $t_i$ ($i=1,2$); and in the second line we defined the notation
\begin{align}  
I\equiv (i,k),\quad \phi^I\equiv\del^k_{t_i}\phi_i(\bar t_W;t_i),\quad \mC_{I_{1}\cdots I_{m}}=\sum_j C_{j;k_{1}\cdots k_{m}}^{i_{1}\cdots i_{m}}\del^{j}g_{W}(t_{12}) \label{eq:mC}
\end{align}
where $\mC$ is a tensor function of $t_{12}$. Note that separate $t_{1,2}$ derivatives on $W_0(t_{1,2})$ in \eqref{new-uen1} and \eqref{new-uen2} are combined to act on the argument of $g_W(t_{12})$ by translation symmetry. In \eqref{eq:93}, the range of the sum for $m$ is from 0 to $\infty$ but the sum for $j,k_s$ could be either finite or infinite for each $m$. For simplicity, we will consider their ranges to be finite at each $m$. For each $k_s$, we choose the range of sum to be the same, i.e. $k_1,\cdots,k_m\in\{0,\cdots,d_m \}$. This choice defines $\phi^I$ in \eqref{eq:mC} as a $2d_m$ dimensional vector at each $m$. It follows that $\mC_{I_1\cdots I_m}$ can be chosen as a symmetric tensor of order $m$ by the permutation symmetry of among all $\phi^I$ in \eqref{ecd}. For $m=0$, we have $C_{j}=\d_{j,0}$ because the leading order of \eqref{eq:93} is just $g_W(t_{12})$. Similarly expansion applies to $\avg{V(t_3)V(t_4)}$ for $\Im t_{34}\in(-2\pi,0)$ with notation
\begin{align}  
\avg{V(t_3)V(t_{4})}_0&=\sum_{\{I_s\},m} \f 1 {m!} \tilde\mC_{I_1\cdots I_m} 
\tilde \phi^{I_1}\cdots\tilde\phi^{I_m} \\
\tilde\phi^I\equiv\del^k_{t_{i+2}}\phi_i(\bar t_V;t_{i+2}),&\quad \tilde\mC_{I_{1}\cdots I_{m}}=\sum_j \tilde C_{j;k_{1}\cdots k_{m}}^{i_{1}\cdots i_{m}}\del^{j}g_{V}(t_{34})
\end{align} 

With four time variables $(t_1,t_2,t_3,t_4)$ in the analytic domain $\mD$, a generic four-point function $\hat\mF$ is
\begin{align}
 \hat \mF  (t_{1},t_{2};t_{3},t_{4}) 
=\sum_{\{I_s,I'_s\},m,m'}\f{\mC_{I_1\cdots I_m}\tilde\mC_{I'_1\cdots I'_{m'}}}{m!m'!}\avg{\hat\mT \phi^{I_1}\cdots\phi^{I_m}\tilde\phi^{I'_1}\cdots\tilde\phi^{I'_{m'}}}
\label{eq:153}
\end{align}
With the effective action being quadratic, the $(m+m')$-point function of $\phi_i$ factorizes into products of two-point functions. We will ignore the self-interaction terms (those pairs of $\phi_i$ with same $\bar t$) because they do not grow in OTOC. It follows that $m'=m$ and 
\be 
\hat \mF(t_{1},t_{2};t_{3},t_{4})\app \sum_{\{I_s,I'_s\},m}\f{\mC_{I_1\cdots I_m}\tilde\mC_{I'_1\cdots I'_{m}}}{m!}\prod_{s=1}^m\avg{\hat\mT \phi^{I_s}\tilde\phi^{I'_s}}\label{eq:162}
\ee
where the the permutation symmetry in Wick contraction gives $m!$ that
cancels one $m!$ in the denominator.

The KMS condition requires $\hat \mF$ invariant under $t_1\ra t_2-2\pi i$ and $t_2\ra t_1$ (and also $t_3\ra t_4-2\pi i$ and $t_4\ra t_3$). Using~\eqref{pkms1}-\eqref{pkms2}, we find coefficients $C$ and $\tilde C$ must satisfy the constraints
\be 
C^{i_1\cdots i_m}_{j;k_1\cdots k_m}=(-)^jC^{\bar i_1\cdots \bar i_m}_{j;k_1\cdots k_m},\quad \tilde C^{i_1\cdots i_m}_{j;k_1\cdots k_m}=(-)^j\tilde C^{\bar i_1\cdots \bar i_m}_{j;k_1\cdots k_m} \label{3.2}
\ee
where the map $i\ra \bar i$ means $1\leftrightarrow 2$. This is equivalent to
\be 
\mC_{I_1\cdots I_m}(t)=\mC_{\bar I_1\cdots \bar I_m}(-t-2\pi i),\quad\tilde\mC_{I_1\cdots I_m}(t)=\tilde\mC_{\bar I_1\cdots \bar I_m}(-t-2\pi i) \label{kmsmC}
\ee
where $\bar I$ means $(i,k)\ra (\bar i,k)$.

We further require the right hand side of~\eqref{ecd} to be invariant under 
shift symmetry \eqref{eq:x44-1}, which lead to
\begin{equation}
\mC_{I_1\cdots I_m} (\del^{k_1}\d\phi_{i_1})\cdots\phi^{I_m}=0,\quad \d\phi_i\equiv(\d\phi_1,\d\phi_2)=(e^{\pm \lambda(t_{1}+i \pi)},-e^{\pm \lambda t_{2}})\label{eq:163}
\end{equation}
for arbitrary $\phi^I$. This is a quite strong constraint. Let us define the $2d_m$ dimensional vector $e^{I}=\del^{k}\d\phi_{i}$ at each $m$ and (\ref{eq:163}) becomes
\be
\sum_{I_{1}}\mC_{I_{1}\cdots I_{m}}e^{I_{1}}=0 \label{6.12x}
\ee  
At each $m$, for the symmetric tensor $\mC$ we can always find a linear independent set of vectors  $\{u^{(m)a}\}$  such that
\be  
\mC_{I_{1}\cdots I_{m}}=\sum_{a\in D_{m}}\xi_{a}^{(m)}u_{I_{1}}^{(m)a}\cdots u_{I_{m}}^{(m)a} \label{5.11C}
\ee  
where $D_m$ is the size of the set, and $\xi_a^{(m)}$ is a $D_m$ dimensional vector function of $t_{12}$. By \eqref{6.12x}, the whole set $\{u^{(m)a}\}$ must be orthogonal  to $e$, i.e. $u^{(m)a}_I e^I=0$ for all $a\in D_m$. Similar decomposition applies to $\tilde \mC$.

From the definition of $\mC$, it is a function
of $t_{12}$, and so is $u_{I}^{(m)a}=u_{I}^{(m)a}(t_{12})$. The orthogonality to $e^I$ for each $u^{(m)a}_I$ leads to
\begin{equation}
\sum_{k}u_{1,k}^{(m)a}(t_{12})(\pm \lambda)^{k}e^{\pm \lambda(t_{1}+i\pi)}=\sum_{k}u_{2,k}^{(m)a}(t_{12})(\pm \lambda)^{k}e^{\pm \lambda t_{2}}\label{eq:169}
\end{equation}
which implies
\begin{equation}
u_{+}^{(m)a}(\pm\lambda,t_{12})=\mp u_{-}^{(m)a}(\pm\lambda,t_{12})\coth\f{\lambda(t_{12}+i\pi)}2 \label{n5.13}
\end{equation}
where we define
\be 
u_{\pm}^{(m)a}(\lambda,t) \equiv\f 12\sum_{k}(u_{1,k}^{(m)a}(t)\pm u_{2,k}^{(m)a}(t))\lambda^{k} \label{n5.14}
\ee
Take transformation $t_1\ra t_2-2\pi i$, $t_2\ra t_1$ and $(i,k)\leftrightarrow (\bar i,k)$ in \eqref{n5.13}. It follows that $u_{\bar{I}}^{(m)a}(-t_{12}-2\pi i)$ is also an orthogonal vector to $e^I$ where $\bar I=(\bar i,k)$. Let us choose the normalization of $u_{I}^{(m)a}$ such that $\xi_a^{(m)}(t_{12})$ is invariant under KMS transformation $t_{12}\ra -t_{12}-2\pi i$. By KMS symmetry of $\mC$ in \eqref{kmsmC}, we must have both $u^{(m)a}_I(t_{12})$ and $u^{(m)a}_{\bar{I}}(-t_{12}-2\pi i)$ summed with the same $\xi_a^{(m)}$ coefficient in \eqref{5.11C}. Sometimes, these two are linear dependent or even the same, in which cases we just need to include one of them.

Taking \eqref{5.11C} to \eqref{eq:162}, we have
\begin{equation}
\hat\mF(t_{1},t_{2};t_{3},t_{4})=\sum_{m}\f 1{m!}\sum_{a\in D_{m},b\in\tilde{D}_{m}}\xi_{a}^{(m)}\tilde{\xi}_{b}^{(m)}\left(\sum_{I,I'}u_{I}^{(m)a}\avg{\hat\mT\phi^{I}\tilde{\phi}^{I'}}\tilde{u}_{I'}^{(m)b}\right)^{m}\label{eq:170}
\end{equation}
Note that the shift symmetry \eqref{eq:169} is in the same form as \eqref{eq:x44}. Therefore, we can use the same technique in Section \ref{sec:2.4} to show that all TOC do not have exponential growth piece. Indeed, the proof in Section \ref{sec:2.4} does not depend on the explicit form of $L_t$. One can complete a similar proof for each $\sum_{I,I'}u_{I}^{(m)a}\avg{\hat\mT\phi^{I}\tilde{\phi}^{I'}}\tilde{u}_{I'}^{(m)b}$ in \eqref{eq:170} by replacing $L_{t_{1}}g_{W}$ with $\sum_{k}u_{1,k}^{(m)a}(t_{12})\del_{t_{1}}^{k}$
and $L_{t_2}g_{W}=\sum_{k}u_{2,k}^{(m)a}(t_{12})\del_{t_{2}}^{k}$
(and doing similar replacement for $L_{t_{3,4}}$) in Section \ref{sec:2.4}.

For OTOC $F_4$ in \eqref{eq:FG2.63}, the replacement still holds. Similar to \eqref{eq:151}, we can write in a symmetric way that
\begin{align}
& \sum_{I,I'}u_{I}^{(m)a}\avg{\phi^{I}\tilde{\phi}^{I'}}\tilde{u}_{I'}^{(m)b} \nn\\
 =& \a e^{\lambda(t_{1}+t_{2}-t_{3}-t_{4}+i\pi)/2}\left(u_{+}^{(m)a}(\lambda,t_{12})\cosh\f{\lambda(t_{12}+i\pi)}2+u_{-}^{(m)a}(\lambda,t_{12})\sinh\f{\lambda(t_{12}+i\pi)}2\right)\nonumber \\
 & \times\left(\tilde{u}_{+}^{(m)a}(-\lambda,t_{34})\cosh\f{\lambda(t_{34}+i\pi)}2-\tilde{u}_{-}^{(m)a}(-\lambda,t_{34})\sinh\f{\lambda(t_{34}+i\pi)}2\right)+(\lam\leftrightarrow -\lam) \nn\\
 =& \a e^{\lambda(t_{1}+t_{2}-t_{3}-t_{4}+i\pi)/2} \f {u_{+}^{(m)a}(\lambda,t_{12})\tilde{u}_{+}^{(m)a}(-\lambda,t_{34})}{\cosh\f{\lambda(t_{12}+i\pi)}2\cosh\f{\lambda(t_{34}+i\pi)}2}+(\lam\leftrightarrow -\lam)  \label{eq:106}
\end{align}
where we defined
\begin{align}
\tilde{u}_{\pm}^{(m)a}(\lambda,t) & \equiv\f 12\sum_{k}(\tilde{u}_{1,k}^{(m)a}(t)\pm\tilde{u}_{2,k}^{(m)a}(t))\lambda^{k}
\end{align}
and used \eqref{n5.13} in the last step for both $u_{\pm}^{(m)a}$ and $\tilde{u}_{\pm}^{(m)a}$.

Since we are only interested in large Lorentzian time separation $t_{1},t_2\gg t_3,t_4$ or $t_{1},t_2\ll t_3,t_4$, we only need to keep the exponentially
growth term in (\ref{eq:106}) because the other term is exponentially
suppressed. Let us take $t_{1},t_2\gg t_3,t_4$, then (\ref{eq:170}) becomes
\begin{equation}
F_4=\sum_{m}\f 1{m!}\sum_{a\in D_{m}}\xi_{a}^{(m)}\left(\f{u_{+}^{(m)a}(\lambda,t_{12})}{\cosh\f{\lambda(t_{12}+i\pi)}2}\right)^{m}\sum_{b\in\tilde{D}_{m}}\tilde{\xi}_{b}^{(m)}\left(\f{\tilde u_{+}^{(m)b}(-\lambda,t_{34})}{\cosh\f{\lambda(t_{34}+i\pi)}2}\right)^{m}X^{m}\label{eq:176}
\end{equation}
with the exponential growth
\be 
X\equiv \a e^{\lambda(t_{1}+t_{2}-t_{3}-t_{4}+i\pi)/2} \label{x521}
\ee
Assume we can do an inverse Mellin transformation to define $h$ and $\tilde h$ as
\begin{align}
\sum_{a\in D_{m}}\xi_{a}^{(m)}\left(\f{u_{+}^{(m)a}(\lambda,t_{12})}{\cosh\f{\lambda(t_{12}+i\pi)}2}\right)^{m} & =\int_{0}^{\infty}dyy^{m}h(t_{12},y)\label{eq:108}\\
\sum_{b\in\tilde{D}_{m}}\tilde{\xi}_{b}^{(m)}\left(\f{\tilde u_{+}^{(m)b}(-\lambda,t_{34})}{\cosh\f{\lambda(t_{34}+i\pi)}2}\right)^{m} & =\int_{0}^{\infty}d\tilde{y}(-\tilde{y})^{m}\tilde{h}(t_{34},\tilde{y}) \label{eq:108-1}
\end{align}
Then we can rewrite (\ref{eq:176}) as
\begin{align}
F_4 & =\int_{0}^{\infty}d\tilde{y}\int_{0}^{\infty}dy\sum_{m}\f 1{m!}h(t_{12},y)(-Xy\tilde{y})^{m}\tilde{h}(t_{34},\tilde{y})\nonumber \\
 & =\int_{0}^{\infty}d\tilde{y}\int_{0}^{\infty}dye^{-Xy\tilde{y}}h(t_{12},y)\tilde{h}(t_{34},\tilde{y})\label{eq:178}
\end{align}
This has exactly the same form as \cite{Stanford:2021bhl, Gu:2021xaj}, and also matches with
the trans-plankian string scattering formula near horizon of a AdS black
hole \cite{Shenker:2014cwa}. In \cite{Stanford:2021bhl}, this scattering formula was conjectured by a heuristic argument and the authors of \cite{Gu:2021xaj} later proved it with a specific structure of Feynmann diagrams. In this work, we show that \eqref{eq:178} holds in a more general scenario since it is just a result of a shift symmetry. 

Note that the assumption of inverse Mellin transformation requires analyticity
in $m$. This is a nontrivial constraint on the vertices because at
each level of $m$ one could have choosen $\xi^{(m)}_a$ and the sets $D_{m}$ and $\tilde{D}_{m}$  quite randomly in a pattern without analyticity
in $m$. However, there is a simple and sufficient way to guarantee analyticity, which requires three parts: 
\begin{enumerate}
 \item The total ways of coupling
between bare operator and effective mode does not change as we increase
the number $m$ of effective modes, which picks $D_{m}$ and $\tilde{D}_{m}$
as some fixed sets for all $m$, in which the vector dimension $2d_m$ is also fixed. 
\item All these types of couplings exist at
any level of $m$, which releases the $m$ dependence of $u_{+}^{(m)a}$.
\item The coefficient $\xi_{a}^{(m)}$
is an analytic function of $m$.
\end{enumerate} 
It is noteworthy that the ordinary ladder diagrams in eikonal scattering obeys these three conditions. Therefore, this explains why the exponentiation in \eqref{eq:178} is also a result of eikonal scattering \cite{Shenker:2014cwa}, in which $e^{-X y \tilde y}$ is the eikonal scattering amplitude, $h$ is the wave function of two $W$'s and $\tilde h$ is the wave function of two $V$'s.

\subsection{An example \label{subsec:An-example}}

In this subsection, we will present a simple example following the general construction in the last subsection. We will solve a simple orthogonal vector $u_{I}$, find the vertices leading to this vector $u_I$, compute the exponentiation formula and compare it with the known result of large $q$ SYK model \cite{Gu:2021xaj}.

Let us first solve the orthogonal vectors $u_{I}^{a}$. Define $q=e^{-\lambda(t_{12}+i\pi)}$
and the equation $u_{I}^{a}e^I=0$ can be written as
\begin{equation}
\sum_{k}u_{1k}^{a}(q)\lambda^{k}=\sum_{k}u_{2k}^{a}(q)\lambda^{k}q,\quad\sum_{k}u_{1k}^{a}(q)(-\lambda)^{k}q=\sum_{k}u_{2k}^{a}(q)(-\lambda)^{k}\label{eq:179}
\end{equation}
Without loss of generality, we can assume $u_I^a$ is a polynomial in $q$
\begin{equation}
u_{I}^{a}=\sum_{n=0}^{p_{a}}c_{I,n}^{a}q^{n} \label{eq:uform}
\end{equation}
up to normalization where $p_{a}$ could be either a finite number
or $\infty$. Then (\ref{eq:179}) leads to
\begin{equation}
\sum_{k}c_{1k,n}^{a}\lambda^{k}=\sum_{k}c_{2k,n-1}^{a}\lambda^{k},\quad\sum_{k}c_{1k,n-1}^{a}(-\lambda)^{k}=\sum_{k}c_{2k,n}^{a}(-\lambda)^{k}
\end{equation}
for $n\geq0$ with definition $c_{I,-1}^{a}=0$. The simplest nontrivial
solution is for $p_{a}=1$ and $k=0,1$ 
\begin{align}
c_{10,0} & =c_{20,1}=-\lambda,\quad c_{20,0}=c_{10,1}=\lambda c\label{eq:182}\\
c_{11,0} & =c_{21,1}=1,\quad c_{21,0}=c_{11,1}=c\label{eq:183}
\end{align}
We can choose the normalization of $u_I$ such that it is invariant under KMS transformation $q\ra 1/q$ and $i\ra \bar i$. In this case, by the discussion below \eqref{n5.14}, we can construct a KMS invariant $\mC$  just using this vector. It is easy to see that the appropriate normalization is $q^{-1/2}$ and the orthogonal vector is
\be 
u_{10}=\f{\lam(cq-1)}{q^{1/2}},\quad u_{11}=\f{1+cq}{q^{1/2}},\quad u_{20}=\f{\lam (c-q)}{q^{1/2}},\quad u_{21}=\f{c+q}{q^{1/2}} \label{usol}
\ee

Then we will solve the vertices with coefficients $C_{j;k_{1}\cdots k_{m}}^{i_{1}\cdots i_{m}}$ leading to this vector
$u_{I}$ for each order $m$. Let us start with $m=1$. We can further impose a simple condition that $j$
only takes values $0$ and $1$ in \eqref{eq:mC}. Define $n$-th order $\del_{t_{12}}$ derivative to bare correlator $g_W(t_{12})=g(q)$ as $g^{n}(q)=(-\lambda q\del_{q})^{n}g(q)$
with $g^{0}(q)=g(q)$. For $c\neq0,\infty$, comparing \eqref{eq:mC} and \eqref{5.11C} leads to
\begin{equation}
C_{0;k}^{i}g(q)+C_{1;k}^{i}g^{1}(q)=\xi^{(1)}(q)u_{ik}(q) \label{diffg}
\end{equation}
Our goal is to solve coefficients $C^i_{j;k}$. However, for arbitrary $C^i_{j;k}$ and $\xi^({1})(q)$, this is also a differential equation for $g(q)$, which will constrain $g(q)$ to a specific form. We solve \eqref{diffg} in detail in Appendix \ref{app:e} and present the result here. The correlation $g(q)$ must be in the form
\begin{equation}
g(q)=(q^{1/2}+q^{-1/2})^{-2\D}
\end{equation}
up to normalization and with a constant $\D$. It is obvious that this $g(q)$ obyes KMS symmetry $g(q)=g(1/q)$. With this solution, the coefficients $C^i_{j;k}$ are 
\begin{align}
C_{0;0}^{1} =\f{\D\lambda(c-1)}{c+1},&\quad C_{1;0}^{1}=1,\quad C_{0;1}^{1}=\D,\quad C_{1,1}^{1}=\f{c-1}{(c+1)\lambda}\\
C_{0;0}^{2} =\f{\D\lambda(c-1)}{c+1},&\quad C_{1;0}^{2}=-1,\quad C_{0;1}^{2}=\D,\quad C_{1;1}^{2}=-\f{c-1}{(c+1)\lambda}
\end{align}
where we choose the normalization such that $C^{1}_{1;0}=1$. One can check that this solution obeys KMS symmetry \eqref{3.2}. Taking them into \eqref{diffg} and using \eqref{usol}, we find
\be  
\xi^{(1)}(q)=\f{2\D g(q)}{(1+c)(q^{1/2}+q^{-1/2})}
\ee 
which is explicitly KMS invariant as expected. 

In the two component vertex form, we have
\begin{align}
\sum_{jk}C_{j;k}^{i}\del^{j}&g(t_{12})\del_{t_{i}}^{k}\phi_{i}(\bar t;t_i)= (\del g\phi_{1}+\D g\del\phi_{1},-\del g\phi_{2}+\D g\del\phi_{2})\nonumber \\
 & +\left(\f{c-1}{c+1}\right)(\D\lambda g\phi_{1}+\lambda^{-1}\del g\del\phi_{1},\D\lambda g\phi_{2}-\lambda^{-1}\del g\del\phi_{2})
\end{align}
where on the RHS we have suppressed the notation and the derivatives only act on the second argument of $\phi_i(\bar t;t_i)$. Note that the first line, namely $c=1$, exactly matches with the vertex
(\ref{pio-2}) of large $q$ SYK model.

To construct the higher order coupling coefficient $C_{j;k_{1}\cdots k_{m}}^{i_{1}\cdots i_{m}}$ 
such that (\ref{5.11C}) holds for all $m$ is not hard for this example. The
point is to note the following feature
\begin{equation}
\f{g^{n}(q)}{g(q)}=\f{\mP_{n}(q)}{(1+q)^{n}}
\end{equation}
where $\mP_{n}(q)$ is a $n$-order polynomial has no factor of $(1+q)$.
Therefore, for any $n$-order polynomial $P_{n}(q)$, we can pick
a linear combination such that
\begin{equation}
\sum_{j=0}^{n}w_{j}\f{g^{j}(q)}{g(q)}=\f 1{(1+q)^{n}}\sum_{j=0}^{n}w_{j}(1+q)^{n-j}\mP_{j}(q)=\f{P_{n}(q)}{(1+q)^{n}}
\end{equation}
Since $q^{1/2}u_{I}(q)$ is a first order polynomial of $q$, the product $q^{m/2}u_{I_{1}}(q)\cdots u_{I_{m}}(q)$ is  a polynomial of $q$
up to order $m$ for each choice of $(I_{1},\cdots,I_{m})$. It follows that we can choose $C_{j,k_{1}\cdots k_{m}}^{i_1,\cdots,i_k}$ such that
\begin{equation}
\mC_{I_1\cdots I_m}=\sum_{j=1}^mC_{j;k_{1}\cdots k_{m}}^{i_1,\cdots,i_k}(-\lambda q\del_{q})^{j}g(q)=\f{z(m)g(q)}{(q^{1/2}+q^{-1/2})^{m}}u_{I_{1}}(q)\cdots u_{I_{m}}(q)\label{eq:138}
\end{equation}
where $z(m)$ is a constant that could depend on $m$. It clear that \eqref{eq:138} obeys KMS symmetry \eqref{kmsmC}. 

If
we only include this type of orthogonal vector $u_{I}$ in $\mC$, we have
\begin{equation}
\xi^{(m)}=\f{z(m)g(q)}{(q^{1/2}+q^{-1/2})^{m}},\quad u_{\pm}(\lambda,t)=\f{\lambda c(q\pm 1)}{q^{1/2}},\quad u_{\pm}(-\lam,t)=-\f{\lam (1\pm q)}{q^{1/2}}
\end{equation}
which by (\ref{eq:108}) and \eqref{eq:108-1} leads to
\begin{align}
\int_{0}^{\infty}dyy^{m}h(t_{12},y)&=\f{z(m)g_W(t_{12})(\lambda c)^{m}}{\left(\cosh\f{\lambda(t_{12}+i\pi)}2\right)^{m}}\\
\int_{0}^{\infty}d\tilde y\tilde y^{m}\tilde h(t_{34},\tilde y)&=\f{\tilde z(m)g_V(t_{34})\lambda^{m}}{\left(\cosh\f{\lambda(t_{34}+i\pi)}2\right)^{m}}
\end{align}
where we assume the vertex of $V$ consists of the same vector $u_I$ but with a possibly different coefficient $\tilde z(m)$. For analytic
functions $z(m)$ and $\tilde z(m)$ that decay fast enough along $\Im m\ra\pm\infty$,
the Mellin inversion theorem guarantees the existence of $h(t_{12},y)$ and $\tilde h(t_{34},\tilde y)$.
However, to determine the exact form of $z(m)$ and $\tilde z(m)$ needs
detailed knowledge of the dynamics of underlying UV model (for example \cite{Gu:2021xaj}).

For large $q$ SYK model, we can choose $W=V$ to be the fundamental Majorana fermion $\psi$, whose conformal weight is $\D=1/q$. The $\a$ parameter in \eqref{x521} is given by \eqref{pio-2}. Its exponentiation exactly falls into above case of $c=1$ with the following choices of $z(m)$ and $\tilde z(m)$
\be  
z(m)=\tilde z(m)=\f{\G(2\D+m)}{2^m\G(2\D)}
\ee 
which leads to
\begin{align}
h(t,y) & =\tilde h (t,y)=g_{\psi}(t)\f{\left(\f{2}{\lam}\cosh\f{\lambda(t+i\pi)}2\right)^{2\D}}{\G(2\D)}y^{2\D-1}\exp\left(-\f{2y}{\lam}\cosh\f{\lambda(t+i\pi)}2\right) \label{5.75}
\end{align}
where $g_\psi(t)$ is the fermion correlation function \eqref{pio-1}. One can check that this exactly matches with the result in \cite{Gu:2021xaj}.

\section{Conclusion and discussion} \label{sec:Discussion-and-conclusion}

In this paper, we constructed an effective field theory to capture the behavior of OTOCs of non-maximal
quantum chaotic systems. While the theory is constructed phenomenologically, we showed that it is constraining enough to predict the general structure of OTOCs both at leading order in the $1/N$ expansion, and after resuming over an infinite number of higher order corrections. These general results agree with those preciously explicitly obtained in specific models. 
We also showed that the general structure of the EFT can in fact be extracted from the large $q$ SYK model, providing further support 
for its validity. There are many future directions to explore, on which we make some brief comments.

\paragraph{Higher dimensional systems}

A most immediate direction is to generalize the current discussion to higher dimensional systems. 
Including spatial directions will make it possible to consider much wider range of physical issues, for example, 
operator growths and scrambling in spatial directions, the behavior of the butterfly velocity \cite{Roberts:2014isa, Khemani:2018sdn,Mezei:2019dfv,Lunkin:2022kvj}, connections between quantum chaos and energy as well as charge transports \cite{Gu:2016oyy,Gu:2017ohj,Blake:2017ris}, and so on. 
In the case of maximal chaos, a phenomenon that connects chaos and energy transport is the so-called 
pole-skipping~\cite{Blake:2017ris,Grozdanov:2017ajz}. Understanding what happens to this phenomenon for non-maximal chaotic systems is of interests. In~\cite{Choi:2020tdj}
it was conjectured that pole-skipping survives in non-maximal system and the location of pole-skipping 
is given by
\begin{equation}
(\w,k)=i(2\pi/\b)(1,1/u_{B}^{(T)})
\end{equation}
where 
$u_{B}^{(T)}$
is an upper bound of the true butterfly velocity $u_{B}$.
An EFT including spatial directions could help check the conjecture and understand connections between energy transport and chaos in more general systems. 

\paragraph{Physical nature of the effective fields and the shift symmetry}

Here we introduced chaos effective fields and shift symmetry on phenomenology ground. In the example of the large-$q$ SYK model, we can identify the effective fields and origin of shift symmetry from the microscopic system. It is, however, not clear whether the understanding obtained in this model can be applied to general systems. 

In maximal chaotic holographic systems, the shift symmetry of the EFT should be related to the existence of a sharp horizon. 
It is an outstanding question regarding the nature of the horizon when including stringy corrections on the gravity side, 
understanding the physical nature and origin of the shift symmetry for non-maximal case could provide hints for this question.

\paragraph{Effective field theories for Reggeons} 

As mentioned in the Introduction (see Fig.~\ref{fig:reg}), there is a close connection between the exponential behavior in non-maximally chaotic systems and scattering amplitudes in the Regge limit. Our formulation of an EFT for non-maximally chaotic systems could provide new ideas for formulating effective field theories for Reggeons. More explicitly, the stringy scattering processes corresponding to OTOCs in holographic systems can be described by the BFKL Pomeron~\cite{Shenker:2014cwa,Brower:2006ea}. 
The effective fields we identified could shed light on an effective description of the Pomeron.

\section*{Acknowledgements} We would like to thank Mark Mezei and Daniel Jafferis for 
stimulating and helpful discussions. PG is supported by the US Department of Defense (DOD) grant  KK2014 and also by the Simons foundation as a member of the {\it It from Qubit} collaboration. HL is supported by the Office of High Energy Physics of U.S. Department of Energy under grant Contract Number DE-SC0012567 and DE-SC0020360 (MIT contract ${\#}$ 578218).

\appendix
\section{A few oversimplified constructions \label{app:a}}
 
In this appendix, we list two oversimplified constructions of EFT for non-maximal chaos, which are slightly generalized from the EFT for maximal chaos \cite{Blake:2017ris,Blake:2021wqj} with one time argument. It turns out that both constructions are only compatible with maximal chaos. The purpose of this section serves as a support for the construction in Section \ref{sec:The-structure-of} as a minimal and sufficient generalization to account for non-maximal chaos. In particular, including two time arguments in the effective modes is necessary.

\subsection{Multiple effective modes with one time argument} \label{app:a.1}
The simplest generalization of  \cite{Blake:2017ris,Blake:2021wqj} is to include more effective modes but still formulated with one time argument. Let us label these effective modes as $\p_\mu$ with $\mu=1,\cdots,D$ for some finite $D$. The four-point function $\hat\mF_{WWVV}(t_1,t_2;t_3,t_4)$ is symmetric under exchange of $t_1 \leftrightarrow t_2$ and $t_3 \leftrightarrow t_4$ respectively. Therefore, to define the coupling between a bare operator $W_0$ and effective modes $\p_\mu$, we must respect this symmetry. There are two simple ways: one is that every $W_0$ couples with all $\p_\mu$, the other is coupling in an ordered way (just like \eqref{new-uen1} and \eqref{new-uen2} for two effective modes). Here we will consider the first choice.

In this case, the coupling in linear order of $\p_\mu$ is 
\be 
W(t)=W_0(t)+L_t^\mu W_0(t)\p_\mu (t),\quad V(t)=V_0(t)+\tilde L_t^\mu V_0(t)\p_\mu(t)
\ee
where $\mu$ is summed from 1 to $D$ and $L_t^\mu$ is a set of differential operators
\be  
L_t^\mu W_0(t)\p_\mu(t)=\sum_{nm}c^\mu_{nm}\del_t^n W_0(t)\del_t^m \p_\mu(t) \label{a.2-L}
\ee
and $\tilde L_t^\mu$ is defined similarly with $c^\mu_{nm}\ra \tilde c^\mu_{nm}$.
To quadratic order of $\p_\mu$, the four-point function is
\begin{align}
\hat\mF_{WWVV}(t_1,t_2;t_3,t_4)=\sum_{i,j=1,2}L_{t_i}^\mu g_W\tilde L_{t_{j+2}}^\nu g_V\avg{\mT\p_\mu (t_i)\p_\nu(t_{j+2})} \label{a.3-F}
\end{align}
where $g_{W,V}$ are short for $g_W(t_{12})$ and $g_V(t_{34})$ and $\avg{\mT \p_\mu \p_\nu}$ is the Euclidean time ordered two-point function in the thermal state. Imposing the same shift symmetry \cite{Blake:2017ris} 
\be  
 \p_\mu(t)\ra\p_\mu(t)+\a e^{\pm \lam t}
\ee 
 in the effective action, we will have the following exponential terms in the Wightman function
\be 
\avg{\p_\mu(t)\p_\nu(0)}=d_{\mu \nu} e^{\lam t}+\bar d_{\mu \nu} e^{-\lam t}
\ee
where $c_{\mu\nu}$ and $\bar c_{\mu \nu}$ are two nonzero constant matrices.

Let us consider OTOC $F_4$ and TOC $G_4$ defined as
\be  
F_{4}=\avg{W(t_{1})V(t_{3})W(t_{2})V(t_{4})},\quad H_{4}=\avg{W(t_{1})W(t_{2})V(t_{3})V(t_{4})} 
\ee  
where $t_1,t_2\gg t_3,t_4$ or $t_1,t_2\ll t_3,t_4$. For TOC, we have
\begin{align}
H_4=&d_{\mu \nu}\left( L_{t_1}^\mu g_We^{\lam t_1}+ L_{t_2}^\mu g_We^{\lam t_2} \right)\left(\tilde L_{t_3}^\nu g_V e^{-\lam t_3}+\tilde L_{t_4}^\nu g_V e^{-\lam t_4} \right)+b.c. \label{a.7}
\end{align}
where $b.c.$ means bar-conjugate which replaces $c_{\mu\nu}$ with $\bar c_{\mu\nu}$ and swaps $e^{\lam t} \leftrightarrow e^{-\lam t}$. For OTOC, we have 
\begin{align}
F_4=H_4+\left[d_{\mu\nu}\left(\tilde L_{t_3}^\mu g_V e^{\lam t_3}L_{t_2}^\nu g_W e^{-\lam t_2}-L_{t_2}^\mu g_We^{\lam t_2}\tilde L_{t_3}^\nu g_V e^{-\lam t_3}\right)+b.c.\right] \label{a.8}
\end{align}

Let us define two vector functions
\be  
v^\mu(t_1,t_2)=L_{t_1}^\mu g_We^{\lam t_1}+ L_{t_2}^\mu g_We^{\lam t_2}, \quad u^\nu(t_3,t_4)=\tilde L_{t_3}^\nu g_V e^{-\lam t_3}+\tilde L_{t_4}^\nu g_V e^{-\lam t_4}
\ee  
Absence of exponential terms in TOC $H_4$ means 
\be
d_{\mu\nu}v^\mu(t_1,t_2)u^\nu(t_3,t_4)=0 \label{a10x}
\ee
and the other equation with bar-conjugate. For \eqref{a10x}, we can first expand $v^\mu$ and write it as
\be  
G_{even}(t_3,t_4;t)=-\tanh \f {\lam t} 2 G_{odd}(t_3,t_4;t) \label{a11x}
\ee  
where we define
\be 
G_{even/odd}(t_3,t_4;t)\equiv \sum_{n~even/odd}\sum_{m} d_{\mu \nu}u^\nu(t_3,t_4)c^\mu_{nm}\del_t^n g_W(t) \lambda^m
\ee

Let us consider a few cases. 
\begin{enumerate}
\item If both sides of \eqref{a11x} are nonzero, we can take a KMS transformation $t\ra-t-2\pi i$ in \eqref{a11x}. Since $g_W(t)=g_W(-t-2\pi i)$ by KMS condition, we have  $G_{even}\ra G_{even}$ and $G_{odd}\ra-G_{odd}$ and thus \eqref{a11x} yields
\be 
\tanh \f {\lam t} 2 =\tanh \f {\lam (t+2\pi i)} 2 \implies \lambda =1 \label{a.13}
\ee

\item If both sides of \eqref{a11x} are zero, this means 
\be  
 d_{\mu \nu}u^\nu(t_3,t_4)L_{t_1}^\mu g_We^{\lam t_1}=0,\quad  d_{\mu \nu}u^\nu(t_3,t_4)L_{t_2}^\mu g_We^{\lam t_2}=0 \label{a14x}
\ee 
Then we can expand $u^\nu$ in the second equation and find a similar equation to \eqref{a11x}
\be  
\tilde G_{even}(t_1,t_2;t)=\tanh \f {\lam t} 2 \tilde G_{odd}(t_1,t_2;t)  \label{a15x}
\ee 
where we define
\be 
\tilde G_{even/odd}(t_1,t_2;t)\equiv \sum_{n~even/odd}\sum_{m} d_{\mu\nu}(L_{t_2}^\mu g_We^{\lam t_2})\tilde c^\nu_{nm}\del_t^n g_W(t) (-\lambda)^m
\ee
If both sides of \eqref{a15x} are nonzero, the KMS transformation in \eqref{a15x} again leads to \eqref{a.13} and maximal chaos $\lam=1$. 

\item If both sides of \eqref{a15x} are zero, we have
\be  
 d_{\mu \nu}L_{t_2}^\mu g_We^{\lam t_2}\tilde L_{t_j}^\nu g_V e^{-\lam t_j}= 0,\quad j=3,4\label{a17x}
\ee  
Similarly, using the first equation of \eqref{a14x}, we will have either $\lam=1$ or
\be  
 d_{\mu \nu}L_{t_1}^\mu g_We^{\lam t_1}\tilde L_{t_j}^\nu g_V e^{-\lam t_j}= 0,\quad j=3,4
\ee  
Then we can consider another TOC $H'_4=\avg{V(t_3)V(t_4)W(t_1)W(t_2)}$. Following a similar analysis for $H_4$ (which simply swaps $t_{1,2}\leftrightarrow t_{3,4}$ and $W\leftrightarrow V$ everywhere), we will have either $\lam=1$ or 
\be  
d_{\mu \nu}\tilde L_{t_j}^\mu g_Ve^{\lam t_j} L_{t_i}^\nu g_W e^{-\lam t_i}= 0,\quad i=1,2,\quad j=3,4 \label{a18x}
\ee 
Taking both \eqref{a17x} and \eqref{a18x} into \eqref{a.8}, we find that the exponential terms proportional to $d_{\mu\nu}$ vanish in $F_4$. Similarly, we can show that the bar-conjugate terms vanish as well. This means that if we require exponential growth of OTOC but no exponential growth in TOC, we must have maximal chaos $\lam=1$ in this model.

\end{enumerate}

\subsection{Two effective modes with one time argument and ordered coupling} \label{sec:a.2}

As we mentioned before, for two effective modes $\p_i$ with $i=1,2$, there is another simple way to couple bare operators with them in a symmetric way. This is essentially the same as our proposal \eqref{new-uen1} and \eqref{new-uen2} in which $W_0(t_S)$ couples with $\p_1(\tau_S)$ and $W_0(t_L)$ couples with $\p_2(t_L)$ where $t_{L,S}$ is the one of $t_1$ and $t_2$ with larger (smaller) imaginary part. The only difference is that here we will consider 
 the effective modes with one time argument. In other words, there is no $\bar t$ argument.

For this model, we can simply substitute $\phi_i(\bar t,t)$ in Section \ref{sec:2.1} with
\be 
\phi_i(\bar t;t)=\p_i(t)
\ee
In particular, the KMS symmetry \eqref{pkms1}-\eqref{pkms2} reduces to
\be
\avg{\hat \mT\p_1(t)\p_i(t')}=\avg{\hat \mT\p_2(t)\p_i(t')}=\avg{\hat \mT\p_1(t+2\pi i)\p_i(t')}\label{a.17}
\ee
This means that the two effective modes $\p_i(t)$ are completely degenerate to a single effective mode $\p(t)$ in a thermal state with inverse temperature $2\pi$. It reduces to the case in \cite{Blake:2017ris,Blake:2021wqj} which is inevitably restricted to the maximal chaos.

\section{Unitary and dynamical KMS conditions} \label{app:dyna}

The periodicity \eqref{yhi} along imaginary time direction can be understood as $\eta_{s,p}$ in a thermal state with imaginary chemical potential. Let as define the charge carried by $\eta_{s,p}(\bar{\tau};t)$ as
$Q=p$, namely 
\begin{equation}
[Q,\eta_{s,p}(\bar{\tau};t)]=p\eta_{s,p}(\bar{\tau};t)
\end{equation}
It follows that the state consistent with the KMS condition \eqref{kmseta} is 
\begin{equation}
\r=\hat{S}e^{-2\pi iQ/3}e^{-\pi H} \label{rho2.47}
\end{equation}
where $\hat{S}$ is the operator that takes the $s$ value of $\eta_{s,p}$,
namely 
\begin{equation}
\hat S\eta_{s,p}(\bar{\tau};t)\hat S^{-1}=s\eta_{s,p}(\bar{\tau};t) 
\end{equation}

For the state given by \eqref{rho2.47}, the time reversal transformation $\bT$ of $\r$ is given by
\begin{equation}
\bT\r\bT^{-1}=\hat{S}e^{2\pi iQ/3}e^{-\pi H}/Z^*=\r^{\dagger}
\end{equation}
where we assumed time reversal invariance of $\hat{S},Q$ and $H$
(and also hermicity of $\hat{S}$ and $Q$). The time reversal transformation
$\bT$ of $\eta_{s,p}(\bar{\tau};t)$ is defined as flipping $t$
and $\bar{\tau}$ simultaneously
\begin{equation}
\bT\eta_{s,p}(\bar{\tau};t)\bT^{-1}=\eta_{s,p}(-\bar{\tau};-t)
\end{equation}
From this definition, $\bT^{2}=1$. In this definition, the periodicity
of $\eta_{s,p}(\bar{\tau};t)$ along $\bar{\tau}$ direction is consistent
\begin{equation}
\bT\eta_{s,p}(\bar{\tau}+\pi;t)\bT^{-1}=\bT e^{2\pi ip/3}\eta_{s,p}(\bar{\tau};t)\bT^{-1}=e^{-2\pi ip/3}\eta_{s,p}(-\bar{\tau};-t)=\eta_{s,p}(-(\bar{\tau}+\pi);-t)
\end{equation}

Let us consider the generating functional on the Keldysh contour
\begin{equation}
e^{W[J_{s,p}^{(1)},J_{s,p}^{(2)}]}=\Tr\left[\r\left(\tilde{\T}e^{-i\int J_{s,-p}^{(2)}\eta_{s,p}^{(2)}}\right)\left(\T e^{i\int J_{s,-p}^{(1)}\eta_{s,p}^{(1)}}\right)\right]\label{eq:d4}
\end{equation}
where $\T$ is time ordering and $\tilde{\T}$ is inverse time ordering. Let us define another generating functional with $\rho^\dagger$ 
\be  
e^{\bar W[J_{s,p}^{(1)},J_{s,p}^{(2)}]}=\Tr\left[\r^\dag\left(\tilde{\T}e^{-i\int J_{s,-p}^{(2)}\eta_{s,p}^{(2)}}\right)\left(\T e^{i\int J_{s,-p}^{(1)}\eta_{s,p}^{(1)}}\right)\right]\label{eq:d4tilde}
\ee  
Suppose the generating functional can be represented by the path integral
of effective field theory
\begin{align}
e^{W[J_{s,p}^{(1)},J_{s,p}^{(2)}]} & =\int D\eta_{s,p}^{(1)}D\eta_{s,p}^{(2)}e^{iI[J_{s,p}^{(1)},\eta_{s,p}^{(1)};J_{s,p}^{(2)},\eta_{s,p}^{(2)}]}\label{eq:d13-1}
\end{align}
Similar formula applies to $\bar W$ with $I$ replaced with $\bar I$. The unitary and dynamical KMS conditions are as follows.

\begin{enumerate}
    \item Taking two sources $J^{(1)}$ and $J^{(2)}$ identical leads to vanishing $W$ and $\bar W$. To satisfy this condition, we impose
\be  
I[J_{s,p},\eta_{s,p};J_{s,p},\eta_{s,p}]=\bar I[J_{s,p},\eta_{s,p};J_{s,p},\eta_{s,p}]=0
\ee 
Turning off the sources, the effective
action for $\eta_{s,p}^{a,r}$ in Keldysh formalism needs to obey
\begin{equation}
S[\eta_{s,p}^{r},\eta_{s,p}^{a}=0]=\bar S[\eta_{s,p}^{r},\eta_{s,p}^{a}=0]=0 \label{b.8x}
\end{equation}
This means that the effective action does not have $K^{r\cdots r}\eta^r\cdots \eta^r$ term. In other words, each term in the action must contain at least one $\eta^a_{s,p}$.

\item 
Taking complex conjugate of \eqref{eq:d4}, assuming $J_{s,p}^*=J_{s,-p}$, it is clear that
\be  
W[J_{s,p}^{(1)},J_{s,p}^{(2)}]^*=\bar W[J_{s,p}^{(2)},J_{s,p}^{(1)}]
\ee
To guarantee this condition, we need to impose 
\be  
I[J_{s,p}^{(1)},\eta_{s,p}^{(1)};J_{s,p}^{(2)},\eta_{s,p}^{(2)}]^*=-\bar I[J_{s,p}^{(2)},\eta_{s,p}^{(2)};J_{s,p}^{(1)},\eta_{s,p}^{(1)}]
\ee 
which written in terms of effective action without source is
\be  
S[\eta_{s,p}^{r},\eta_{s,p}^{a}]^*=-\bar S[\eta_{s,p}^{r},-\eta_{s,p}^{a}] \label{b.11x}
\ee

\item Under time reversal transformation, the generating functional $W$
becomes
\begin{align}
&e^{W[J_{s,p}^{(1)}(\bar{\tau};t),J_{s,p}^{(2)}(\bar{\tau};t)]}  =\Tr\left[\r^{\dagger}\left(\T e^{i\int J_{s,-p}^{(2)*}(\bar{\tau};t)\eta_{s,p}^{(2)}(-\bar{\tau};-t)}\right)\left(\tilde{\T}e^{-i\int J_{s,-p}^{(1)*}(\bar{\tau};t)\eta_{s,p}^{(1)}(-\bar{\tau};-t)}\right)\right]^{*}\nonumber \\
 &~~~~~ =\Tr\left[\r\left(\T e^{i\int J_{s,-p}^{(1)}(\bar{\tau};t)\eta_{s,-p}^{(1)}(-\bar{\tau};-t)}\right)\left(\tilde{\T}e^{-i\int J_{s,-p}^{(2)}(\bar{\tau};t)\eta_{s,-p}^{(2)}(-\bar{\tau};-t)}\right)\right] \nn\\
 &~~~~~ =\Tr\left[\left(\T e^{i\int J_{s,-p}^{(1)}(\bar{\tau};t)\eta_{s,-p}^{(1)}(-\bar{\tau};-t+i\pi)se^{2\pi ip/3}}\right)\r\left(\tilde{\T}e^{-i\int J_{s,-p}^{(2)}(\bar{\tau};t)\eta_{s,-p}^{(2)}(-\bar{\tau};-t)}\right)\right]\nonumber \\
 &~~~~~ =\Tr\left[\r\left(\tilde{\T}e^{-i\int J_{s,p}^{(2)}(\bar{\tau};t)\eta_{s,p}^{(2)}(-\bar{\tau};-t)}\right)\left(\T e^{i\int J_{s,p}^{(1)}(\bar{\tau};t)\eta_{s,p}^{(1)}(-\bar{\tau};-t+i\pi)se^{-2\pi ip/3}}\right)\right]\nonumber \\
 &~~~~~ =e^{W[se^{2\pi ip/3}J_{s,-p}^{(1)}(-\bar{\tau};-t+i\pi),J_{s,-p}^{(2)}(-\bar{\tau};-t)]}\label{eq:d4-1}
\end{align}
where $J_{s,p}^{*}(\bar{\tau};t)$ is the complex conjugate source
of $J_{s,p}(\bar{\tau};t)$. Applying (\ref{eq:d4-1}) twice maps
back to original $W[J_{s,p}^{(1)}(\bar{\tau};t),J_{s,p}^{(2)}(\bar{\tau};t)]$.

Define the notation
\begin{align}
\tilde{J}_{s,p}^{(1)}(\bar{\tau};t) & =se^{2\pi ip/3}J_{s,-p}^{(1)}(-\bar{\tau};-t+i\pi),\quad\tilde{J}_{s,p}^{(2)}(\bar{\tau};t)=J_{s,-p}^{(2)}(-\bar{\tau};-t)\label{d13-0}\\
\tilde{\eta}_{s,p}^{(1)}(\bar{\tau};t) & =se^{2\pi ip/3}\eta_{s,-p}^{(1)}(-\bar{\tau};-t+i\pi),\quad\tilde{\eta}_{s,p}^{(2)}(\bar{\tau};t)=\eta_{s,-p}^{(2)}(-\bar{\tau};-t)\label{eq:d13}
\end{align}
and (\ref{eq:d4-1}) can be written as
\begin{equation}
W[J_{s,p}^{(1)},J_{s,p}^{(2)}]=W[\tilde{J}_{s,p}^{(1)},\tilde{J}_{s,p}^{(2)}]\label{eq:dKMS}
\end{equation}
In terms of $a$-$r$ fields, we can rewrite (\ref{eq:d13}) as 
\begin{align}
\tilde{\eta}_{s,p}^{a}(\bar{\tau};t) & =\sD_{+}^{p}\eta_{s,-p}^{a}(-\bar{\tau};-t)-2\sD_{-}^{p}\eta_{s,-p}^{r}(-\bar{\tau};-t)\label{eq:d17}\\
\tilde{\eta}_{s,p}^{r}(\bar{\tau};t) & =\sD_{+}^{p}\eta_{s,-p}^{r}(-\bar{\tau};-t)-\f 12\sD_{-}^{p}\eta_{s,-p}^{a}(-\bar{\tau};-t)\label{eq:d18-1}
\end{align}
where we defined two operators
\begin{equation}
\sD_{\pm}^{p}=\f 12(1\pm se^{2\pi ip/3}e^{-i\pi\del_{t}})
\end{equation}
Taking (\ref{eq:dKMS}) into (\ref{eq:d13-1}), we can derive the
dynamical KMS condition
\begin{equation}
I[J_{s,p}^{(1)},\eta_{s,p}^{(1)};J_{s,p}^{(2)},\eta_{s,p}^{(2)}]=I[\tilde{J}_{s,p}^{(1)},\tilde{\eta}_{s,p}^{(1)};\tilde{J}_{s,p}^{(2)},\tilde{\eta}_{s,p}^{(2)}]
\end{equation}
Turning off the sources, the effective
action for $\eta_{s,p}^{a,r}$ in Keldysh formalism needs to obey
\begin{equation}
S[\eta_{s,p}^{r},\eta_{s,p}^{a}]=S[\tilde{\eta}_{s,p}^{r},\tilde{\eta}_{s,p}^{a}]\label{eq:d18}
\end{equation}
For $\bar W$, the time reversal symmetry just changes the factor $e^{2\pi i p/3}$ to $e^{-2\pi ip/3}$ in \eqref{eq:d4-1}. This leads to a slightly different dynamical KMS condition
\be  
\bar I[J_{s,p}^{(1)},\eta_{s,p}^{(1)};J_{s,p}^{(2)},\eta_{s,p}^{(2)}]=\bar I[\bar{J}_{s,p}^{(1)},\bar{\eta}_{s,p}^{(1)};\bar{J}_{s,p}^{(2)},\bar{\eta}_{s,p}^{(2)}]
\ee 
where $\bar J_{s,p}$ and $\bar \eta_{s,p}$ are the same as \eqref{d13-0} and \eqref{eq:d13} except replacing $e^{2\pi i p/3}$ with $e^{-2\pi ip/3}$. Turning off the sources, the effective action obeys 
\be  
\bar S[\eta_{s,p}^{r},\eta_{s,p}^{a}]=\bar S[\bar{\eta}_{s,p}^{r},\bar{\eta}_{s,p}^{a}] \label{b.22x}
\ee
where 
\begin{align}
\bar{\eta}_{s,p}^{a}(\bar{\tau};t) & =\sD_{+}^{-p}\eta_{s,-p}^{a}(-\bar{\tau};-t)-2\sD_{-}^{-p}\eta_{s,-p}^{r}(-\bar{\tau};-t)\label{eq:d17bar}\\
\bar{\eta}_{s,p}^{r}(\bar{\tau};t) & =\sD_{+}^{-p}\eta_{s,-p}^{r}(-\bar{\tau};-t)-\f 12\sD_{-}^{-p}\eta_{s,-p}^{a}(-\bar{\tau};-t)\label{eq:d18-1bar}
\end{align}

\item The imaginary part of both $I$ and $\bar I$ needs to be nonnegative to guarantee convergent of path integral. Turning off the sources, the effective actions should obey
\be  
\Im S[\eta_{s,p}^{r},\eta_{s,p}^{a}] \geq 0, \quad \Im \bar S[\eta_{s,p}^{r},\eta_{s,p}^{a}] \geq 0 \label{b.25x}
\ee  

\end{enumerate}

These four unitary and KMS conditions need to be consistent with each other. First, we should check if the dynamical KMS conditions \eqref{eq:d18} and \eqref{b.22x} are consistent with \eqref{b.11x}. Using  \eqref{b.22x} and \eqref{b.11x}, we can write down a new KMS-type symmetry for $S$
\be  
S[\eta^r_{s,p},-\eta^a_{s,p}]=-\bar S[\eta_{s,p}^{r},\eta_{s,p}^{a}]^*=-\bar S[\bar\eta_{s,p}^{r},\bar\eta_{s,p}^{a}]^*= S[\bar\eta_{s,p}^{r},-\bar\eta_{s,p}^{a}]
\ee  
Combine with \eqref{eq:d18}, we have one more condition
\be  
S[\eta^r_{s,p},\eta^a_{s,p}]=S[se^{2\pi ip/3}e^{-i\pi\del_t}\eta^r_{s,p},se^{2\pi ip/3}e^{-i\pi\del_t}\eta^a_{s,p}]
\ee  
which holds for any local, time $t$ translational invariant and (both $\hat S$ and $Q$) charge conserved action $S$. In other words, each term in the action should be in the form of
\be  
\int K^{\a_1\cdots \a_k}_{s_1,p_1,\cdots,s_k,p_k}(\del_{\bar\tau},\del_t)\eta^{\a_1}_{s_1,p_1}(\bar\tau,t)\cdots \eta^{\a_k}_{s_k,p_k}(\bar\tau,t), \quad (\a_1,\cdots\a_k\in\{a,r\}) \label{b.28x}
\ee  
with $\prod_{i=1}^k s_i=1$ and $\sum_{i=1}^k p_i=3\Z$. 

The second case we need to check is the consistency between \eqref{b.11x} and \eqref{b.25x}. It is easy to see that they together lead to another inequality to $S$
\be  
\Im S[\eta_{s,p}^{r},-\eta_{s,p}^{a}]=-\Im \bar S[\eta^r_{s,p},\eta^a_{s,p}]^*=\Im \bar S[\eta^r_{s,p},\eta^a_{s,p}]\geq 0
\ee
Given each term in the effective action in the form of \eqref{b.28x}, this inequality together with \eqref{b.25x} leads to further constraint to the terms with odd numbers of $\eta^a_{s,p}$, namely 
\be  
K^{\a_1\cdots \a_k}_{s_1,p_1,\cdots,s_k,p_k}(\del_{\bar\tau},\del_t)=\left(K^{\a_1\cdots \a_k}_{s_1,-p_1,\cdots,s_k,-p_k}(\del_{\bar\tau},\del_t)\right)^* \label{b.30x}
\ee 
for $\a_1,\cdots,\a_k$ contain odd numbers of $a$.

Let us focus on the quadratic action, it is given by \eqref{eq:2.53}, which is copied
here
\begin{equation}
S[\eta_{s,p}^{r},\eta_{s,p}^{a}]=\sum_{s,p}\int_{0}^{\pi}d\bar{\tau}\int_{-\infty}^{\infty}dt\left[\eta_{s,-p}^{a}K_{s,p}^{ar}(\del_{\bar{\tau}},\del_{t})\eta_{s,p}^{r}+\f 12\eta_{s,-p}^{a}K_{s,p}^{aa}(\del_{\bar{\tau}},\del_{t})\eta_{s,p}^{a}\right]\label{eq:d20}
\end{equation}
where no pure $\eta^r$ term due to \eqref{b.8x}. By definition, we have 
\begin{equation}
K_{s,p}^{aa}(\del_{\bar{\tau}},\del_{t})=K_{s,-p}^{aa}(-\del_{\bar{\tau}},-\del_{t})
\end{equation}
By \eqref{b.30x} and \eqref{b.25x}, we have
\be  
K^{ar}_{s,p}(\del_{\bar\tau},\del_t)=K^{ar}_{s,-p}(\del_{\bar\tau},\del_t)^*,\quad \Im \sum_{s,p}\eta_{s,-p}^{a}K_{s,p}^{aa}(\del_{\bar{\tau}},\del_{t})\eta_{s,p}^{a}\geq 0
\ee  
To solve the dynamical KMS condition, taking the transformation (\ref{eq:d17}) and (\ref{eq:d18-1}) into
RHS of (\ref{eq:d18}), we will generally have all types $aa$, $ar$
and $rr$ terms. Requiring these three terms all match with LHS leads
to a single condition
\begin{equation}
K_{s,p}^{ar}(\del_{\bar{\tau}},\del_{t})-K_{s,-p}^{ar}(-\del_{\bar{\tau}},-\del_{t})=-2is\left(\tan\pi(p/3-\del_{t}/2)\right)^{s}K_{s,p}^{aa}(\del_{\bar{\tau}},\del_{t})
\end{equation}
In non-dissipation case, we have $K^{aa}=0$, which implies 
\begin{equation}
K_{s,p}^{ar}(\del_{\bar{\tau}},\del_{t})=K_{s,-p}^{ar}(-\del_{\bar{\tau}},-\del_{t})=K_{s,p}(-\del_{\bar\tau},-\del_t)^*
\end{equation}
Taking this back to (\ref{eq:d20}), one can easily see that $S[\eta_{s,p}^{r},\eta_{s,p}^{a}]$
factorizes as
\begin{equation}
S[\eta_{s,p}^{r},\eta_{s,p}^{a}]=S_{f}[\eta_{s,p}^{(1)}]-S_{f}[\eta_{s,p}^{(2)}]
\end{equation}
where 
\begin{equation}
S_{f}[\eta_{s,p}]=\f 12\sum_{s,p}\int_{0}^{\pi}d\bar{\tau}\int_{-\infty}^{\infty}dt\eta_{s,-p}K_{s,p}^{ar}(\del_{\bar{\tau}},\del_{t})\eta_{s,p}
\end{equation}
is a real action.

\section{Generalization to polynomial-exponential case} \label{app:b}

The differential operator $K_{s,p}(i\del_t)$  can be generalized to contain higher powers of $\del_t^2-\lam^2$, which would lead to polynomial-exponential behaviors $t^ke^{\pm\lambda t}$ in correlation functions. However, it turns out that most of these polynomial-exponential terms are excluded by the self-consistency condition \eqref{mcond}: there are two cases $\Im \bar t_W<\Im \bar t_V$ and $\Im \bar t_W>\Im \bar t_V$ in the TOC $G_4$, which must smoothly match each other at $\Im \bar t_W=\Im \bar t_V$.


If we only have pure exponential terms, it is automatically compatible with this condition because $G_4$ does not grow exponentially in both cases due to $G^{rr}_+=0$ (up to non-exponential terms) as explained in Section \ref{sec:2.4}. However, as we show in this section that this matching condition of $G_4$ becomes nontrivial when we include polynomial-exponential terms. It turns out that only pure exponential and linear-exponential terms are allowed in correlation functions, which implies that $K_{s,p}(i\del_t)$ can at most contain quadratic $\del_t^2-\lam^2$. Furthermore, we will solve the most general Wightman functions of effective modes that obey all three constraints \eqref{sm}, \eqref{mcond} and \eqref{Grr=0}.

\subsection{Wightman functions} \label{wightmanf}

It is convenient to work directly with Wightman functions. Let us assume for $p=0,\pm$ that
\begin{equation}
G_{s,p}^{>}(\bar{\tau};t)=\begin{cases}
h_{s,p}(t) & \bar{\tau}\in[0,\pi]\\
e^{- 2\pi i p/3}h_{s,p}(t) & \bar{\tau}\in[-\pi,0]
\end{cases}\label{eq:nx26}
\end{equation}
where each $h_{s,p}(t)$ contains polynomial-exponential pieces
\be
h_{s,p}(t)=\sum_{k=0}^n \g^k_{s,p}t^k e^{\lam t}+s \g^k_{s,-p} (-t-i\pi)^k e^{-\lam(t+i\pi)}=sh_{s,-p}(-t-i\pi) \label{hspB2}
\ee
where $n$ is finite number and the coefficient choice manifests KMS condition \eqref{eq:x17}. It follows that
\begin{align}
G_{s,p}^{ra}(\bar{\tau};t)=\t(t)\begin{cases}
\D_{s,p}(t) & \bar{\tau}\in[0,\pi]\\
e^{- 2\pi i p/3}\D_{s,p}(t) & \bar{\tau}\in[-\pi,0]
\end{cases}\label{eq:nx21}
\end{align}
with 
\be
\D_{s,p}(t)=h_{s,p}(t)-e^{2\pi i p/3} h_{s,-p}(-t) \label{Dhsprel}
\ee
The smoothness condition \eqref{sm} can be written in this case as
\be 
\D_{+,\pm}(t)-\D_{-,\pm}(t)=-e^{\pm\pi i /3}(\D_{+,0}(t)-\D_{-,0}(t) ) \label{Drelation}
\ee

Let us explicitly implement the requirement that $G_4$ is smoothly defined at $\Im \bar t_W=\Im \bar t_V$. For $\Im \bar t_W<\Im \bar t_V$, $G_4$ is given by \eqref{eq:g4}. For $\Im \bar t_W>\Im \bar t_V$, $G_4$ is given by flipping the sign of the first argument of $G_{s,p}^>$, namely
\begin{align}
G_{4}-g_{W}g_{V}=&\f 1 2  \sum_{s,p}L_{t_{1}}\tilde{L}_{t_{3}}\left[g_{W}g_{V}G_{s,p}^>(-;t_{31})\right]+e^{2\pi ip/3}L_{t_{2}}\tilde{L}_{t_{3}}\left[g_{W}g_{V}G_{s,p}^>(-;t_{32})\right]\nonumber \\
 & +e^{2\pi ip/3}L_{t_{1}}\tilde{L}_{t_{4}}\left[g_{W}g_{V}G_{s,p}^>(+;t_{14})\right]+L_{t_{2}}\tilde{L}_{t_{4}}\left[g_{W}g_{V}G_{s,p}^>(+;t_{24})\right] \label{eq:g4>}
\end{align}

The reason that $G^{rr}_+=0$ leads to $G_4=0$ in \eqref{eq:g4} is that the shift symmetry of vertex \eqref{eq:x44} transforms the four pure exponential terms in \eqref{eq:g4} to just one exponential function of $t_2$ and $t_4$ in \eqref{eq:g4-2.87}. However, for polynomial-exponential terms, we do not have an extended symmetry transforming, say, $t_1^k e^{\lambda t_1}$ to $t_2^k e^{\lambda t_2}$.  Therefore, to make sure \eqref{eq:g4} matches with \eqref{eq:g4>} at $\Im \bar t_W=\Im \bar t_V$ for the quadratic and higher polynomial-exponential pieces, we must require the four terms in both \eqref{eq:g4} and \eqref{eq:g4>} match with each other separately 
\be
\sum_{s,p} h_{s,p}(t)\simeq \sum_{s,p} e^{-2\pi i p/3}h_{s,p}(t)\simeq \sum_{s,p} e^{2\pi i p/3}h_{s,p}(t)\label{2.111}
\ee
where  ``$\simeq$" means that the equation hold for $t^k e^{\pm\lambda t}$ terms with $k>1$. The $k=1$ case is a little bit different and will be discussed later. It follows that
\be 
\sum_s h_{s,p}(t)\simeq 0,\quad p=\pm \label{b4}
\ee 
where there is no constraint to $p=0$ piece simply because $G^>_{s,0}(t)$ is independent on $\bar \tau$. Using ansatz \eqref{hspB2}, we can easily show order by order from \eqref{b4} for $p=\pm$ that 
\be 
\g^k_{s,\pm}=0,\quad k=2,\cdots,n
\ee
Then by $G_+^{rr}=0$  \eqref{G+rr}, we have 
\be
h_{+,0}(t)+h_{+,0}(-t)\simeq 0\implies \g_{+,0}^k=0,\quad k=2,\cdots,n \label{h+0}
\ee
On the other hand, using \eqref{eq:nx26}, \eqref{Dhsprel}, \eqref{Drelation} and \eqref{h+0} that
\be 
h_{-,0}(t)-h_{-,0}(-t)\simeq 0\implies \g_{-,0}^k=0,\quad k=2,\cdots,n \label{b5}
\ee
In this way, the consistency conditions and shift symmetry kill all quadratic and higher polynomial-exponential terms.

For the linear-exponential term, the shift symmetry of vertex \eqref{eq:x44} does help a bit because, for example,
\be  
L_{t_1}[g_W (t_1-t_2)e^{\pm \lambda(t_1+i\pi)}]=L_{t_1}[g_W t_1 e^{\pm \lambda(t_1+i\pi)}]-t_2 L_{t_2}[g_W e^{\pm \lambda t_2}]
\ee
However, this is still not enough to save nontrivial linear-exponential term in $h_{s,\pm}(t)$. After some algebra, matching the linear-exponential pieces in \eqref{eq:g4} and \eqref{eq:g4>} leads to
\be 
\g_{+,+}^1=\g_{+,-}^1=0,\quad \g_{-,+}^1=\f{\cos(\pi(\lam/2+1/3))}{\cos(\pi(\lam/2-1/3))}\g_{-,-}^1
\ee
Taking this into \eqref{G+rr} leads to $\g_{+,0}^1=0$. Using \eqref{Drelation}, we can solve
\be 
\g_{-,\pm}^1=-\f{\cos(\pi\lam/2)}{\cos\pi(\lam/2\mp1/3)}\g_{-,0}^1
\ee
In particular, we should have $\g^1_{-,0}=\g^1_{-,+}=0$ and $\g^1_{-,-}\neq 0$ when $\lam=1/3$. Taking these solutions to $G_4$, one can see that no linear-exponential term survives.

With the existence of linear-exponential growth term, the constraint to the pure exponential term is slightly different. Comparing the pure exponential piece in \eqref{hspB2} and \eqref{exphsp}, we should identify
\be 
A_{s,p}=\g^0_{s,p},\quad B_{s,p}=s(\g_{s,-p}^0 -i\pi\g_{s,-p}^1)e^{-i\lam \pi}
\ee
Taking this into \eqref{G+rr} and \eqref{Drelation} leads to
\begin{align}
\g^0_{+,\pm}&=-\f{\cos{(\pi\lam/2)}}{\cos\pi(\lam/2\mp 1/3)}\g_{+,0}^0\\
\g_{-,\pm}^0&=-\f{\cos{(\pi\lam/2)}}{\cos\pi(\lam/2\mp 1/3)}\g^0_{-,0}\pm \f{\sqrt{3}(2i\g^0_{+,0}+\pi\g^1_{-,0})}{4\cos^2\pi(\lam/2\mp 1/3)}
\end{align}
Taking these solutions into \eqref{C1} and \eqref{C2}, we will find both $C_1$ and $C_2$ vanish. One can also check \eqref{eq:g4>} vanishes as well. This confirms the result in Section \ref{sec:2.4} that $G_+^{rr}=0$ and smoothness condition \eqref{Drelation} imply vanishing pure exponential terms in TOC for both ordering of $\Im \bar t_{W,V}$.

In conclusion, the consistency conditions require $h_{+,p}(t)$ only contain pure exponential terms and $h_{-,p}(t)$ only contain up to linear-exponential terms. With the expression \eqref{hspB2}, there are only three independent parameters $\g_{+,0}^0$, $\g_{-,0}^0$ and $\g_{-,0}^1$. Other parameters are determined by the following relation
\begin{align}
\g_{+,p}^0&=(-)^p \f{\cos{(\pi\lam/2)}}{\cos\pi(\lam/2-p/3)}\g_{+,0}^0 \label{grl1} \\
\g_{-,p}^0&=(-)^p \f{\cos{(\pi\lam/2)}}{\cos\pi(\lam/2-p/3)}\g^0_{-,0}+p \f{\sqrt{3}(2i\g^0_{+,0}+\pi\g^1_{-,0})}{4\cos^2\pi(\lam/2-p/3)}\\
\g_{-,p}^1&=(-)^p \f{\cos(\pi\lam/2)}{\cos\pi(\lam/2-p/3)}\g_{-,0}^1 \label{grl3}
\end{align}

\subsection{The effective action} \label{app:genact}

The above most general consistent Wightman functions correspond to the effective actions in the form of \eqref{leadS} with 
\be 
K_{+,p}(i\del_t)=(\del_t^2-\lam^2)k_{+,p}(i\del_t),\quad K_{-,p}(i\del_t)=(\del_t^2-\lam^2)^2 k_{-,p}(i\del_t) \label{generalact}
\ee
where one should note that $K_{-,p}(x)$ has a double zero at $\pm i\lam$. By symmetry \eqref{Kp}, we have $k_{s,p}(x)=(-)^p k_{s,-p}(-x)$. Since $K_{+,p}$ is linear in $\del_t^2-\lam^2$, the computation \eqref{G+rr} for $G_+^{rr}=0$ up to non-exponential terms still hold and leads to the same condition \eqref{eq:x49}:
\be  
k_{+,\pm}(i\lam)=\mp \f{ik_{+,0}(i\lam)}{\sqrt{3}}\tan\left(\f{\pi}{2}(\lam\pm 1/3)\right)\tan \f {\pi \lam}{2} \label{c22}
\ee 
The smoothness condition \eqref{eq:2.76xx} at $\w=i\lam$ leads to
\begin{align}  
k_{-,\pm}(i\lam)&=\pm \f{i k_{-,0}(i\lam)}{\sqrt{3}} \\
k'_{-,\pm}(i\lam)&=\pm \f{\sqrt{3}}{3}\left[ \f{2\lam k_{-,0}(i\lam)^2}{k_{+,0}(i\lam)}\left(1+\f 1 {\tan\left(\f{\pi}{2}(\lam\pm 1/3)\right)\tan \f {\pi \lam}{2}}   \right)+i k'_{-,0}(i\lam) \right] \label{c24}
\end{align}
which allows three undetermined parameters $k_{+,0}(i\lam)$, $k_{-,0}(i\lam)$ and $k'_{-,0}(i\lam)$. From \eqref{2.74x-1} and \eqref{2.74x}, we have
\begin{align}
\D_{+,0}(t)=&\f{i}{2\lam k_{+,0}(i\lam)}(e^{\lam t}-e^{-\lam t}) \label{c25x}\\
\D_{+,\pm}(t)=&\pm \f {e^{\pm i \pi/3}}{2\sqrt{3}\lambda}\left(\f{e^{\lambda t}}{k_{+,\pm}(i\lambda)}+\f{e^{-\lambda t}}{k_{+,\mp}(i\lambda)}\right) \\
\D_{-,0}(t)=&\f{\lam k_{-,0}'(i\lam)-ik_{-,0}(i\lam)}{4\lam^3 k_{-,0}(i\lam)^2}(e^{\lam t}-e^{-\lam t})+\f{i}{4\lam^2 k_{-,0}(i\lam)}t(e^{\lam t}+e^{-\lam t}) \\
\D_{-,\pm}(t)=&\pm \f {e^{\pm i \pi/3}}{4\sqrt{3}\lambda^2}\left(\f{\lam k_{-,\pm}'(i\lam)-ik_{-,\pm}(i\lam)}{i\lam k_{-,\pm}(i\lam)^2}e^{\lam t}+ \f{\lam k_{-,\mp}'(i\lam)-ik_{-,\mp}(i\lam)}{i\lam k_{-,\mp}(i\lam)^2}e^{-\lam t} \right. \nn\\
&\left.+t\left(\f{e^{\lambda t}}{k_{-,\pm}(i\lambda)}-\f{e^{-\lambda t}}{k_{-,\mp}(i\lambda)}\right)\right) \label{c28x}
\end{align}
where we see linear-exponential terms appear because of the double poles in $1/K_{-,p}(\w)$ at $\w=\pm i\lam$.

Take ansatz \eqref{hspB2} with $n=0$ for $s=+$ and $n=1$ for $s=-$. Using \eqref{Dhsprel} and comparing with \eqref{c25x}-\eqref{c28x}, we find 
\begin{align} 
\g^0_{+,0}&=\f 1 {2i\lam (e^{-i\pi \lam}-1)k_{+,0}(i\lam)} \label{c25}\\
\g^1_{-,0}&=\f i {4\lam^2 (e^{-i\pi \lam}+1)k_{-,0}(i\lam)}\\
\g^0_{-,0}&=\frac{\left(1+e^{i \pi  \lambda }\right) \lambda  k'_{-,0}(i\lam)-i \left(1+e^{i \pi  \lambda }-i \pi  \lambda \right) k_{-,0}(i\lam)}{16 \lambda ^3  \cos ^2\frac{\pi  \lambda }{2} k_{-,0}(i\lam)^2} \label{c31}
\end{align} 
and other $\g^k_{s,p}$ are given by \eqref{grl1}-\eqref{grl3}. By the conclusion from last subsection, this implies that the action with \eqref{generalact} and three constraints \eqref{c22}-\eqref{c24} lead to absence of exponential terms in TOC.

Using \eqref{c25x}-\eqref{c28x} and the first line of \eqref{delta2.91}, we find that
\be
\D(t)=\f{3}{4\lam (1/2-\cos\pi\lam)\sin \f{\pi\lam}{2}k_{+,0}(i\lam)}(e^{i\pi\lambda/2}e^{\lambda t}+e^{-i\pi\lam/2}e^{-\lambda t}) \label{Dtgen}
\ee
which exactly matches with \eqref{delta2.91}. This means that for any four-point functions, $k_{-,0}(i\lam)$ and $k'_{-,0}(i\lam)$ are irrelevant parameters in the action. What we discussed in Section \ref{sec:2.4} with pure exponential case is indeed the most general situation for four-point functions.

\section{Correlation functions of effective modes in the large $q$ SYK model}\label{app:d}

In this appendix, we first ignore the prefector $N/(8q^2)$ in the action \eqref{eq:10}. In the end, this prefactor simply adds a $8/(N\D^2)$ factor to any two-point function.

\subsection{Canonical quantization trick}

As explained in Section \ref{sec:quantimode}, we will take the mathematical trick of canonical quantization to solve the Euclidean two-point function of $\e(\bar\tau,x)$. First, we need to solve the equations of motion $\mL g_{m}(\bar\tau,x)=0$ for wave function $g_m(\bar\tau,x)$ with UV condition $g_m(\bar\tau,0)=g_m(\bar\tau,2\pi)=0$ from \eqref{eq:4.11} and expand the quantized field $\hat\e(\bar\tau,x)$ in terms of
\begin{equation}
\hat\e(\bar\tau,x)=\sum_m g_{m}\hat a_{m}+g_{m}^{*}\hat a_{m}^{\dagger} \label{ephat-app}
\end{equation}
where $\hat a_{m}$ and $\hat a_{m}^{\dagger}$ are annihilation and creation
operators obeying canonical commutation relation $[\hat a_{m},\hat a_{m'}^{\dagger}]=\d_{m,m'}$, and $g_{m}$ is well-normalized under Klein-Gordon
inner product, which is defined in \eqref{eq:16}. Note that the quantum number $m$ must be discrete because the spatial direction $x$ is finite.
Due to translation symmetry in $\bar\tau$, we can choose the positive energy wave function 
\be  
g_m(\bar\tau,x)\propto e^{-im\bar\tau}, \quad m\geq 0
\ee
for which the Hamiltonian $H$ has eigen value $m$ and
\be  
[H,\hat a^\dagger_m]=m \hat a^\dagger_m,\quad [H,\hat a_m]=-m \hat a_m
\ee 

Since $g_m(\bar\tau,x)$ is on-shell, following from \eqref{eq:43} it can be expanded as 
\be  
g_m(\bar\tau,x)=\f 1{\D G_0(\tau_{12})} \sum_{i=1,2} L_{\tau_i}G_0(\tau_{12}) \chi_{i,m}(\bar\tau;\tau_i) \label{4.37-app}
\ee 
where $\bar\tau$ in $\chi_{i,m}(\bar\tau;\tau_i)$ is just a dummy argument whose dependence is trivial. It follows that we can rewrite \eqref{ephat-app} as
\begin{align}
\hat \e (\bar\tau,x)&=\f 1{\D G_0(x)} \sum_{i=1,2} L_{\tau_i}G_0(x) \hat \chi_i(\bar\tau;\tau_i) \label{4.38n-app}\\
\hat\chi_i(\bar\tau;\tau_i)&\equiv \sum_m\chi_{i,m}(\bar\tau;\tau_i)\hat a_m+\chi_{i,m}^*(\bar\tau;\tau_i)\hat a_m^\dag \label{4.34x-app}
\end{align}
where $L_\tau$ is defined in \eqref{x4.47}. By the fundamental domain $\mD_\e$, the defining domain for $\hat\chi_1(\bar\tau;\tau)$ is $\{\bar\tau\in[0,\pi], \tau-\bar\tau\in[0,\pi]\}$ and that for  $\hat\chi_2(\bar\tau;\tau)$ is $\{\bar\tau\in[0,\pi], \tau-\bar\tau\in[-\pi,0]\}$.

Note that the path integral is defined on the finite spacetime $\mD_\e$. To compute any correlation function of quantized field $\hat \e$ on $\mD_\e$, we must first specify the states on time slice $\bar\tau=0,\pi$ respectively. This is dictated by the boundary condition of $\e(\bar\tau,x)$ in the Euclidean path integral on $\mD_\e$ at $\bar\tau=0,\pi$. From the first equation of \eqref{eq:4.11}, we see that any configurations of $\e(0,x)$ is identified with $\e(\pi,2\pi-x)$. For our canonical quantization trick, this implies that we need to trace over all states at $\bar\tau=0,\pi$ with a reflection $x\ra 2\pi-x$. More explicitly, we consider the following Wightman function
\begin{equation}
W(\bar\tau,x;\bar\tau',x')\equiv\f 1 Z\Tr \left[P e^{-i\pi H}\hat \e(\bar\tau,x)\hat \e(\bar\tau',x')\right],\quad Z\equiv \Tr\left[P e^{-i\pi H}\right]
\end{equation}
where the time evolution for $\bar\tau=\pi$ is present and $P$ is the reflection operator
\be  
P \hat \e (\bar\tau,x)P^{-1}=\hat \e(\bar\tau,2\pi-x) \label{4.28xxx}
\ee 
Therefore, we will define the expectation $\avg{\cdots}_{\rm c.q.}$ as 
\be 
\avg{\cdots}_{\rm c.q.}\equiv\f 1 Z \Tr\left[P e^{-i\pi H}\cdots\right] \label{dfnavgL}
\ee 
Expanding in $\hat a_{m}$ and $\hat a_{m}^{\dagger}$, we have
\begin{align}
W(\bar\tau,x;\bar\tau',x')=&\sum_m\f 1{N_{m}}\left(g_{m}(\bar\tau,x)g_{m}^{*}(\bar\tau',x')\avg{\hat a_{m}\hat a_{m}^{\dagger}}_{\rm c.q.}\right.\nn\\
&\left.+g_{m}^{*}(\bar\tau,x)g_{m}(\bar\tau',x')\avg{\hat a_{m}^{\dagger}\hat a_{m}}_{\rm c.q.}\right)\label{eq:18}
\end{align}
Since the Euclidean two-point function $\avg{\e(\bar\tau,x)\e(\bar\tau',x')}$ is symmetric under exchange of $\bar\tau,x$ with $\bar\tau',x'$, this corresponds to the Feynman propagator of $\hat \e(\bar\tau,x)$, i.e.
\begin{align}
\avg{\e(\bar\tau,x)\e(\bar\tau',x')}&=-i\avg{\bar\mT \hat \e(\bar\tau,x)\hat \e(\bar\tau',x')}_{\rm c.q.} \label{e2-pt}\\
&=-i[\t(\bar{\tau}-\bar{\tau}')W(\bar\tau,x;\bar\tau',x')+\t(\bar{\tau}'-\bar{\tau})W(\bar\tau',x';\bar\tau,x)]\label{eq:19}
\end{align}
where $\bar \mT$ is the time ordering of $\bar\tau$ and $\bar\tau-\bar\tau'$ is restricted to $[-\pi,\pi]$. We emphasize that the LHS of \eqref{e2-pt} is a Euclidean two-point function and the RHS of \eqref{e2-pt} is just a mathematical trick to compute it in terms of a Feynman propagator of quantized field $\hat\e$ by regarding $\bar\tau$ as ``time". The quantized field $\hat \e$ and the corresponding Hilbert space are part of the trick and have no physical meaning.

On the domain $\mD_\e$, taking above equations into \eqref{n4.7} leads to
\begin{align}  
\mF_\psi(\tau_1,&\tau_2,\tau_3,\tau_4)=G_0(\tau_{12})G_0(\tau_{34}) \nn\\
&+\sum_{i,j=1,2}L_{\tau_i}L_{\tau_{j+2}}\left[G_0(\tau_{12})G_0(\tau_{34})\left(-i\avg{\bar\mT \hat \chi_i(\bar\tau_a;\tau_i)\hat\chi_j(\bar\tau_b;\tau_{j+2})}_{\rm c.q.}\right)\right] \label{F4.36}
\end{align}
where one should note that the dummy argument $\bar\tau$ plays a role in the time ordering $\bar\mT$ though its explicit dependence in \eqref{4.37-app} is trivial. Given \eqref{4.38n-app}, it is noteworthy that the reflection \eqref{4.28xxx} acts nontrivially on $\hat\chi_i$. The transformation $(\bar\tau,x)\ra (\bar\tau,2\pi-x)$ is equivalent to $(\tau_1,\tau_2)\ra (\tau_2+\pi,\tau_1-\pi)$, which implies
 \be  
P \hat \chi_1(\bar\tau,\tau)P^{-1}=\hat \chi_2(\bar\tau,\tau-\pi),\quad P \hat \chi_2(\bar\tau,\tau)P^{-1}=\hat \chi_1(\bar\tau,\tau+\pi)
 \ee 
Taking this back to \eqref{F4.36} and using definition \eqref{dfnavgL}, we have the following KMS conditions
\begin{align}  
\avg{\bar\mT \hat \chi_1(0;\tau)\hat\chi_j(\bar\tau';\tau')}_{\rm c.q.}&=\avg{\bar\mT \hat \chi_2(\pi;\tau)\hat\chi_j(\bar\tau';\tau')}_{\rm c.q.} \label{kmschi1}\\
\avg{\bar\mT \hat \chi_2(0;-\tau)\hat\chi_j(\bar\tau';\tau')}_{\rm c.q.}&=\avg{\bar\mT \hat \chi_1(\pi;2\pi-\tau)\hat\chi_j(\bar\tau';\tau')}_{\rm c.q.} \label{kmschi2}
\end{align} 
where $\tau\in[0,\pi]$ by the defining domain of $\hat\chi_i$.

It is very interesting that the Euclidean four-point function \eqref{F4.36} now has the same structure as \eqref{n2.16}. Therefore, we would like to define the Euclidean two-point function of two effective fields $\phi^E_{1,2}(\bar\tau;\tau)$ such that
\be  
\avg{\phi^E_i(\bar\tau;\tau)\phi^E_j(\bar\tau';\tau')}\equiv-i\avg{\bar\mT \hat \chi_i(\bar\tau;\tau)\hat\chi_j(\bar\tau';\tau')}_{\rm c.q.} \label{eq:4.41df}
\ee 
Take this definition into \eqref{F4.36} and analytically continue $\bar\tau_k\ra i t_k$. Comparing with \eqref{n2.16}, we will identify the Euclidean fields $\phi^E_i$ as $\phi_i$ after analytic continuation. In particular, the two KMS conditions \eqref{kmschi1} and \eqref{kmschi2} in terms of $\phi^E_i$ are equivalent to the KMS conditions of $\phi_i$ \eqref{pkms1}-\eqref{pkms2}.

\subsection{Solve discrete quantum numbers\label{subsec:quantN}}

From the general solution \eqref{eq:43} to $\mL \e=0$, let us expand $\chi_i$ in Fourier modes
\begin{equation}
\chi_1(\tau)=\sum_m A_{m}e^{-im\tau},\quad \chi_2(\tau)=\sum_m B_{m}e^{-im\tau}
\end{equation}
The UV condition (\ref{eq:4.11}) for $\tau_{1}=\tau_{2}$ ($x=0$) leads to
\begin{equation}
-\lambda(A_{m}-B_{m})\tan\f{\lambda\pi}2-im(A_{m}+B_{m})=0
\end{equation}
and for $\tau_{1}=\tau_{2}+2\pi$ ($x=2\pi$)  leads to
\begin{equation}
\lambda(A_{m}e^{-2mi\pi}-B_{m})\tan\f{\lambda\pi}2-im(A_{m}e^{-2mi\pi}+B_{m})=0
\end{equation}
Combining these two equations, we have
\begin{equation}
\f{A_{m}}{B_{m}}=\f{\lambda\tan\f{\lambda\pi}2-im}{\lambda\tan\f{\lambda\pi}2+im}=\f{\lambda\tan\f{\lambda\pi}2+im}{\lambda\tan\f{\lambda\pi}2-im}e^{2mi\pi}\label{eq:186}
\end{equation}
which leads to
\begin{equation}
\lambda\tan\f{\lambda\pi}2+im=\pm e^{-\pi im}(\lambda\tan\f{\lambda\pi}2-im),\quad A_{m}=\pm B_{m} e^{im\pi }\label{eq:51}
\end{equation}
These two cases can be solved as two sets of positive quantum numbers $m$
\begin{align}
\mM^+=\{m>0|\lambda\tan\f{\lambda\pi}2+m\cot\f{m\pi}2 & =0\}\\
\mM^-=\{m>\lam|\lambda\tan\f{\lambda\pi}2-m\tan\f{m\pi}2 & =0\}
\end{align}
where $m=\lam$ is excluded in $\mM^-$ because it leads to trivial $\e$. This is exactly the shift symmetry \eqref{eq:44} of the $\e_{\rm on-shell}$ in \eqref{eq:43}. Note that here we only choose positive $m$ because it is indeed the positive frequency of the canonical quantization in \eqref{ephat}.

It follows that in \eqref{4.37-app} we can take 
\be  
\chi_{1,m}(\bar\tau;\tau)=A_m e^{-im \tau},\quad \chi_{2,m}(\bar\tau;\tau)=B_m e^{-im \tau},\quad m\in\mM^\pm
\ee 
For convenience, we will add a superscipt ``$\pm$"  to distinguish the two types of solutions with $m\in\mM^\pm$ respectively. Let us take
\begin{equation}
A_{m}^{+}=B_{m}^{+} e^{im\pi }=i\f{e^{im\pi/2}}{2\lambda\sqrt{N_m^+}},\quad A_{m}^{-}=-B_{m}^{-} e^{im\pi }=\f{e^{im\pi/2}}{2\lambda\sqrt{N_m^-}} \label{d10x}
\end{equation}
which leads to wave functions
\begin{align}
g_{m}^{+}(\bar\tau,x) & =\f{e^{-im\bar{\tau}}}{\sqrt{N_m^+}}\left[\f m{\lambda}\cos\f m2(\pi-x)+\sin\f m2(\pi-x)\tan\f{\lambda}2(\pi-x)\right]\\
g_{m}^{-}(\bar\tau,x) & =\f{e^{-im\bar{\tau}}}{\sqrt{N_m^-}}\left[\f m{\lambda}\sin\f m2(\pi-x)-\cos\f m2(\pi-x)\tan\f{\lambda}2(\pi-x)\right]
\end{align}
where $g^\pm_m$ is even/odd under $x\ra 2\pi-x$. Taking these solutions into Klein-Gordon inner product \eqref{eq:16} leads to 
\be
N^\pm_m =(m-\lambda)(m+\lambda)(m\pi\mp\sin m\pi)/\lambda^{2}
\ee
It follows from \eqref{4.34x-app} that
\begin{align}
\hat{\chi}_1(\bar\tau;\tau) & =\sum_{s=\pm}\sum_{m\in\mM^{s}}A_{m}^{s}e^{-im\tau}\hat a_{m}^{s}+A_{m}^{s*}e^{im\tau}\hat a_{m}^{s\dagger}\label{chi1hat}\\
\hat{\chi}_2(\bar\tau;\tau) & =\sum_{s=\pm}\sum_{m\in\mM^{s}}B_{m}^{s}e^{-im\tau}\hat a_{m}^{s}+B_{m}^{s*}e^{im\tau}\hat a_{m}^{s\dagger}\label{chi2hat}
\end{align}

\subsection{Correlations\label{subsec:Correlations-in-terms}}

To compute the correlation functions, we need to first evaluate the expectation value of $\hat a_{m}^{s}\hat a_{m}^{s\dagger}$. Taking \eqref{chi1hat} and \eqref{chi2hat} into the conditions \eqref{kmschi1} and \eqref{kmschi2} leads to
\begin{equation}
\avg{\hat a_{m}^{s}\hat a_{m}^{s\dagger}}_{\rm c.q.}=se^{im\pi}\avg{\hat a_{m}^{s\dagger}\hat a_{m}^{s}}_{\rm c.q.}
\end{equation}
By canonical quantization, we have $[\hat a_{m}^s,\hat a_{m}^{s\dagger}]=1$, which yields\footnote{We can also derive these equations by the definition of these operators in the Fock space generated by $\hat a_m^{s\dagger}$ if we add an infinitesimal negative imaginary part to all $m$.}
\be
\avg{\hat a_{m}^{s}\hat a_{m}^{s\dagger}}_{\rm c.q.} =\f{1}{1-se^{-im\pi}},\quad\avg{\hat a_{m}^{s\dagger}\hat a_{m}^{s}}_{\rm c.q.}=\f{se^{-im\pi}}{1-se^{-im\pi}}\label{eq:172}
\ee

In the following, we will compute the correlations by summing over Matsubara frequencies $m$. The method is the Sommerfeld-Watson transformation used in \cite{Choi:2019bmd} and we will only restrict ourselves to exponential pieces. 
Since the $\bar\tau$ dependence is only through time ordering, in the following we will suppress the argument $\bar\tau$ in $\hat\chi_i(\bar\tau;\tau)$ in the computation of Wightman functions with specified ordering of fields, e.g.
\begin{align}
\avg{\hat\chi_1(\tau_1)\hat\chi_1(\tau_2)}_{\rm c.q.}&\equiv\avg{\hat\chi_1(\bar\tau;\tau_1)\hat\chi_1(0;\tau_2)}_{\rm c.q.}~~~~(\bar\tau>0)\nn\\
& =\sum_{s=\pm}\sum_{m\in\mM^{s}}A_{m}^{s}A_{m}^{s*}\left[e^{-im\tau_{12}}\avg{a_{m}^{s}a_{m}^{s\dagger}}_{\rm c.q.}+e^{im\tau_{12}}\avg{a_{m}^{s\dagger}a_{m}^{s}}_{\rm c.q.}\right]\nonumber \\
 & =\f{1}4\sum_{s=\pm}\sum_{m\in\pm\mM^{s}}\f{e^{-im\tau_{12}}}{(1\mp e^{-im\pi})(m-\lambda)(m+\lambda)(m\pi\mp\sin\pi m)}\label{eq:60-1}
\end{align}
where in the second line, we extend the sum over $m$ to both positive
and negative $\mM^{\pm}$ due to the two terms in the first line.

\begin{figure}
\begin{centering}
\subfloat[]{\begin{centering}
\includegraphics[height=2.5cm]{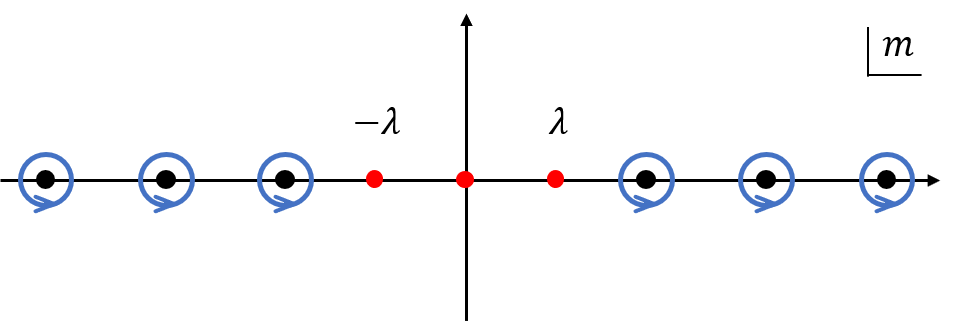} 
\par\end{centering}

}
\subfloat[]{\begin{centering}
\includegraphics[height=2.5cm]{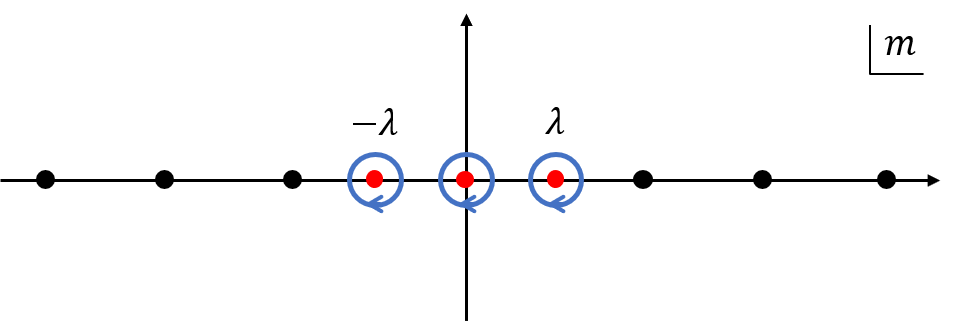} 
\par\end{centering}
}

\par\end{centering}
\caption{(a) The contour of $C_+$ that circles around all $\pm\mM^+$ (black dots) anticlockwise. The three poles (red dots) at $0,\pm\lam$ are not circled. (b) We can deform the contour $C_+$ through infinity to just three clockwise circles around $0,\pm \lam$. \label{fig:Ce}}

\end{figure}

Let us compute $s=\pm$ separately. For $s=+$, we insert 
\begin{equation}
\f 1{2\pi i}\f{(\pi m-\sin m\pi)/(\cos m\pi-1)}{m\cot m\pi/2+\lambda\tan\lambda\pi/2}\label{eq:61-1}
\end{equation}
into the sum and change the sum over $m$ to anticlockwise contour
integral along small circles of $m$ around all $\pm\mM^{+}$, which
we denoted as $C_{+}$ (see Fig. \ref{fig:Ce}). The denominator of (\ref{eq:61-1}) pickes
out the residue at these points and the numerator is the reciprocal
of residues. Our purpose is to deform the contour to small circles
around $0,\pm\lambda$. However, the extra $1/(\cos m\pi-1)$ gives
additional unwanted poles at even integers. To replace it with an
equivalent expression, we use (\ref{eq:51}) to write 
\begin{equation}
\f 1{\cos m\pi-1}=\f{(\lambda\tan\lambda\pi/2)^{2}+m^{2}}{-2m^{2}} \label{n4.67}
\end{equation}
at all $m\in\pm\mM^{+}$. Putting them all together, we have
\begin{equation}
\avg{\hat{\chi}_1(\tau_{1})\hat{\chi}_1(\tau_{2})}_{\rm c.q.}^+=-\f{1}4\ointctrclockwise_{C_{+}}\f{dm}{2\pi i}\f{e^{-im\tau_{12}}((\lambda\tan\lambda\pi/2)^{2}+m^{2})}{2m^{2}(1-e^{-im\pi})(m-\lambda)(m+\lambda)(m\cot m\pi/2+\lambda\tan\lambda\pi/2)}\label{eq:63}
\end{equation}
Note that though $1-e^{-im\pi}=0$ for even $m$, they are not poles
because $m\cot m\pi/2$ is also divergent for even $m$ and cancels it out. For $\tau_{12}\in(0,\pi)$,
the integral over infinite arc on upper and lower half planes are
both zero. Then the integral reduces to residues at $0,\pm\lambda$
\begin{equation}
\avg{\hat{\chi}_1(\tau_{1})\hat{\chi}_1(\tau_{2})}_{\rm c.q.}^+=\f{1}4\text{Res}{}_{0,\pm\lambda}\left[\f{e^{-im\tau_{12}}((\lambda\tan\lambda\pi/2)^{2}+m^{2})}{2m^{2}(1-e^{-im\pi})(m-\lambda)(m+\lambda)(m\cot m\pi/2+\lambda\tan\lambda\pi/2)}\right]
\end{equation}
If $\tau_{12}\in(-\pi,0)$, we need simply multiply both numerator
and denominator of (\ref{eq:63}) by $e^{-im\pi}$ and replace the
$e^{-im\pi}$ in denominator by $(\lambda\tan\lambda\pi/2+im)/(\lambda\tan\lambda\pi/2-im)$
due to (\ref{eq:51}). This leads to
\begin{align}
\avg{\hat{\chi}_1(\tau_{1})\hat{\chi}_1(\tau_{2})}_{\rm c.q.}^+ & =-\f{1}4\ointctrclockwise_{C_{+}}\f{dm}{2\pi i}\f{e^{-im(\tau_{12}+\pi)}(\lambda\tan\lambda\pi/2-im)^{2}}{2m^{2}(1-e^{-im\pi})(m-\lambda)(m+\lambda)(m\cot m\pi/2+\lambda\tan\lambda\pi/2)}\nonumber \\
 & =\f{1}4\text{Res}{}_{0,\pm\lambda}\left[\f{e^{-im(\tau_{12}+\pi)}(\lambda\tan\lambda\pi/2-im)^{2}}{2m^{2}(1-e^{-im\pi})(m-\lambda)(m+\lambda)(m\cot m\pi/2+\lambda\tan\lambda\pi/2)}\right]
\end{align}
If $\tau_{12}\in(\pi,2\pi)$, we will multiply  both numerator
and denominator of (\ref{eq:63}) by $e^{im\pi}$ and replace the
$e^{im\pi}$ in denominator by $(\lambda\tan\lambda\pi/2-im)/(\lambda\tan\lambda\pi/2+im)$
instead. This leads to
\begin{equation}
\avg{\hat{\chi}_1(\tau_{1})\hat{\chi}_1(\tau_{2})}_{\rm c.q.}^+=\f{1}4\text{Res}{}_{0,\pm\lambda}\left[\f{e^{-im(\tau_{12}-\pi)}(\lambda\tan\lambda\pi/2+im)^{2}}{2m^{2}(1-e^{-im\pi})(m-\lambda)(m+\lambda)(m\cot m\pi/2+\lambda\tan\lambda\pi/2)}\right]
\end{equation}

Similarly, for $s=-$, we need to consider the contour $C^-$ that circles around all points in $\mM^-$, and insert
\begin{equation}
\f 1{2\pi i}\f{(\pi m+\sin m\pi)/(\cos m\pi+1)}{m\tan m\pi/2-\lambda\tan\lambda\pi/2}\simeq\f 1{2\pi i}\f{(\pi m+\sin m\pi)((\lambda\tan\lambda\pi/2)^{2}+m^{2})/(2m^{2})}{m\tan m\pi/2-\lambda\tan\lambda\pi/2}
\end{equation}
where ``$\simeq$" means equal at the poles $\mM^-$ by \eqref{n4.67}. Following a similar computation, this leads to
\begin{equation}
\avg{\hat{\chi}_1(\tau_{1})\hat{\chi}_1(\tau_{2})}_{\rm c.q.}^-=-\f{1}4\text{Res}{}_{0,\pm\lambda}\left[\f{e^{-im\tau_{12}}((\lambda\tan\lambda\pi/2)^{2}+m^{2})}{2m^{2}(1+e^{-im\pi})(m-\lambda)(m+\lambda)(m\tan m\pi/2-\lambda\tan\lambda\pi/2)}\right]
\end{equation}
for $\tau_{12}\in(0,\pi)$, and
\begin{equation}
\avg{\hat{\chi}_1(\tau_{1})\hat{\chi}_1(\tau_{2})}_{\rm c.q.}^-=\f{1}4\text{Res}{}_{0,\pm\lambda}\left[\f{e^{-im(\tau_{12}+\pi)}(\lambda\tan\lambda\pi/2-im)^{2}}{2m^{2}(1+e^{-im\pi})(m-\lambda)(m+\lambda)(m\tan m\pi/2-\lambda\tan\lambda\pi/2)}\right]
\end{equation}
for $\tau_{12}\in(-\pi,0)$, and
\begin{equation}
\avg{\hat{\chi}_1(\tau_{1})\hat{\chi}_1(\tau_{2})}_{\rm c.q.}^-=\f{1}4\text{Res}{}_{0,\pm\lambda}\left[\f{e^{-im(\tau_{12}-\pi)}(\lambda\tan\lambda\pi/2+im)^{2}}{2m^{2}(1+e^{-im\pi})(m-\lambda)(m+\lambda)(m\tan m\pi/2-\lambda\tan\lambda\pi/2)}\right]
\end{equation}
for $\tau_{12}\in(\pi,2\pi)$. 

Note that $\hat\chi_2$ is not completely independent from $\hat\chi_1$ due to (\ref{d10x}),
above formula indeed covers all correlations between $\hat\chi_1$ and $\hat\chi_2$. Explicitly, with translation symmetry we have 
\begin{align} 
\avg{\hat{\chi}_1(\tau)\hat{\chi}_1(0)}_{\rm c.q.}&=\avg{\hat{\chi}_2(\tau)\hat{\chi}_2(0)}_{\rm c.q.}=\avg{\hat{\chi}_1(\tau)\hat{\chi}_1(0)}_{\rm c.q.}^++\avg{\hat{\chi}_1(\tau)\hat{\chi}_1(0)}_{\rm c.q.}^- \\
\avg{\hat{\chi}_1(\tau)\hat{\chi}_2(0)}_{\rm c.q.}&=\avg{\hat{\chi}_1(\tau-\pi)\hat{\chi}_1(0)}_{\rm c.q.}^+-\avg{\hat{\chi}_1(\tau-\pi)\hat{\chi}_1(0)}_{\rm c.q.}^- \\
\avg{\hat{\chi}_2(\tau)\hat{\chi}_1(0)}_{\rm c.q.}&=\avg{\hat{\chi}_1(\tau+\pi)\hat{\chi}_1(0)}_{\rm c.q.}^+-\avg{\hat{\chi}_1(\tau+\pi)\hat{\chi}_1(0)}_{\rm c.q.}^-
\end{align}
Evaluation of residues is straightforward and leads to
\begin{align}
\avg{\hat{\chi}_1(\tau)\hat{\chi}_1(0)}_{\rm c.q.}=&\avg{\hat{\chi}_2(\tau)\hat{\chi}_2(0)}_{\rm c.q.}=\avg{\hat{\chi}_1(2\pi-\tau)\hat{\chi}_2(0)}_{\rm c.q.}=\avg{\hat{\chi}_2(-\tau)\hat{\chi}_1(0)}_{\rm c.q.} \label{4.80chi}\\
=&i\times\begin{cases}
\hat m_{+}e^{i\lam \tau}+\hat m_1 \tau e^{i\lam \tau}+c.c., &\tau\in[0,2\pi] \\
\hat m_{-}e^{i\lam \tau}+\hat m_1 \tau e^{i\lam \tau}+c.c., &\tau\in[-\pi,0]
\end{cases} \label{4.81chi}
\end{align}
where star means complex conjugate and
\begin{align}
\hat m_{+} & =i\f{2\pi\lambda-\sin2\pi\lambda+2(\pi\lambda+\sin\pi\lambda)(2\pi i\lambda+3-e^{-i\pi\lambda})}{32\lambda^{2}(\pi\lambda+\sin\pi\lambda)^{2}(1+e^{i\pi\lambda})},\\
\hat m_{-} & =\hat m_{+}+\f 1{4\lambda^{2}(1+e^{i\pi\lambda})},\quad \hat m_{1}=\f 1{8\lambda(\pi\lambda+\sin\pi\lambda)(1+e^{i\pi\lambda})}
\end{align}
As a consistent check, one can show that the correlator \eqref{4.80chi} obeys the KMS condition \eqref{kmschi1} and \eqref{kmschi2}. It is noteworthy that even if the computations for even and odd modes are different for the ranges $\tau\in[0,\pi]$ and $\tau\in[\pi,2\pi]$, we still have smoothness of $\avg{\hat{\chi}_1(\tau)\hat{\chi}_1(0)}$ at $\tau=\pi$ in \eqref{4.81chi}, which is consistent with item \ref{item:4b} of our effective theory.

Using \eqref{eq:4.41df} and restoring the $8/(N\D^2)$ factor in \eqref{4.80chi}, we can derive the two-point function of $\phi^E_i$  as
\begin{align}
\avg{\phi^E_i(\bar\tau;\tau)\phi^E_j(0;0)}&=\t(\bar\tau)M_{ij}(\tau)+\t(-\bar\tau)M_{ji}(-\tau)\label{phiEcor}\\
M_{11}(\tau)&=M_{22}(\tau)=M_{12}(2\pi-\tau)=M_{21}(-\tau) \label{4.42chi}\\
&=\f{8}{N\D^2}\times\begin{cases}
\hat m_{+}e^{i\lam \tau}+\hat m_1 \tau e^{i\lam \tau}+c.c., &\tau\in[0,2\pi] \\
\hat m_{-}e^{i\lam \tau}+\hat m_1 \tau e^{i\lam \tau}+c.c., &\tau\in[-\pi,0]
\end{cases} \label{4.43chi}
\end{align}
where the argument ranges in \eqref{phiEcor} are
\begin{align}  
i=j&:\{(\bar\tau,\tau)|\bar\tau\in[-\pi,\pi],\bar\tau-\tau\in[-\pi,\pi]\} \\
i=1,j=2&:\{(\bar\tau,\tau)|\bar\tau\in[-\pi,\pi],\bar\tau-\tau\in[-2\pi,0]\}  \\
i=2,j=1&:\{(\bar\tau,\tau)|\bar\tau\in[-\pi,\pi],\bar\tau-\tau\in[0,2\pi]\} 
\end{align}  

\subsection{The effective action} \label{app:d4}

The effective action will be still formulated in $\eta_{s,p}(\bar\tau;t)$ variables and the correlation functions of $\phi_i$ will be given by \eqref{eq:x29-2} to \eqref{eq:phi21}. Since the correlation function \eqref{4.43chi} contains linear-exponential terms, the effective action needs to contain quadratic factor of $\del_t^2-\lam^2$. It turns out that we need to take the action \eqref{leadS} with
\be 
K_{+,p}(i\del_t)=(\del_t^2-\lam^2)k_{+,p}(i\del_t),\quad K_{-,p}(i\del_t)=(\del_t^2-\lam^2)^2 k_{-,p}(i\del_t) \label{sykact}
\ee
where $K_{-,p}(x)$ has a double zero at $\pm i\lam$. By symmetry \eqref{Kp}, we have $k_{s,p}(x)=(-)^p k_{s,-p}(-x)$. This effective action has been thoroughly discussed in Appendix \ref{app:genact}.

To match the correlation functions \eqref{phiEcor} with this EFT action, let us first compute $\avg{\hat\mT\phi_i(\bar t;t)\phi_j(0;0)}$ correlation functions by \eqref{eq:x29-2}-\eqref{eq:phi21} with the most general consistent Wightman functions solved in Appendix \ref{wightmanf}, which are also results of the action \eqref{sykact} by the analysis in Appendix \ref{app:genact}. Then we compare $\avg{\hat\mT\phi_i(\bar t;t)\phi_j(0;0)}$ with \eqref{phiEcor} after analytic continuation. 
Since these correlation functions obey the same KMS conditions, matching one of them is sufficient. In the following, we will take $i=j=1$ and assume $\Im t\in [-2\pi,0]$.

The most general consistent Wightman function in Appendix \ref{wightmanf} takes the form \eqref{eq:nx26} and \eqref{hspB2} where $h_{+,p}$ only contains pure exponential terms and $h_{-,p}$ only contains up to linear-exponential terms. The coefficients $\g^k_{s,p}$ are constrained by \eqref{grl1}-\eqref{grl3}. Using these solutions in \eqref{eq:x29-2} we have 
\be   
\avg{\hat\mT \phi_1(\bar t;t)\phi_1(0;0)}=\begin{cases}
(m_{+}+m_1 t) e^{- \lam  t}+(\bar m_{+}+\bar m_1 t) e^{\lam  t}, &\Im \bar t\in[-\pi,0] \\
(m_{-}-\bar m_1 t) e^{- \lam  t}+(\bar m_{-}- m_1 t) e^{\lam  t}, &\Im \bar t\in[0,\pi]
\end{cases}
\ee
where $\Im \bar t-\Im t\in[0,\pi]$ and 
\begin{align}
\bar m_+&=\f {3((\g^0_{-,0}+\g^0_{+,0})(1-2\cos \pi \lam)-2(\pi\g^1_{-,0}+2i\g^0_{+,0})\sin \pi \lam)}{2(1-2\cos\pi\lam)^2} \\
m_+&=\f{3e^{-i\pi \lam}(2\g^0_{+,0}+i\pi \g^{1}_{-,0})}{2(1-2\cos \pi \lam)}-e^{-i\pi \lam}\bar m_+\\
m_-&=\bar m_+ - \f{6\g^0_{+,0}}{1-2\cos \pi \lam} \\
\bar m_-&=m_+ - \f{6e^{-i\pi \lam}\g^0_{+,0}}{1-2\cos \pi \lam} \\
\bar m_1&= e^{i \pi \lam} m_1= \f {3 \g^1_{-,0}}{2(1-2\cos \pi \lam)}
\end{align}
Comparing with \eqref{phiEcor}-\eqref{4.43chi} with analytic continuation  $\phi^E_j(\bar\tau;\tau)\ra \phi_j(\bar t;t)=-i\phi^E_j(i\bar t;it)$, in which $\t(\bar\tau)\ra\t(-\Im \bar t)$ and $\tau\ra i t$, we should identify
\begin{align}  
\g^1_{-,0}&=\f{2i(2\cos \pi \lam-1)}{3\lam N \D^2(\pi\lam+\sin \pi \lam)(1+e^{-i\pi \lam})} \\
\g^0_{+,0}&=\f{1-2\cos \pi \lam}{3\lam^2N \D^2(1+e^{-i\pi \lam})} \\
\g^0_{-,0}&=-\frac{(\pi  \lambda +4 i) \sin 2 \pi  \lambda -(\pi  \lambda +7 i) (\pi  \lambda +\sin \pi  \lambda ) +\pi  \lambda  (2 \pi  \lambda +9 i) \cos \pi  \lambda +i \pi  \lambda  \cos 2 \pi  \lambda }{3 \lambda ^2 N \D^2\left(1+e^{-i \pi  \lambda }\right)  (\pi  \lambda +\sin \pi  \lambda )^2}
\end{align}  

Moreover, in Appendix \ref{app:genact} we show that the most general consistent Wightman functions in Appendix \ref{wightmanf} can also be reproduced by the effective action \eqref{generalact}. Given the parameter relations \eqref{c25}-\eqref{c31}, we can derive
\begin{align} 
k_{+,0}(i\lam)&=\f{3\lam N \D^2\cot \f {\pi \lam}{2}}{2(1-2\cos\pi \lam)} \label{k0-app}\\
k_{-,0}(i\lam)&=\frac{3 N \D^2 (\pi  \lambda +\sin \pi  \lambda )}{8\lambda (2  \cos \pi  \lambda-1 )} \\
k'_{-,0}(i\lam)&=-\frac{3 i N \D^2\left(\left(2 \pi ^2 \lambda ^2-5\right) \sin \pi  \lambda +2 \sin 2 \pi  \lambda +8 \pi  \lambda  \cos \pi  \lambda -\pi  \lambda  \left(3 \pi  \lambda  \tan \frac{\pi  \lambda }{2}+7\right)\right)}{16 \lambda ^2 (1-2 \cos \pi  \lambda )^2} \label{km-app}
\end{align}
In conclusion, the action \eqref{leadS} with \eqref{sykact} and \eqref{k0-app}-\eqref{km-app} fully captures the correlation of the effective modes $\phi^E_i$ after analytic continuation.

\section{Solve $m=1$ vertex} \label{app:e}
In matrix form the equation \eqref{diffg} is
\begin{equation}
\begin{pmatrix}C_{0;0}^{1}g(q)+C_{1;0}^{1}g^{1}(q) & C_{0;1}^{1}g(q)+C_{1;1}^{1}g^{1}(q)\\
C_{0;0}^{2}g(q)+C_{1;0}^{2}g^{1}(q) & C_{0;1}^{2}g(q)+C_{1;1}^{2}g^{1}(q)
\end{pmatrix}=\f{\xi^{(1)}(q)}{q^{1/2}}\begin{pmatrix}-\lambda(1-cq) & 1+cq\\
\lambda(c-q) & c+q
\end{pmatrix}\label{eq:189}
\end{equation}
Note that each entry on RHS has zero at $\pm c$ or $\pm1/c$ respectively.
Taking $q=\pm c,\pm1/c$ on LHS, we can immediately derive
\begin{equation}
C_{1;0}^{1}=-\f{C_{0;0}^{1}g(1/c)}{g^{1}(1/c)},\quad C_{1;1}^{1}=-\f{C_{0;1}^{1}g(-1/c)}{g^{1}(-1/c)},\quad C_{1;0}^{2}=-\f{C_{0;0}^{2}g(c)}{g^{1}(c)},\quad C_{1;1}^{2}=-\f{C_{0,1}^{2}g(-c)}{g^{1}(-c)} \label{5.33x}
\end{equation}
Taking this back to (\ref{eq:189}), we have
\begin{align}
-C_{0;0}^{1}\f{g^{1}(1/c)g(q)-g(1/c)g^{1}(q)}{g^{1}(1/c)\lambda(1-cq)} & =C_{0;1}^{1}\f{g^{1}(-1/c)g(q)-g(-1/c)g^{1}(q)}{g^{1}(-1/c)(1+cq)}\nonumber \\
=C_{0;0}^{2}\f{g^{1}(c)g(q)-g(c)g^{1}(q)}{g^{1}(c)\lambda(c-q)} & =C_{0;1}^{2}\f{g^{1}(-c)g(q)-g(-c)g^{1}(q)}{g^{1}(-c)(c+q)}=\f{\xi^{(1)}(q)}{q^{1/2}}\label{eq:191}
\end{align}
Using KMS symmetry $g^{n}(q)=(-)^{n}g^{n}(1/q)$, comparing the first line and the second line of above equations lead to
\begin{equation}
\xi^{(1)}(q)/C_{0;0}^{1}=\xi^{(1)}(1/q)/C_{0;0}^{2},\quad \xi^{(1)}(q)/C_{0;1}^{2}=\xi^{(1)}(1/q)/C_{0;1}^{1} \label{5.35x}
\end{equation}
Then we just need to solve the first line of (\ref{eq:191}), which is  a first order differential equation
\begin{equation}
\f{(C_{0;0}^{1}+\lambda C_{0;1}^{1})+(C_{0;0}^{1}-\lambda C_{0;1}^{1})cq}{(C_{0;0}^{1}A+\lambda C_{0;1}^{1}B)+(C_{0;0}^{1}A-\lambda C_{0;1}^{1}B)cq}=\f{g^{1}(q)}{g(q)},\quad A=\f{g(1/c)}{g^{1}(1/c)},\quad B=\f{g(-1/c)}{g^{1}(-1/c)}\label{eq:193}
\end{equation}
The solution of above type equation is
\begin{equation}
\f{\a+\b q}{\g+\t q}=\f{g^{1}(q)}{g(q)}\implies g(q)=q^{-\a/(\g\lambda)}(\g+\t q)^{(\a\t-\b\g)/(\g\t\lambda)}
\end{equation}
Requiring KMS symmetry $g(q)=g(1/q)$ leads to
\begin{equation}
\g=\t,\quad\b=-\a,\quad g(q)=(q^{1/2}+q^{-1/2})^{-2\D},\quad\D=\b/(\g\lambda)\label{eq:195}
\end{equation}
Using this solution of $g(q)$, we have
\begin{equation}
\f{g(q)}{g^{1}(q)}=-\f{1+q}{\D\lambda(1-q)}\implies A=\f{1+c}{\D\lambda(1-c)},\quad B=\f{1-c}{\D\lambda(1+c)}
\end{equation}
Plugging this back to \eqref{eq:193} and using \eqref{eq:195}, we only find one independent relation $(1+c)C^{1}_{0;0}+\lam (1-c)C^1_{0;1}=0$. Taking this back to \eqref{5.33x} leads to 
\begin{align}
C_{0;0}^{1} =\f{\D\lambda(c-1)}{c+1},\quad C_{1;0}^{1}=1,\quad C_{0;1}^{1}=\D,\quad C_{1,1}^{1}=\f{c-1}{(c+1)\lambda}
\end{align}
where we have chosen the normalization $C_{1;0}^{1}=1$. Taking them back to \eqref{eq:189}, we can solve 
\be  
\xi^{(1)}(q)=\f{2\D g(q)}{(1+c)(q^{1/2}+q^{-1/2})}
\ee 
which is explicitly KMS invariant as expected. Taking this to \eqref{5.35x} and \eqref{5.33x}, we have
\be  
C_{0;0}^{2} =\f{\D\lambda(c-1)}{c+1},\quad C_{1;0}^{2}=-1,\quad C_{0;1}^{2}=\D,\quad C_{1;1}^{2}=-\f{c-1}{(c+1)\lambda}
\ee 
One can check that this solution also obeys KMS symmetry \eqref{3.2}.


\bibliographystyle{JHEP.bst}
\bibliography{main.bib}

\end{document}